\documentclass[aps,prd,showpacs,tightenlines,nofootinbib,floatfix,preprint,superscriptaddress]{revtex4-1}
\usepackage{graphicx}
\usepackage{amsmath,amssymb,mathtools,amsfonts}
\usepackage{hyperref}
\usepackage{microtype}
\usepackage{subfig}
\usepackage[font=small,skip=2pt]{caption}
\usepackage{multirow}
\usepackage{tikz}
\usepackage{float}
\usepackage[normalem]{ulem}
\usepackage{multirow}

\DeclareMathOperator{\udhh}{ud \bar{h}\bar{h}}
\DeclareMathOperator{\udbh}{ud \bar{b}\bar{h}}
\DeclareMathOperator{\lshh}{ls \bar{h}\bar{h}}
\DeclareMathOperator{\lsbh}{ls \bar{b}\bar{h}}
\DeclareMathOperator{\udbb}{ud \bar{b}\bar{b}}
\DeclareMathOperator{\lsbb}{ls \bar{b}\bar{b}}
\DeclareMathOperator{\udbc}{ud \bar{b}\bar{c}}
\DeclareMathOperator{\udcc}{ud \bar{c}\bar{c}}

\graphicspath{{./figs/}}

\newcommand\glasgow{SUPA, School of Physics and Astronomy,
University of Glasgow, Glasgow, G12 8QQ, UK}

\newcommand\york{Department of Physics and Astronomy, York University, Toronto, Ontario, M3J 1P3, Canada}

\newcommand\nycu{Institute of Physics, National Yang Ming Chiao Tung University, 30010 Hsinchu, Taiwan}

\newcommand\kmyork{Department of Mathematics and Statistics, York University, Toronto, Ontario M3J 1P3, Canada}

\newcommand\adelaide{CSSM, University of Adelaide, Adelaide, SA, 5005, Australia}
\begin{document}

\newcommand\darmstadt{GSI Helmholtzzentrum für Schwerionenforschung, 64291 Darmstadt, Germany}

\title{Improved analysis of strong-interaction-stable doubly-bottom tetraquarks on the lattice}

\author{B. Colquhoun}
\affiliation{\glasgow}
\author{A. Francis}
\affiliation{\nycu}
\author{R. J. Hudspith}
\affiliation{\darmstadt}
\author{R. Lewis}
\affiliation{\york}
\author{K. Maltman}
\affiliation{\kmyork}
\affiliation{\adelaide}
\author{W. G. Parrott}
\affiliation{\york}

\date{\today}

\begin{abstract}  
We update earlier lattice results for the 
binding energies of the flavor antitriplet of 
strong-interaction-stable doubly bottom, $J^P=1^+$
tetraquarks, employing an extended sink construction 
which produces significantly improved ground-state
effective-mass plateaus, as well as new, larger-volume
ensembles which reduce possible finite-volume effects 
at lighter pion masses. The updated bindings 
are $115(17)$ MeV for the $I=0$ member of the
antitriplet and $47(8)$ MeV for its $I=1/2$ partner.
We also provide an update of our earlier study of the 
variable heavy mass dependence of binding in the
$1^+$ channel and new results on this dependence for
binding in the $0^+$ channel, accessible when the 
two heavy quarks have unequal masses. Implications of
these results of potential relevance to experimental
searches for signals of the production of doubly bottom
tetraquarks and/or a possible bottom-charm partner of 
the $T_{cc}$ are also discussed.

\end{abstract}

\maketitle

\section{Introduction}\label{introduction}
It has long been known that 
strong-interaction-stable manifestly exotic 
doubly-antiheavy-doubly-light tetraquark states exist 
in QCD in the limit of arbitrarily large heavy quark
mass~\cite{Ader:1981db,Heller:1986bt,Manohar:1992nd}.
The source of this binding is the color Coulomb
interaction between heavy antiquarks in a color
$3_c$ configuration, which produces a binding 
contribution proportional to the reduced mass of 
the heavy antiquark pair. Additional attraction 
is present in channels where the accompanying 
light degrees of freedom (which necessarily have 
color $\bar{3}_c$,) are in the flavor
antisymmetric, color $\bar{3}_c$,
light-quark spin $J_l =0$ ``good light diquark'' 
configuration, known to be attractive from the
observed splittings in the heavy baryon spectrum.
Neither the heavy antidiquark color-Coulomb 
attraction nor the spin-dependent, 
good-light-diquark attraction are accessible to 
a system of two well-separated heavy mesons. 
The doubly heavy channels most favorable to
tetraquark binding are thus those with 
$J^P=1^+$ and either $I=0$ (light quark 
flavor $ud$) or $I=1/2$ (light quark flavor 
$ls$, with $l=u,\, d$) for the case of
identical heavy antiquarks, and $J^P=0^+$ or 
$1^+$ and either $I=0$ or $I=1/2$ for the case of
non-identical heavy antiquarks. The discussion
below focuses on these channels.

While bound doubly heavy tetraquark states 
necessarily exist for sufficiently large heavy 
antiquark mass,{\footnote{see \cite{Richards:1990xf,Mihaly:1996ue,Green:1998nt,Stewart:1998hk,Michael:1999nq,Pennanen:1999xi,Cook:2002am,Doi:2006kx,Detmold:2007wk} for (early) lattice studies in this direction.}} whether the physical $\bar{b}$ 
and $\bar{c}$ masses are large enough for 
this binding to be realized in nature is a
dynamical question. Since the heavy antidiquark
reduced mass, and hence the $3_c$ color Coulomb 
attraction, decreases with decreasing heavy quark 
mass, one expects such binding, if any, to be 
largest in doubly bottom channels. Numerous 
studies using QCD-inspired models predict binding
in the doubly bottom $I=0$, $J^P=1^+$ channel, 
with additional doubly heavy bound states
predicted in other channels in a number of these
works~\cite{Ader:1981db,Heller:1986bt,Carlson:1987hh,Zouzou:1986qh,Lipkin:1986dw,SilvestreBrac:1993ss,Semay:1994ht,Pepin:1996id,Brink:1998as,Vijande:2003ki,Janc:2004qn,Vijande:2007rf,Ebert:2007rn,Zhang:2007mu,Vijande:2009kj,Yang:2009zzp,Carames:2011zz,Silbar:2013dda,Karliner:2017qjm,Caramees:2018oue,Deng:2018kly,Park:2018wjk,Yang:2019itm,Hernandez:2019eox,Tan:2020ldi,Lu:2020rog,Cheng:2020wxa,Meng:2020knc,Noh:2021lqs}.
The possibility of bound, doubly heavy 
tetraquarks has also been investigated in 
heavy-quark-symmetry-based 
approaches~\cite{Manohar:1992nd,Ohkoda:2012hv,Czarnecki:2017vco,Eichten:2017ffp,Braaten:2020nwp}
as well as in implementations of the 
Borel-transformed QCD sum rule framework employing 
the approximate, 
single-narrow-resonance-plus-OPE-continuum SVZ
ansatz~\cite{Shifman:1978bx,Shifman:1978by} for the
relevant spectral functions~\cite{Navarra:2007yw,Du:2012wp,Chen:2013aba,Wang:2017uld,Agaev:2018khe,Agaev:2019kkz,Agaev:2019wkk,Tang:2019nwv,Wang:2020jgb,Agaev:2020zag}.
The sum rule analyses produce much larger errors 
for the ground state masses, with, in addition, 
results from different groups not in agreement, 
even within these larger errors, for a number of 
the channels. This situation is reviewed in more 
detail in Ref.~\cite{Hudspith:2020tdf}.

Recent lattice studies exist for all of the doubly
bottom, bottom-charm and doubly charm channels. All
analyses concur on the existence of deeply bound 
strong-interaction-stable doubly bottom, $I=0$, 
$J^P=1^+$~\cite{Bicudo:2015kna,Bicudo:2016ooe,Bicudo:2017szl,Francis:2016hui,Junnarkar:2018twb,Leskovec:2019ioa,Mohanta:2020eed,Hudspith:2020tdf,Pflaumer:2020ogv,Pflaumer:2021ong,Wagner:2022bff,Pflaumer:2022lgp,Colquhoun:2022dte,Colquhoun:2022sip,Aoki:2023nzp,Hudspith:2023loy,Mueller:2023wzd,Alexandrou:2024iwi}
and $I=1/2$, $J^P=1^+$~\cite{Francis:2016hui,Junnarkar:2018twb,Hudspith:2020tdf,Pflaumer:2020ogv,Pflaumer:2021ong,Colquhoun:2022dte,Colquhoun:2022sip,Mueller:2023wzd,Alexandrou:2024iwi}
states. The situation is less clear for the $I=0$
$J^P=0^+$ and $1^+$ bottom-charm channels.
Refs.~\cite{Hudspith:2020tdf,Pflaumer:2020ogv,Pflaumer:2021ong,Meinel:2022lzo,Colquhoun:2022dte,Colquhoun:2022sip} 
find finite-volume ground-state energies near or just 
above threshold and hence no deep binding, but are 
unable to rule out the possibility that these 
correspond to weakly bound states in the continuum.  
Ref.~\cite{Mathur:2021gqn}, in contrast, found a
preliminary result of $\sim 20-40$ MeV for binding 
in the $1^+$ channel, though with the caveat that 
a finite-volume study would be required to determine 
whether this represented a true continuum bound state 
signal. A L\"uscher scattering analysis employing 
only ground-state energies as input and ensembles 
with partially quenched valence 
$m_\pi\gtrsim 500$ MeV~\cite{Padmanath:2023rdu}, also
reports an extrapolated physical-point $B^*\bar{D}$ 
scattering length corresponding to a $1^+$ binding of
$43\left({}^{+7}_{-6}\right) \left({}^{+24}_{-14}\right)$ MeV. 
The analogous study of the $0^+$ channel
by the same authors~\cite{Radhakrishnan:2024ihu}, 
found a physical-point $B\bar{D}$ scattering 
length corresponding to 
$39\left({}^{+6}_{-4}\right) \left({}^{+18}_{-8}\right)$ MeV $0^+$ binding.
A more recent L\"uscher scattering 
analysis~\cite{Alexandrou:2023cqg}, however, with 
significantly lighter $m_\pi\simeq 220$ MeV and larger
operator basis (including operators designed to have 
improved overlap with continuum meson-meson scattering 
states), on two $a\simeq 0.12$ fm ensembles with
different volumes, finds instead very weakly bound 
genuine bound states (though compatible within 
statistical errors with being virtual bound states)
$2.4(2.9)$ and $0.5(9)$ MeV, respectively, below 
$B^*\bar{D}$, respectively $B\bar{D}$, threshold in 
the $I=0$, $J^P=1^+$ and $0^+$ channels.

In the doubly charmed sector, the 2021 LHCb discovery
of the $T_{cc}$~\cite{LHCb:2021vvq,LHCb:2021auc} 
established the existence of a very weakly bound 
doubly charmed $I=0$ state lying essentially right at 
$DD^*$ threshold. This observation was in keeping 
with expectations from then-recent lattice studies, 
with Ref.~\cite{Cheung:2017tnt} providing compelling
evidence against the existence of a doubly charmed
$I=0$ $J^P=1^+$ state with significant binding at 
the heavier-than-physical pion mass $m_\pi = 391$ 
MeV, and the results of the variable-$b$-mass study 
of Ref.~\cite{Francis:2018jyb}, in combination 
with the negative bottom-charm results of
Ref.~\cite{Hudspith:2020tdf}, disfavoring the
possibility of significant binding at physical $m_\pi$.
The only lattice study leaving open the possibility of 
a moderately bound doubly charmed $I=0$, $J^P=1^+$ 
tetraquark had been that of
Ref.~\cite{Junnarkar:2018twb},
which reached $m_\pi$ as low as $257$ MeV for the 
coarsest of the three ensembles considered, and found 
a $J^P=1^+$, $I=0$ ground state lying $23\pm 11$
MeV below $DD^*$ threshold, though with no additional
state (such as might correspond to the continuum
$DD^*$ threshold scattering state) nearby. As the 
authors noted, given the proximity of the two-meson 
threshold, a dedicated finite-volume study would have 
been required to clarify whether this result reflected
the existence of an actual weakly bound ground
state or, instead, corresponded to a continuum 
scattering state shifted below threshold by finite
volume effects. The mass of the doubly charmed, 
strange, $I=1/2$, $J^P=1^+$ ground state in this 
study was also found to correspond within errors 
to that of the lowest-lying two-meson threshold, 
$D^*D_s$~\cite{Junnarkar:2018twb}. With the
weakness of binding in the $T_{cc}$ channel
established experimentally, lattice efforts in this
channel have turned to dedicated finite-volume 
scattering studies~\cite{Padmanath:2022cvl,Chen:2022vpo,Lyu:2023xro,Ortiz-Pacheco:2023ble,Collins:2024sfi,Whyte:2024ihh}.
These studies, mostly at heavier-than-physical
$m_\pi$, clearly establish the attractive nature
of the $I=0$ $J^P=1^+$ $DD^*$ interaction and
agree on the existence of genuine or virtual 
$DD^*$ bound states just below $DD^*$ threshold
for the $m_\pi$ considered, in semi-quantitative 
agreement with LHCb results. Additional studies
at lighter $m_\pi$ and on finer lattices are
expected in the near future. 

As discussed in Ref.~\cite{Hudspith:2020tdf}, 
existing lattice results rule out all 
implementations of the chiral quark model 
framework we are aware of in the literature.
Predictions from a number of non-chiral
models\cite{Carlson:1987hh,Zouzou:1986qh,SilvestreBrac:1993ss,Semay:1994ht,Brink:1998as,Janc:2004qn,Vijande:2007rf,Ebert:2007rn,Vijande:2009kj,Yang:2009zzp,Karliner:2017qjm,Park:2018wjk,Hernandez:2019eox,Cheng:2020wxa,Meng:2020knc,Noh:2021lqs},
in contrast, are in reasonable agreement with 
existing lattice results, and would require 
improved lattice determinations to test them 
further. The heavy-quark-symmetry analyses of
Refs.~\cite{Eichten:2017ffp,Braaten:2020nwp},
which rely on phenomenological input for leading
finite heavy mass corrections, including
model~\cite{Eichten:2017ffp}, respectively
lattice~\cite{Braaten:2020nwp}, results for doubly
heavy baryon masses not yet measured 
experimentally, also predict binding only in the 
two doubly bottom channels, with no binding of 
bottom-charm or doubly charmed states.

While recent lattice studies concur on the 
existence of an $SU(3)_F$ antitriplet of bound, 
strong-interaction-stable, doubly bottom $J^P=1^+$
tetraquarks, results for the 
physical-point-extrapolated
binding energies show some disagreements. In the
$I=0$ channel, two wall-source, local-sink studies 
from 2016 and 2018, Ref.~\cite{Francis:2016hui}
and \cite{Junnarkar:2018twb}, quote $189\pm 10$ MeV 
and $143\pm 34$ MeV, respectively, compatible with the
result of the later study, Ref.~\cite{Mohanta:2020eed},
which found $189(18)$ MeV from the finest ensemble and
$167(19)$ MeV averaged over all three ensembles 
considered. These results, however, are significantly 
higher than those of other recent lattice studies, with
Ref.~\cite{Leskovec:2019ioa} reporting $128\pm 26$ MeV,
the preliminary one-ensemble $m_\pi\simeq 192$ MeV 
study of Ref.~\cite{Hudspith:2020tdf} $\sim 113$ MeV,
Refs.~\cite{Pflaumer:2022lgp,Wagner:2022bff} $103(8)$
MeV, the HAL QCD study of Ref.~\cite{Aoki:2022xxq} 
$83(10)(20)$ MeV, Ref.~\cite{Hudspith:2023loy} 
$112(13)$ MeV and Ref.~\cite{Alexandrou:2024iwi}
$100(10)\left( {}^{+43}_{-30}\right)$ MeV. The situation is
similar in the $I=1/2$ channel where binding energies
$46(12)$ MeV and $30(3)\left({}^{+31}_{-11}\right)$ MeV, 
found in the two most recent 
studies~\cite{Hudspith:2023loy} and 
\cite{Alexandrou:2024iwi}, are significantly lower 
than those of other recent lattice studies, which
produced results $98(7)$ MeV~\cite{Francis:2016hui}, 
$87(32)$ MeV~\cite{Junnarkar:2018twb}, a preliminary 
result $\sim 100(40)$ MeV~\cite{Pflaumer:2020ogv}, and
$86(22)(10)$ MeV~\cite{Pflaumer:2021ong,Meinel:2022lzo}.
A preliminary result, $\sim 36$ MeV, also favoring a 
lower value for the binding was also obtained in the 
multi-channel, but single-ensemble ($m_\pi\simeq 192$
MeV) study of Ref.~\cite{Hudspith:2020tdf}.

The most deeply bound of the lattice results for 
the binding in the doubly bottom, $J^P=1^+$, $I=0$
and $I=1/2$ channels, those of 
Ref.~\cite{Francis:2016hui}, came 
from an analysis using gauge-fixed wall sources and 
local sinks. Effective masses in such ``wall-local''
analyses typically approach the asymptotic ground 
state plateau from below, signalling negative 
excited state contamination. The fact that the 
ground state plateaus in this analysis were short, 
and reached at later Euclidean times, raised the 
possibility that the ground state masses were
underestimated, and hence the associated bindings
overestimated. This possibility was investigated in
Ref.~\cite{Hudspith:2020tdf}, at a single $m_\pi$
($\simeq 191$ MeV), using an improved ``Box-Sink''
construction designed to reduce excited-state
contamination. The construction was shown to extend
the ground state effective mass plateaus to much
earlier Euclidean times and to significantly 
improve the ground state mass determinations. 
The results of that study confirmed that, at the 
single $m_\pi$ considered, the $I=0$ and $I=1/2$
binding energies were indeed overestimated in 
Ref.~\cite{Francis:2016hui}. A similar conclusion
was reached in Ref.~\cite{Hudspith:2023loy}, also
employing improved sink-side operators, designed to 
have better ground state overlap. For the $I=0$ 
channel, this conclusion is in keeping with the 
result obtained in Ref.~\cite{Leskovec:2019ioa}, 
which included two non-local ``scattering operators'' 
(products of two heavy-meson-like operators each 
projected to zero spatial momentum), though on the
source side only, to avoid having to compute 
all-to-all propagators. These operators were expected
to have improved overlap with the threshold $BB^*$
scattering state (presumably the first excited state 
in the channel), and hence, as a result, to also 
improve the ground-state signal. These expectations 
were born out by the pattern of relative overlaps of
the source operators with the lowest few GEVP 
Eigenstates. Refs.~\cite{Pflaumer:2022lgp,Wagner:2022bff,Alexandrou:2024iwi} 
took further advantage of this improvement by
including the non-local scattering operators at 
the sink as well as at the source, producing a 
reduced error on the ground-state $I=0$ binding
energy. In this paper, we follow the alternate 
strategy of employing sink-side operators designed
specifically to improve overlap with the ground 
state, extending the improved box-sink analysis 
of Ref.~\cite{Hudspith:2020tdf} to ensembles with a
range of $m_\pi$, allowing an extrapolation of the
results for binding in the doubly bottom, $J^P=1^+$,
$I=0$ and $I=1/2$ channels to physical $m_\pi$. 
Preliminary versions of this analysis were reported
in Refs.~\cite{Colquhoun:2022dte,Colquhoun:2022sip}.
We also revisit the variable heavy mass study of 
Ref.~\cite{Francis:2018jyb} using the improved 
box-sink construction, and consider implications of the
updated results for possible tetraquark binding in 
the $J^P=1^+$ bottom-charm channel.

In light of the above discussion, we 
focus on doubly heavy states containing two heavy
antiquarks, $\bar{Q}$ and $\bar{Q}^\prime$, and 
two light valence quarks, $q$ and $q^\prime$, 
with $Q,\, Q^\prime$ either $b$ or a variable-mass
version thereof (denoted $h$), and $qq^\prime$
either $ud$ or $ls$, with $l=u,\, d$. The compressed 
flavor notations $\udbb$, $\udbh$, $\udhh$, $\lsbb$, 
$\lsbh$ and $\lshh$ will be used throughout to refer 
to such states.

The rest of this paper is organized as follows. In
Sec.~\ref{improved} we detail improvements to our
earlier analysis, Ref.~\cite{Francis:2016hui},
implemented in the current work, outlining, in
particular, the box-sink construction introduced
in Ref.~\cite{Hudspith:2020tdf}. 
Sec.~\ref{latticedetails} provides details of our 
lattice simulations, including information on new,
larger-volume ensembles. These ensembles allow us
to reach near-physical $m_\pi$ while keeping 
$m_\pi L>3.6$. 
This provides better 
control over possible finite volume effects than was 
the case in Ref.~\cite{Francis:2016hui}, where
the ensemble with the lightest $m_\pi$ had
$m_\pi L=2.4$. In Sec.~\ref{results}, we present
the results of these improved analyses for the 
binding energies of the doubly bottom, $J^P=1^+$, 
$I=0$ and $I=1/2$ states together with an updated 
version of our earlier variable heavy mass study. A 
summary of these results, together with a discussion 
of possible implications for the $J^P=1^+$ $\udbc$
channel, is provided in Sec.~\ref{discussion}.
\section{Improvements to the previous analysis}\label{improved}

\subsection{The box sink construction}
As in our previous study of the doubly bottom
$J^P=1^+$ channels~\cite{Francis:2016hui}, 
we work throughout with Coulomb 
gauge-fixed wall sources. The gauge condition has 
been implemented to high
precision using the Fourier-accelerated
conjugate gradient algorithm of
Ref.~\cite{Hudspith:2014oja} and applied 
over the full lattice volume. In order 
to avoid the larger errors noisier 
wall-wall correlators would have produced,
our previous study, Ref.~\cite{Francis:2016hui},
employed local, rather than wall, sinks.

However, while source-sink symmetry ensures
that excited state contributions enter
wall-wall correlators with positive 
weights, and hence that ground state  
effective mass plateaus are approached 
from above, the effective masses for 
wall-local correlators typically
approach their ground state plateaus from
below, signalling negative amplitudes for
at least the first excited state. When, as
found in Ref.~\cite{Francis:2016hui}, 
ground state effective mass plateaus are 
short, and reached only at later Euclidean 
times, $t$, the possibility that the 
ground state signal has not yet fully
plateaued leaves open the possibility 
that the ground state mass has been 
underestimated, and hence that,
for channels in which the ground state is
bound, the ground-state binding energy
has been overestimated.

In Ref.~\cite{Hudspith:2020tdf} we addressed
this issue by introducing the ``box sink''
construction. In contrast to a wall-wall
correlator, which is constructed using propagators,
$S^W(t)$, obtained by summing over the spatial
sites of the sink time slice,
\begin{equation}
S^W(t)=\sum_x S(x,t)
\end{equation}
a ``wall-box'' correlator is constructed
using propagators obtained by restricting
the sum to points lying within a specified
radius, $R$, of the reference sink point,
$x$. The resulting propagators have the form
\begin{equation}\label{eq:box_sink}
S^{B;R}(x,t)={\frac{1}{N}}\sum_{r^2\leq R^2}
S(x+r,t)\, ,
\end{equation}
and, as $R^2$ is varied between the minimum
and maximum values, $0$ and $3L^2/4$, produce
wall-box correlators which interpolate
continuously between the wall-local and wall-wall
limits~\cite{Hudspith:2020tdf}. The construction is
predicated on the expectation that a range
of intermediate $R$ should exist for which
excited-state contamination will be
intermediate between the positive
contamination of wall-wall correlators
and negative contamination of wall-local
correlators. Excited-state contamination
should then be small, producing ground-state
effective-mass plateaus that extend to
considerably lower Euclidean $t$.
One would expect a box-sink radius of
roughly physical ground-state hadron size
to optimize this improvement. Both
this expectation, and the existence of
significantly extended ground-state
effective-mass plateaus, were confirmed
in the single-ensemble study of
Ref.~\cite{Hudspith:2020tdf}. The connection of this 
method to sink smearing was further elucidated in \cite{Hudspith:2021iqu}.
\subsection{Increased number of ensembles}
Another improvement to our earlier analysis is the inclusion of a larger number of ensembles, with different pion masses, including two newly generated ensembles.

In our initial study,
Ref.~\cite{Francis:2016hui}, we employed
three $a\simeq 0.09$ fm, $32^3\times 64$
Wilson-clover ensembles, generated by
the PACS-CS collaboration, having pion
masses $\sim 164$, $299$ and $416$
MeV~\cite{Aoki:2008sm,Aoki:2009ix}.
$m_\pi L$ is $>4$ for the latter two
ensembles, but uncomfortably low
($\sim 2.4$) for the ensemble with
the lightest pion mass.
To ensure that
potential finite-volume effects are
under better control at lighter pion
mass, and to improve the reliability of
the extrapolation to physical $m_\pi$,
we have generated additional, larger-volume,
$48^3\times 64$ ensembles at the same
lattice spacing, with pion masses $\sim 191$
and $\sim 165$ MeV corresponding to
$m_\pi L =4.2$ and $m_{\pi}L=3.6$, respectively.
To more fully map out the light-quark
mass dependence of the doubly bottom
channel binding energies, we have also
carried out box-sink analyses of two
additional PACS-CS ensembles with
heavier pion masses, $\sim 575$
and $707$ MeV, in addition to the
box-sink updates of our earlier analyses
of the three previously studied
$32^3\times 64$ PACS-CS ensembles. Details
of all ensembles used are given in the next Section~\ref{sec:opsandcorrs},
and they are listed in Table~\ref{ensembledetails}. 

\subsection{Operator basis and multi-exponential fits}

Finally, to help better determine the
ground-state signal, we have expanded
our operator bases beyond the two-operator basis used in~\cite{Francis:2016hui}, and hence the number 
of correlators entering our analyses. We also apply the box-sink analysis described above (Eq.~\eqref{eq:box_sink}) to all operators, leading to an increased operator basis, as for each operator, we simultaneously fit data with several box sink radii. Further details of the operator bases used in the various channels are discussed in Appendix~\ref{app:op_basis}.

We also perform multi-exponential fits in place of the GEVP previously used, which will be explained in detail in Section~\ref{corrfit}. This is not necessarily an improvement in itself, but rather a different approach to the problem, as the GEVP and multi-exponential fits are both equally valid methods to extract energies from correlator data. One potential advantage of the multi-exponential fit is that it does not require a complete matrix of all source sink combinations, however, in our case we include all combinations anyway. Another is that for a tower of very finely spaced excited states, the fit can more easily approximate the areas of spectral density over a number of states. Pion masses in Table~\ref{ensembledetails} are also obtained from multi-exponential fits, agreeing well with previous values obtained on these ensembles~\cite{PACS-CS:2008bkb}.
 
\section{Details of the lattice simulations}\label{latticedetails}
\subsection{Ensembles and correlator calculation}\label{sec:opsandcorrs}
\begin{table}[!b]
  \begin{tabular}{cccccc}
    \hline
    \hline
    Label &$V$ & $\kappa_l$ & $N_{\mathrm{conf}}\times N_{\mathrm{src}}$ &$am_{\pi}$& $m_{\pi} L$ \\
    \hline
    E1 & $32^3\times 64$ & $0.13700$ &  $399\times4$ &0.32205(18)& 10.3 \\
    E2 & $32^3\times 64$ & $0.13727$ &  $400\times4$ &0.26193(19)& 8.4 \\
    E3 & $32^3\times 64$ & $0.13754$ &  $400\times4$ &0.18960(29)& 6.1 \\
    E5 & $32^3\times 64$ & $0.13770$ &  $800\times4$ &0.13622(27)& 4.4 \\
    E7 & $48^3\times 64$ & $0.13777$ &  $94\times8$ &0.08719(47)& 4.2 \\
    E9 & $48^3\times 64$ & $0.13779$ &  $88\times4$  &0.07536(58)& 3.6 \\
    \hline
    \hline
  \end{tabular}
  \caption{Details of the lattice 
  ensembles used in this study. The ensemble labels
  and volume, $V$ (in lattice units), are given, as
  well as $\kappa_{l}$ values, 
  the statistics in terms of configurations and 
  sources per configuration, the pion mass in lattice 
  units ($am_{\pi}$) and the $am_{\pi}L$, where $L$ 
  is the spatial extent in lattice units. In all cases $\kappa^{\mathrm{sea}}_s=0.13640$ and $\kappa^{\mathrm{valence}}_s=0.13666$~\cite{Lang:2014yfa}. Inverse
  lattice spacing in each case has been determined to 
  be $a^{-1}=2.194(10)\;\mathrm{GeV}$ from the 
  $\Omega$ mass at the physical 
  point~\cite{PACS-CS:2011ngu}.}
  \label{ensembledetails}
\end{table}

As noted above, our calculations are performed on 
nonperturbatively-improved Wilson-Clover 
PACS-CS ensembles~\cite{Aoki:2009ix}, and additional 
ensembles generated as an extension to this set. 
All ensembles were generated with the same gauge 
action, $\beta$ and $c_{\mathrm{SW}}$, and we 
take the inverse lattice spacing to be 
$a^{-1}=2.194(10)~\mathrm{GeV}$ as determined 
from the $\Omega$ mass at the physical
point~\cite{PACS-CS:2011ngu}. 

The configurations were generated using the HMC algorithm implemented in openQCD version 1.6, see \cite{Luscher:2012av,openQCD} and references quoted therein. This package includes a number of beneficial techniques: such as the deflated SAP solver \cite{Luscher:2007se}, mass-preconditioning \cite{Hasenbusch:2002ai}, chronological inversions \cite{Brower:1995vx}, and multiple time-scale integrators. 
In this work we have extended the existing set of PACS-CS 
ensembles by generating configurations with the same 
action parameters albeit with  larger spatial volumes, 
$L/a=32\rightarrow 48$, as well as new light quark 
masses. For consistency with PACS-CS the strange quark
mass parameter is kept fixed at its PACS-CS value, 
$\kappa_s=0.13640$, while adding two additional 
light quark values, $\kappa_l=0.13777$ (E7) and $0.13779$ (E9). 
The corresponding pion masses and $m_\pi L$ are listed 
in Table~\ref{ensembledetails}. 
There are some algorithmic differences worth noting
in our ensemble generation, compared to that of PACS-CS: 
For the strange quark, the RHMC \cite{Clark:2003na} was 
used, as opposed to the PHMC \cite{Frezzotti:1998eu}. In 
this work, the number of poles in the rational
approximation was kept large to ensure a stable 
generation process and allow for the corresponding 
reweighting factors to be dropped for the level of 
accuracy needed. Additionally a Hasenbusch 
preconditioning chain \cite{Hasenbusch:2002ai} with four 
intermediate masses $\mu_i$ was used for the light 
quarks. This is similar to the PACS-CS setup, however, 
the $\mu_i$ are different and the last is set to zero 
$\mu_0=0$, therefore no light quark reweighting is required.
Measurements were performed every 16 MDU for E7 and every 8 MDU for E9.
For the light quark propagator inversions the deflated SAP solver supplied in the same software package \cite{openQCD} was used. 
For the heavy quarks we used the same 
tadpole-improved $O(v^4)$ prescription detailed 
in the Appendices of \cite{Hudspith:2020tdf}. 

We will use the following meson operators  
in determining our heavy-light meson masses:
\begin{equation}
\mathcal{O}_H = (\bar{h}(x)\gamma_5 q(x) ), \quad \mathcal{O}_{H^*}=(\bar{h}(x)\gamma_i q(x)),
\end{equation}
where $q$ will be either an $l$ or $s$ quark. As 
we treat the source (wall) and sink (box) differently
this will naturally need to be incorporated in our 
notation.

For the $I=0$, $J^P=1^+$, $T_{\udbb}$ and $T_{\lsbb}$
tetraquarks we use the same set of local operators 
as in Ref.~\cite{Hudspith:2020tdf}, listed in Tab.~I
of that reference. 

This same set (with $b$ replaced by $h$) is used
for the local operators of the 
double-variable-heavy-mass $T_{\udhh}$ and $T_{\lshh}$
channels. The operators for the $I(J^P)=0(1^+)/0(0^+)$
and $\frac{1}{2}(1^+)/\frac{1}{2}(0^+)$ channels
having one physical-b-mass and one variable-heavy-mass 
antiquark, $T_{\udbh}$ and $T_{\lsbh}$, are obtained 
from the ``$udcb$" and ``$uscb$" rows of the same
table by replacing the ``$c$" (which in 
Ref.~\cite{Hudspith:2020tdf} denoted a relativistic
charm antiquark) with $h$ (denoting here a
NRQCD heavy antiquark whose mass we allow to vary
over a range including values both heavier and 
lighter than physical $m_b$).\footnote{With NRQCD 
used for the heavy quarks, the variable heavy 
mass, $m_h$, must, of course, remain in the region of
validity of the use of NRQCD. The range thus does not 
extend down to $m_h$ as low as physical $m_c$.
} Appendix~\ref{app:op_basis} lists the operators used in our case explicitly.
For all ensembles except E7 we have, for each
sink operator, four box radii squared\footnote{Owing to limited resources, for E7 we only have two box radii squared.}. In lattice
units, these are in the range 30-49 for the 
tetraquarks, and 16-49 for the mesons (for which we 
will only use one source and sink operator per box 
radius). Our strategy is to 
use multiple sinks in a simultaneous fit to 
reliably extract the ground- and excited-state energies, as will be detailed below.

\subsection{Correlator fitting}\label{corrfit}
We use a standard Bayesian fitting approach, as 
outlined in~\cite{Lepage:2001ym}, to simultaneously 
fit meson and tetraquark two point correlation 
functions to a multi-exponential fit form,
\begin{equation}\label{C2fitform}
  C_2^{\mathcal{O}_{\mathrm{src}}^i\mathcal{O}_{\mathrm{snk}}^j}(t) = \sum_{n=0}^{N}a_n^{\mathrm{src}_i}a_n^{\mathrm{snk}_j}(e^{-E_nt} \pm e^{-E_n(T-t)}),
\end{equation}
obtaining correlated values for the ground state 
energies $E_0$, where $T$ is the length of the 
(periodic) lattice in the time dimension. Our 
data contains different source and sink
combinations, (see Sec.~\ref{sec:opsandcorrs}), and 
these translate into different values for the 
amplitudes $a_n^{\mathrm{src}/\mathrm{snk}}$, with 
a common set of energy levels $E_n$. 

The fits are performed using the 
\textit{corrfitter}, \textit{lsqfit} and
\textit{gvar} python packages~\cite{peter_lepage_2021_5733391,peter_lepage_2023_7931361,peter_lepage_2023_8025535}. 
The software calculates two measures of the 
goodness of fit, the $\chi^2$ per degree of 
freedom (d.o.f.), and the log of the Gaussian 
Bayes Factor, log(GBF). A good fit should have a 
$\chi^2$/d.o.f. close to unity. The log(GBF) gives 
a measure of the likelihood that the fitting ansatz
gave rise to the results. This measure is useful in
relative comparisons of different fit ansatzes, and
penalises over-fitting. 

As described in Appendix D 
of~\cite{Dowdall:2019bea}, unless statistics are many 
times larger than the number of data points, a 
Singular Value Decomposition (SVD) cut is required 
in order to faithfully represent the uncertainties 
in fit results. In this analysis, our statistics are 
limited and so we use an SVD cut in our correlator 
fits. An SVD cut is a conservative move, 
increasing uncertainty in the final result to
reflect poorly determined eigenvalues in the 
covariance matrix. The required cut, which we 
shall always use, is calculated by 
\textit{gvar}~\cite{peter_lepage_2023_8025535},
using the method described 
in~\cite{Dowdall:2019bea}. A downside of applying
an SVD cut is that it artificially reduces the 
$\chi^2$ of the fit, so SVD 
noise~\cite{Dowdall:2019bea} is added as a 
final test to ensure that the unbiased 
$\chi^2$/d.o.f. is acceptable.  

\subsubsection{Variable fit parameters}\label{fitparams}
There are several input parameters which we 
choose in our fits. The first is the 
number of exponentials, $N$, appearing 
in Eq.~\eqref{C2fitform}, which may be 
different for the mesons and tetraquarks 
($N^{\mathrm{mes}}$,$N^{\mathrm{tet}}$). 
Next, once the correlators have been 
folded (they are either periodic or 
anti-periodic), we can discard 
data at early and late times by
choosing $t_{\mathrm{min}}$ and 
$t_{\mathrm{max}}$ and working in the
window $t_{\mathrm{min}}\leq t\leq
t_{\mathrm{max}}$, where, again,
$t_{\mathrm{min}}$ and $t_{\mathrm{max}}$
can be different for mesons and 
tetraquarks. The other fit parameters we 
must choose are the prior values 
$\mathcal{P}['parameter']$ which must 
be provided for each $a_n$ and $E_n$ in 
Eq.~\eqref{C2fitform}. We estimate the
ground state energy prior, $\mathcal{P}
[E_0]$, using an effective mass calculation
\begin{equation}\label{Meff}
    M_{\mathrm{eff}} = \lim_{t\to\infty}\cosh^{-1}\Bigg(\frac{C_2(t-1)+C_2(t+1)}{2C_2(t)}\Bigg),
\end{equation}
and vary the uncertainty given to 
$\mathcal{P}[E_0]$ in the fit. Similarly, for the effective amplitude, we have
\begin{equation}\label{Aeff}
    (a_0^{\mathrm{src}}a_0^{\mathrm{snk}})_{\mathrm{eff}} = \lim_{t\to\infty}\frac{C_2(t)}{e^{-M_{\mathrm{eff}}t}},
\end{equation}
where we can set all $n$ $a_n^{\mathrm{src}_0}$ 
parameters in Eq.~\eqref{C2fitform} to $=1$ for any 
one of the source operators (in this case operator 
$\mathcal{O}_{\mathrm{src}}^0$) without loss of 
generality. We are left with the parameters
$\mathcal{P}[a^{\mathrm{snk}_j}_0]$, (for sink 
operators $\mathcal{O}_{\mathrm{snk}}^j$)
again with a variable uncertainty. We can then use 
these to find effective amplitudes for the other 
source operators, which we have not set to 1. 
The priors for excited state amplitudes 
$\mathcal{P}[a_{n\neq0}]$ (we drop the `src' and 
`snk' label from now on) are another choice varied 
in our fit.

For the meson excited state energy splitting priors, 
with excited state splittings expected to be 
of order $\Lambda_{\mathrm{QCD}}$, we take for
$\mathcal{P}[E^{\mathrm{mes}}_{n\neq0}]$
the range $500\pm250\mathrm{\ MeV}$. Since, in 
contrast, tetraquark/two-heavy-meson excited states
are expected to be very closely spaced, we take 
as our prior for these splittings the value 
$30\pm30\mathrm{\ MeV}$. In both cases logarithmic
priors are used to ensure the splittings are
positive. This means that the 99.7\% confidence interval 
for the meson splitting prior is between 100 MeV 
and 2 GeV, while that for the tetraquark is 
between 1.5 and 600 MeV.    

\subsubsection{Binning}\label{sec:binsize} 
In each of our ensembles, we have correlator data 
on $N_{\mathrm{cfg}}$ uncorrelated configurations 
(see Table~\ref{ensembledetails}). On each 
configuration, we take 4 (8 for E7) evenly spaced
source times, $t_0$, for the correlators. 
Furthermore, for the vector meson and {\bf{$1^+$}} 
tetraquarks, all three spatial components are
calculated. For the pseudoscalar meson there is 
of course only one component. For the purposes of 
this discussion, let's assume this component is 
copied three times for each $t_0$ on each
configuration, such that for each correlator,
on each configuration, there are 12 (24 for E7) 
measurements, which are not necessarily uncorrelated.

To obtain averaged data and a full covariance 
matrix from $N_s$ samples, we use 
\textit{gvar}~\cite{peter_lepage_2023_8025535}, 
which calculates the covariance matrix, and 
required SVD cut, as per appendix D of 
Ref.~\cite{Dowdall:2019bea}. This method relies 
upon the different samples being uncorrelated in 
order to give a true measure of the uncertainties. 
For this reason, we would typically average (bin) 
the 12 measurements for each configuration above 
(which could be correlated), before calculating 
the covariance matrix with the resulting 
$N_s=N_{\mathrm{cfg}}$ samples. However, if we 
can be sure that these 12 measurements are only
weakly correlated, we can treat them as independent
samples. This effective increase in statistics will
not change the central value or uncertainties on 
individual correlators, but should lead to a better 
determination of the covariance matrix, and thus a 
lower SVD cut, ultimately leading to smaller 
uncertainties in final fit parameters. To test 
what size of bin is permissible, we bin the data 
in sets of size 1, 3, 6, 12, and 24, with 
12 being the default\footnote{On E7, there are twice as many $t_0$ values, so 24 is the default.}, and 24 produced by combining
the results from pairs of adjacent configurations.
For each bin size, we average the data and generate 
the covariance matrix. Then we take each timeslice 
of each correlator and compare the relative change 
in the uncertainties from the default case (bin 
size 12), which we know is uncorrelated. If the 
samples in the new binning regime 
are also uncorrelated, we would expect the relative 
change in uncertainty to form a Gaussian 
distribution, with a mean of zero. In the case 
that they are correlated, we would expect the 
average standard deviation to be significantly 
different from the default (uncorrelated) binning 
for different bin sizes.     
\begin{figure}
    \centering
    \includegraphics[width=0.45\textwidth]{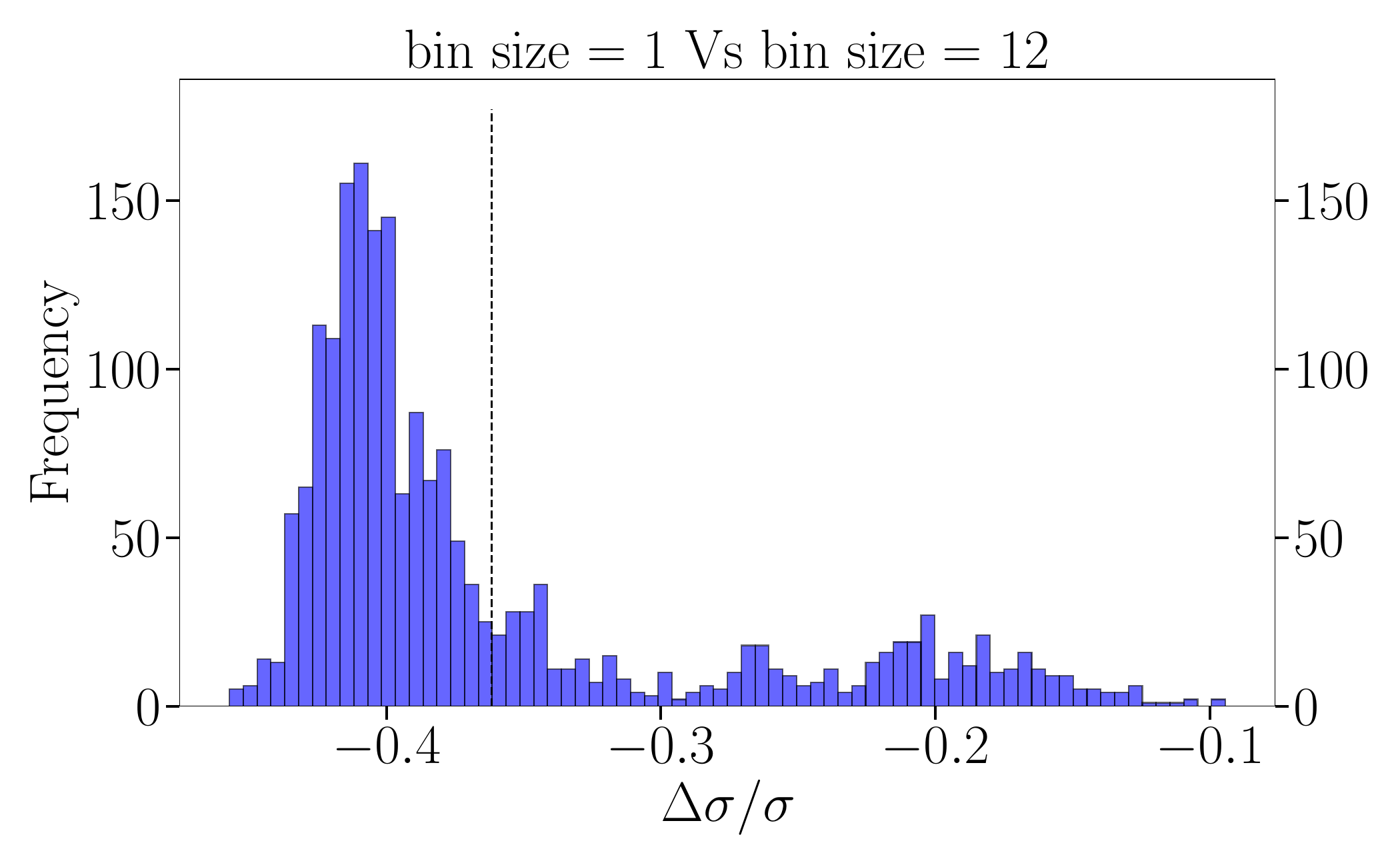}
    \includegraphics[width=0.45\textwidth]{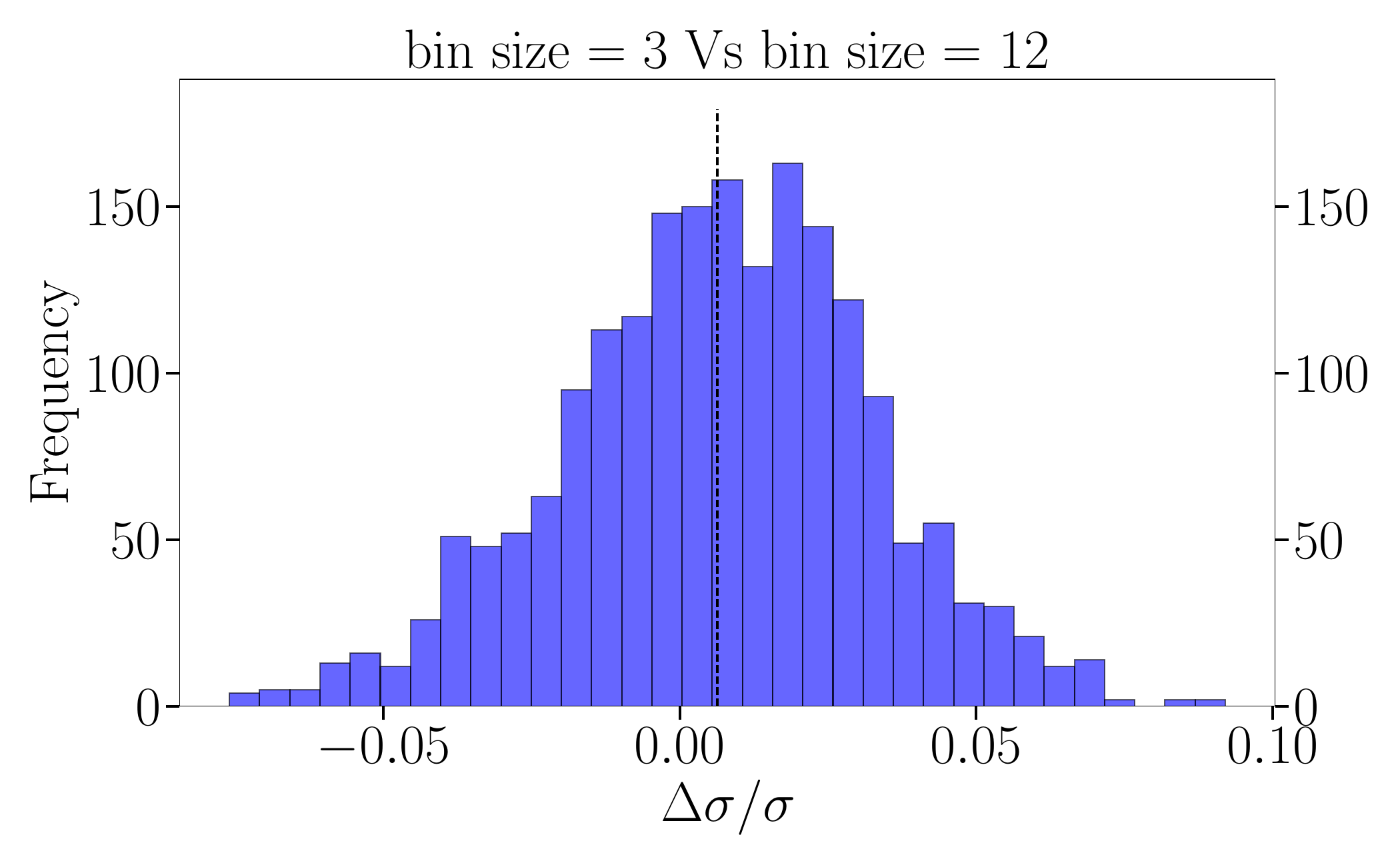}
    \includegraphics[width=0.45\textwidth]{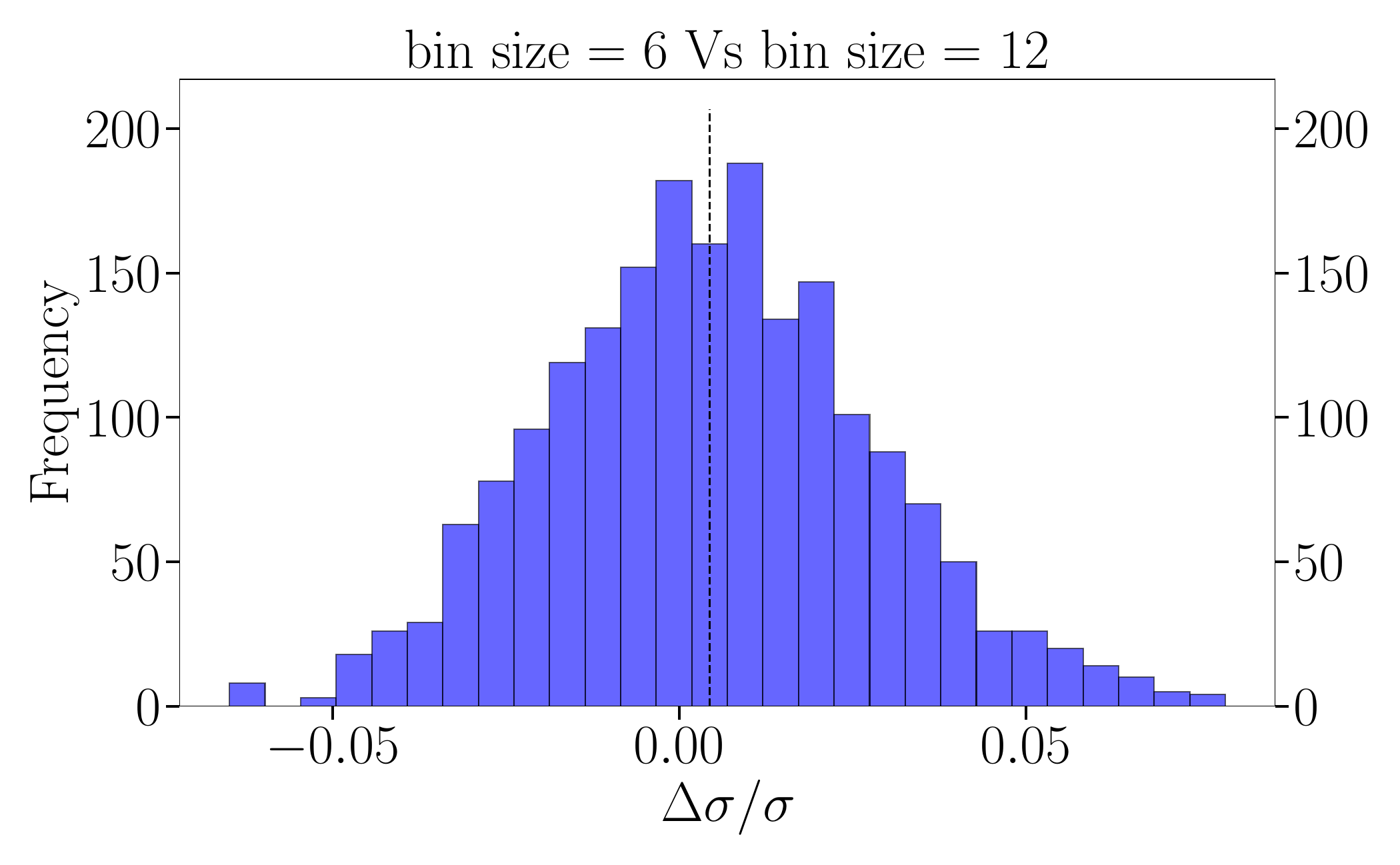}
    \includegraphics[width=0.45\textwidth]{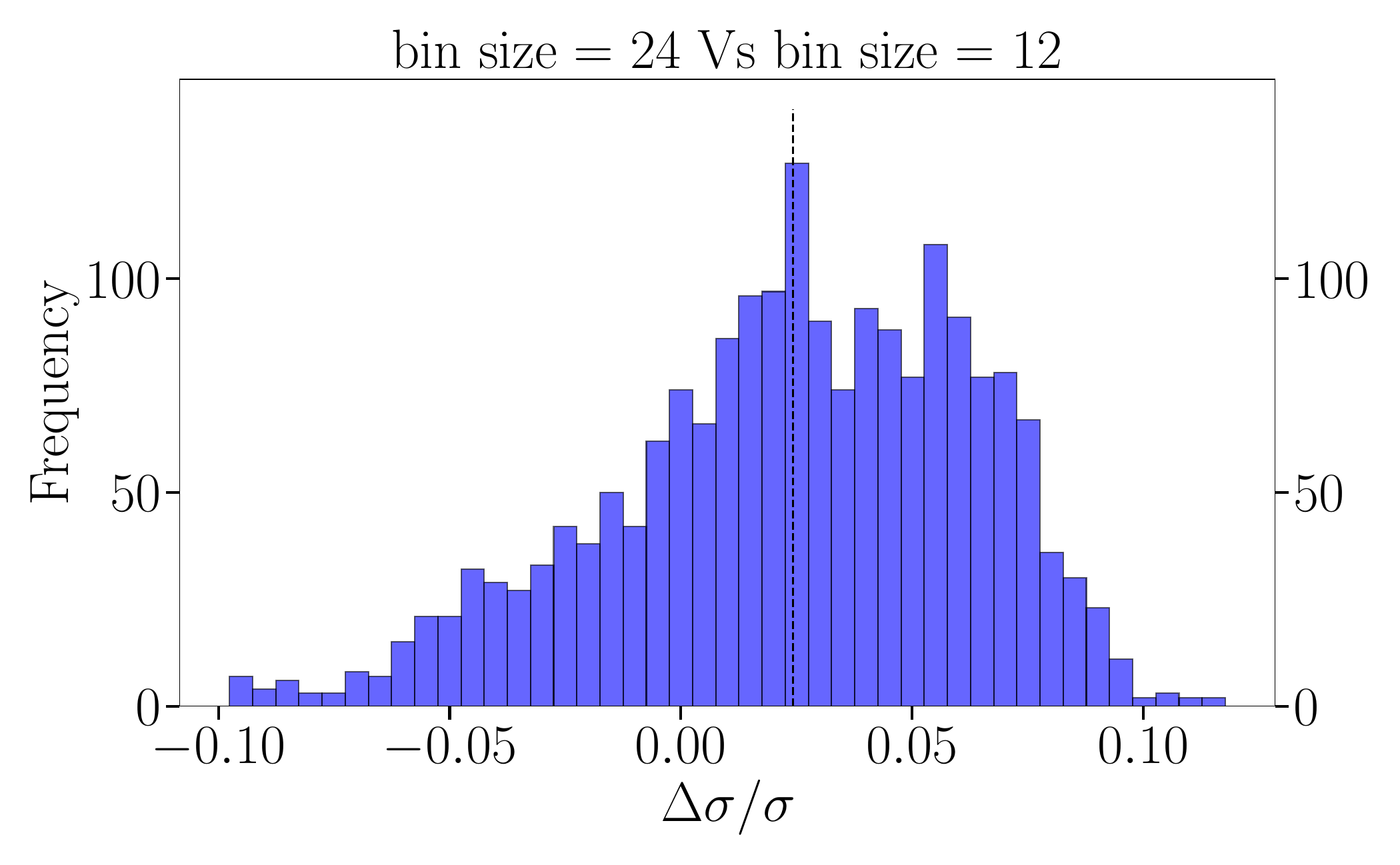}
    \caption{The distribution of relative changes in uncertainty on correlators for binnings of size 1, 3, 6, and 24, relative to the default of 12. The black dotted line indicates the mean. }
    \label{fig:binning}
\end{figure}

Figure~\ref{fig:binning} shows this distribution of
the change in relative uncertainty for binnings of 
size 1, 3, 6, and 24, relative to the default
binning of 12 for $\udbb$ data on the E3 ensemble.
We see that the mean of the distribution is largely 
unchanged for binnings of 3 or 6, while the 
more conservative binning of 24 produces a 
mean increase of about 2\% in uncertainty. For 
completeness, we include the binning of size
1, that is where individual vector components (or 
for the pseudoscalar 3 identical measurements) are 
treated as uncorrelated. As we expect, this results
in a large decrease in uncertainty. All of our 
ensembles produce very similar results, where 
binnings of 3 or 6 would be acceptable. Taking a
more conservative approach, we use a binning of 6 
throughout, giving $N_s=2N_{\mathrm{cfg}}$ ($4N_{\mathrm{cfg}}$ on E7). In our
testing of fit stability, which will be described 
below, we check the effect of this binning on the 
final result.   

\subsubsection{Fitting strategy}
\begin{figure}
    \centering
    \includegraphics[width=0.98\textwidth]{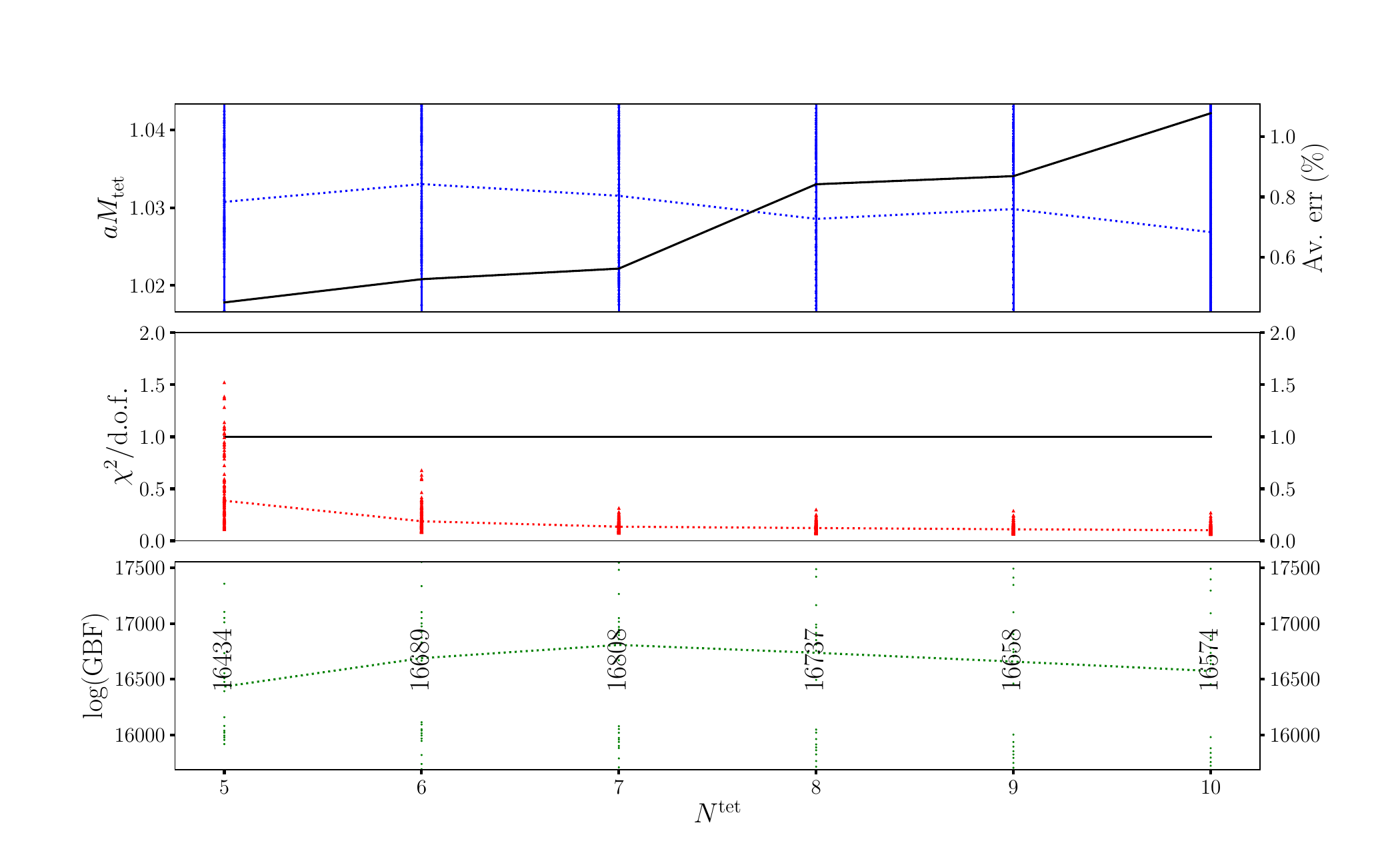}
    \caption{Fit results for the 
    $\udbb$ tetraquark ground-state mass as a 
    function of $N^{\mathrm{tet}}$ on the E1 ensemble. The right-hand 
    vertical axis and solid black line in the top 
    panel give the corresponding mean relative
    uncertainty. The mean values (median for $\chi^2$ 
    and log(GBF)) are given by the dashed lines,
    with the value explicitly printed in the case of log(GBF). The spread of the data reflects the variation of all other parameters.
    \label{fig:E1udbb_tet_N_t}}
\end{figure}

The fitting strategy employed here on each 
ensemble is designed to be as thorough and 
agnostic as possible. First, taking just
the mesons, we select a generous range of 
choices for the variable parameters 
discussed above, and perform a fit for all 
possible combinations of these choices. We 
then plot the resulting ground state 
energy, $aM_{\mathrm{mes}}$, the 
$\chi^2$/d.o.f. (recalling that these are 
artificially reduced by the SVD cut, so 
only meaningful in a relative sense), and
the log(GBF) for each of the fits, against 
each variable in turn, giving a spread of 
results for those choices of that 
variable. We then repeat exactly this 
process for just the tetraquarks. An 
example of such a plot in the tetraquark 
case, varying the number, 
$N^{\mathrm{tet}}$, of exponentials
for the $\udbb$ fit on ensemble E1, 
is given in Fig.~\ref{fig:E1udbb_tet_N_t}. 
In this case, we see that the central value of the 
mass (the dotted lines represent the averages of
the displayed points) is relatively stable against 
changes in $N^{\mathrm{tet}}$, but the $\chi^2$ 
begins to grow below $N^{\mathrm{tet}}=7$. 
As expected, there is a larger relative uncertainty 
for larger values of $N^{\mathrm{tet}}$. In this 
case, we can see that all $N^{\mathrm{tet}}>6$ are 
reasonable choices, with stable $aM_{\mathrm{tet}}$ 
values and small, stable $\chi^2$ values. We 
select $N^{\mathrm{tet}}=7$, as it has the 
largest log(GBF), suggesting 
that including further exponentials beyond this 
would be over-fitting the data.

Once we have completed this process for
the mesons and tetraquarks independently,
we combine them in one simultaneous,
correlated fit. There are now too many
variable parameters to feasibly test all 
possible combinations, so we take the 
values we have already arrived at in the
independent fits, and vary these
one at a time. We check the stability of
the result, and make changes to the 
choices if necessary (within the values
which are acceptable from our above 
analysis), in order to find a fit which is
stable. In general we found $\udbb$ fits to
be much more stable than those for $\lsbb$, likely 
due to the greater density of nearby threshold states 
in the latter.
\begin{figure}
    \centering
    \includegraphics[width=0.98\textwidth]{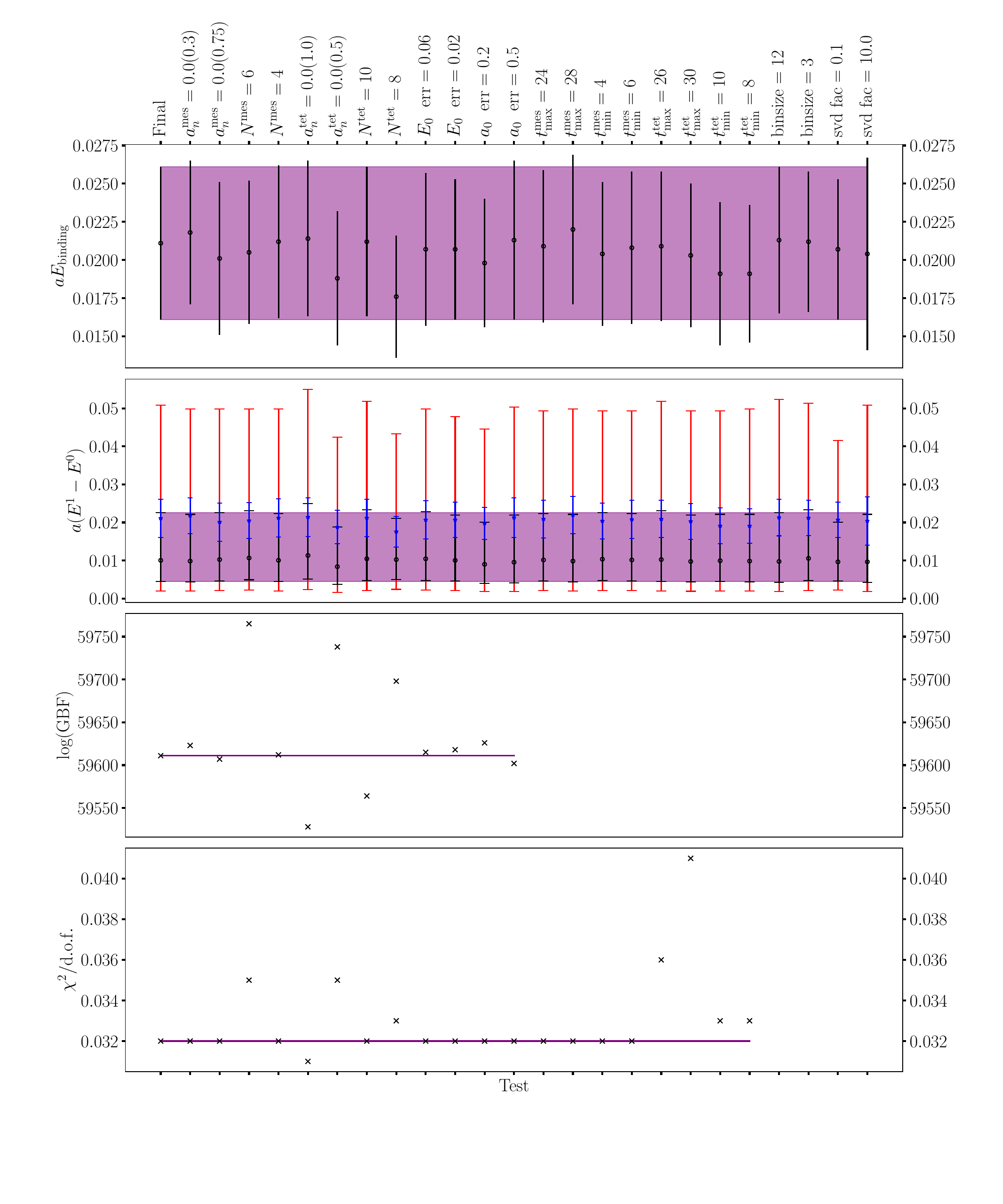}
    \caption{Stability tests for $\lsbb$ on ensemble E3. We vary all parameters discussed in Sec.~\ref{fitparams}, one at a time, both increasing and decreasing them. `Final' is our final fit result. We vary the number of exponentials, $N$, the priors 
    $\mathcal{P}[a_{n\neq0}]$ (labelled $a_n$), and the relative uncertainty on $\mathcal{P}[a_{0}]$ (Eq.~\eqref{Aeff}, labelled $a_0$ err) and $\mathcal{P}[E_{0}]$ (Eq.~\eqref{Meff}, labelled $E_0$ err). We also change $t_{\mathrm{min}/\mathrm{max}}$, change the bin size, (see Sec.~\ref{sec:binsize}) and, finally, multiply the recommended SVD cut by 0.1 and 10, respectively. Note that log(GBF) is not plotted for some tests, as the relative values are meaningful only when the raw data 
    is unchanged. Similarly, 
    $\chi^2$ is affected by changes to the SVD cut and is only meaningful relatively. We plot the binding energy, and the gaps between the ground and first excited state, showing the 1(2)$\sigma$ error bars in black (red). In the case of this first splitting, $a(E^1-E^0)$, the binding energy from the row above is included in blue for comparison.}
    \label{fig:E4lsbb_stability}
\end{figure}

Fig.~\ref{fig:E4lsbb_stability} shows the results 
of this exploration for the $\lsbb$ channel on the 
E3 ensemble. A list of the quantities varied is
provided in the caption. The fit is seen to display 
good stability with respect to these variations, with
generally small induced shifts in the $\chi^2$/d.o.f. 
and log(GBF), except when either the number of 
exponentials, $N$, or the uncertainty on 
a prior, is too small. In the figure, we display 
such variations for completeness, even when the
increases in log(GBF) they have produced
are relatively large. The figure also
includes the results for the splittings of
the first excited tetraquark state, $aE^1-aE^0$. The 
priors for these splittings are logarithmic to 
enforce positive splittings, which results in the 
asymmetric error bars shown. In this case, the error
bars for 1(2)$\sigma$ (i.e. the 68\% and 95\% 
confidence limits) are shown in black (red) and 
the binding energy is included in blue for comparison.  

\subsubsection{
\texorpdfstring{$\udhh$}{udhh}, 
\texorpdfstring{$\lshh$}{lshh}, 
\texorpdfstring{$\udbh$}{udbh} and 
\texorpdfstring{$\lsbh$}{lsbh} fits}
In updating the earlier
study~\cite{Francis:2018jyb} of the
heavy-quark-mass dependence of tetraquark binding, 
we follow Ref.~\cite{Francis:2018jyb} and employ 
the highest-statistics ensemble available, E5,
which has $m_\pi = 299$ MeV and $m_\pi L =4.4$. 
The approach followed is similar to that described
above, though with some modifications required to 
handle the larger data set that results from the
variation of the heavy-quark mass. We begin by
performing a stability analysis identical to that
above, for each tetraquark channel and each 
heavy $am_h$ value in turn, determining the
corresponding bindings from stable versions of 
these individual fits. As we ultimately wish to 
fit the $\udhh$, $\udbh$, $\lshh$ and $\lsbh$ binding 
energies for different $am_h$ to a fit form with common
parameters, we would ideally like to fit all 
of the tetraquark correlators, for all
$am_h$, together with the corresponding threshold
meson correlators in one large, correlated fit.
However, we find that, owing to its
large size, such a combined fit is not
feasible. To break it up, we first performed a number of tests examining the strength of correlations between different fit
parameters. We found it possible to perform 
a simultaneous fit for the doubly heavy 
$1^+$ $\udhh$ tetraquark and associated 
heavy meson, $H$ and $H^*$, channels for 
all $am_h$. The maximum correlation 
between $\udhh$ bindings, $\Delta E(am_h)$, 
in this fit was found to be 0.12, with most 
correlations roughly 0.05. It is reasonable 
to assume the correlations between 
$am_h$-dependent tetraquark bindings in 
the other (singly heavy and/or strange) 
tetraquark channels will be similarly small. 
Next, for each $am_h$, we performed a 
simultaneous fit to extract $aM_{\udhh}$, 
$aM_{\udbh}$, $aM_{H}$ and $aM_{H^*}$. 
Correlations between meson and tetraquark
masses were found to be small 
($\lesssim 0.1$), but those between 
$aM_{\udhh}$ and $aM_{\udbh}$ were more 
significant (typically $\approx 0.25$). 
Similar results were found for the $ls$ 
analogues. These inter-tetraquark 
correlations are further enhanced when we 
turn from the tetraquark masses to the
corresponding binding energies, where the 
effect of shared threshold meson masses comes 
into play and we find correlations between the
bindings of $\udhh$ and $\udbh$ tetraquarks of 
$\gtrsim 0.5$ (and similarly for $\lshh$ and
$\lsbh$).

Based on these observations, we adopt the
following strategy. First, we
simultaneously fit all meson correlators for 
all $am_h$ to extract $aM_H$, $aM_{H^*}$, 
$aM_{H_s}$ and $aM_{H_s^*}$. These will be 
taken in various combinations to provide the 
two meson thresholds for each tetraquark. 
Second, we simultaneously fit all
tetraquarks for \textit{each} $am_h$ to 
extract $aM_{\udhh}$, $aM_{\udbh}$, $aM_{\lshh}$, 
$aM_{\lsbh}$. This ensures we have fits which
are tractable, while also preserving
correlations where they are significant. 
Finally, we ensure that the resulting bindings 
from this correlated method agree with those
found when we optimised and stabilised the 
individual $am_h$ fits above. 
\section{Results}\label{results}
\subsection{Light-quark mass dependence and the physical-point $\udbb$ and $\lsbb$ binding energies}
\label{light_mass_dep}

\begin{figure}[!h]
  \includegraphics[width=0.98\textwidth]{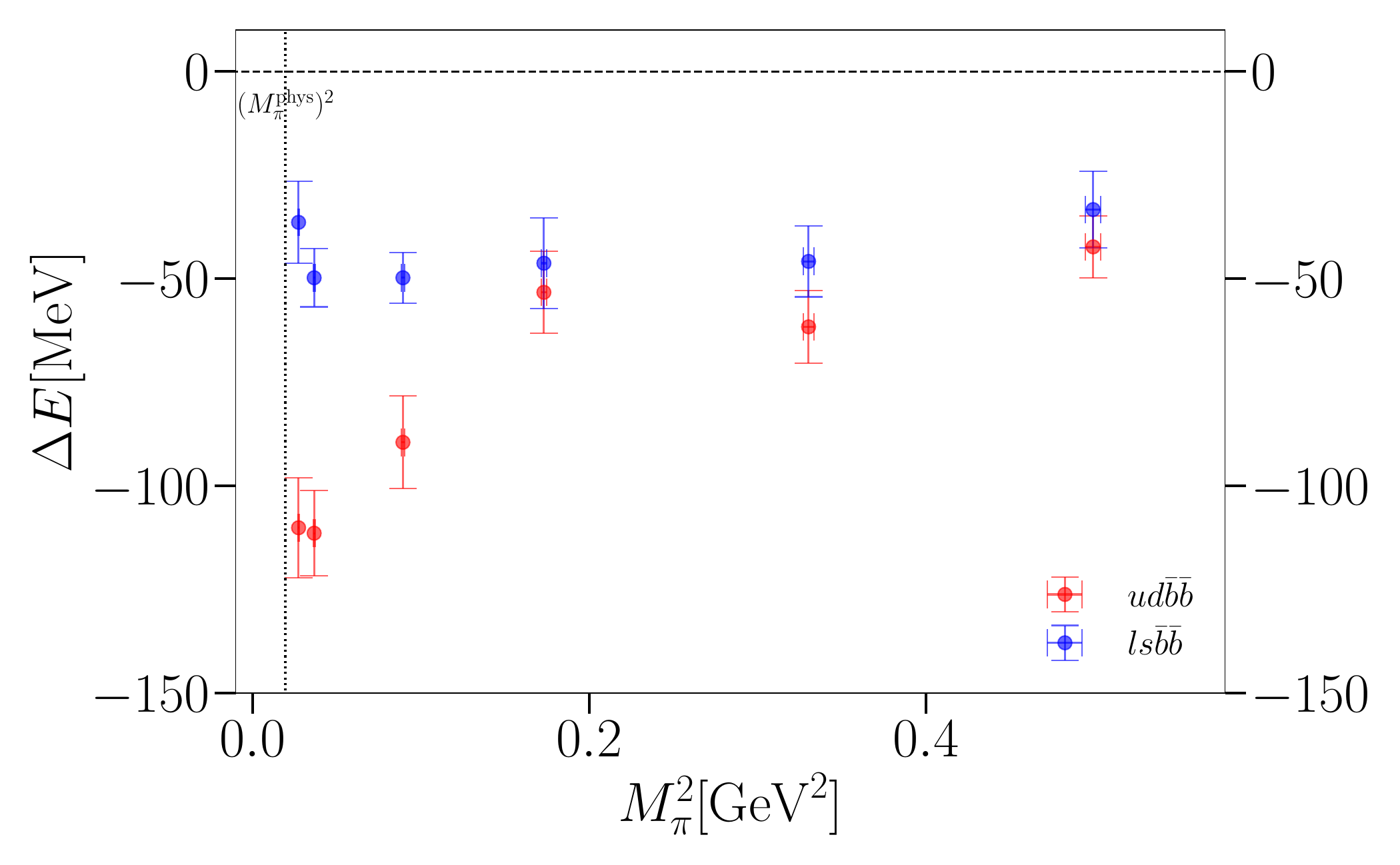}
  \caption{ Tetraquark binding in the 
  $J^P=1^+$ $\udbb$ and $\lsbb$ channels versus
  $m_\pi^2$ for $m_\pi\approx 700,\, 575,\, 415,\, 
  299,\, 192$ and $165$ MeV.}
  \label{fig:mpi_dependence}
\end{figure}

Numerical results for the binding energies of the
$J^P=1^+$ $\udbb$ and $\lsbb$ tetraquarks on the 
six ensembles studied are 
presented in Table~\ref{tab:mpi_dependence}, and 
plotted as a function of $m_\pi^2$ in 
Fig.\ref{fig:mpi_dependence}. Results are given in 
lattice units and MeV.

\begin{table}[!h]
  \begin{tabular}{c | c c c c }
    \hline
    $m_\pi\;~[\mathrm{MeV}]$ &$-a\Delta E_{ud\bar{b}\bar{b}}$ & $-a\Delta E_{ls\bar{b}\bar{b}}$  & $-\Delta E_{ud\bar{b}\bar{b}}\;~[\mathrm{MeV}]$ & $-\Delta E_{ls\bar{b}\bar{b}}\;~[\mathrm{MeV}]$\\
    \hline
    $707$ &0.0193(34)& 0.0152(42)& 42.3(7.5) & 33.3(9.2) \\
    $575$ &0.0281(40)& 0.0209(39)& 61.7(8.8) & 45.9(8.6) \\
    $416$ &0.0243(45)& 0.0211(50)& 53.3(9.9) & 46(11) \\
    $299$ &0.0408(51)& 0.0227(28)& 90(11) & 49.8(6.1) \\
    $191$ &0.0508(47)& 0.0227(32)& 111(10) & 49.8(7.0) \\
    $165$ &0.0502(55)& 0.0166(45)& 110(12) & 36.4(9.9)\\
    \hline
  \end{tabular}
  \caption{Binding energies in the $J^P=1^+$
  $\udbb$ and $\lsbb$ channels as a function of
  $m_\pi$.}
  \label{tab:mpi_dependence}
\end{table}

To extrapolate these results to physical $m_\pi$, 
we employ fits with the expected leading 
linear-in-$m_\pi^2$ chiral behavior and, as in
Ref.~\cite{Hudspith:2023loy}, 
a finite volume (FV) term exponential in $m_{\pi}L$,
where $L$ is the spatial extent of the lattice. 
The resulting fit form is
\begin{equation}\label{Eq:chiral_fit}
    a\Delta E = A_0(1 + A_1 (am_{\pi})^2 + A_2e^{-m_{\pi}L})\, ,
\end{equation}
with the physical result recovered by setting 
$m_{\pi}\to m_{\pi}^{\mathrm{phys}}$ and $L\to\infty$.
We have made $a$ explicit for clarity, noting that all
$A$s are dimensionless\footnote{We perform the fit in 
lattice units to avoid D'Agostini (normalisation) 
bias~\cite{DAGOSTINI1994306}. Results are 
converted to physical units after fitting.}.  
For $A_2$, the range of $m_\pi L$ for the 
ensembles we consider extends down only to 
$\simeq 3.6$, giving us limited leverage for 
fitting FV effects and preventing a meaningful 
fit for this quantity. Since a significantly wider
range of $m_\pi L$, extending down to $\simeq 2.7$, 
was available to the authors of 
Ref.~\cite{Hudspith:2023loy}, we have used the 
results of the fit from that reference,
provided to us by the authors, as priors for our
fit, taking $A_2=-3.1(7)$ in the $\udbb$ case.
\footnote{We thank the authors of for providing
us with this information, which was not explicitly
quoted in Ref.~\cite{Hudspith:2023loy}.} 
With our $m_{\pi}L$ values in the range $3.6$ to 
$10.3$, this turns out to give only a modest 
correction to the final values. In the $\lsbb$ case, 
we expect, in addition to somewhat smaller effects 
proportional to $e^{-m_{\pi} L}$, possible 
additional effects proportional to $e^{-m_K L}$. 
The former are dealt with by employing a FV term
$A_2^{ls} e^{-m_{\pi} L}$ in the $\lsbb$ analogue of
Eq.~\ref{Eq:chiral_fit}, taking a conservative prior of 
$0(3)$ for $A_2^{ls}$. The latter are expected to be 
very small, and hence neglected, given the much larger
values of $m_K L$  ($m_K L>8.7$) for the ensembles
considered. These fits are again carried out using
\textit{lsqfit}~\cite{peter_lepage_2023_7931361}.
Since the range of pion masses considered extends 
up to $\simeq 707$ MeV, a value likely well 
beyond the range of validity of the fit ansatz
Eq.~\eqref{Eq:chiral_fit}, we perform a number of
such fits, starting with one involving the two 
ensembles with $m_\pi$ closest to the physical
value (E7 and E9), and adding, one at a time, the
ensemble with the next lightest $m_\pi$ to the fit.
\begin{figure}[!h]
  \includegraphics[width=0.98\textwidth]{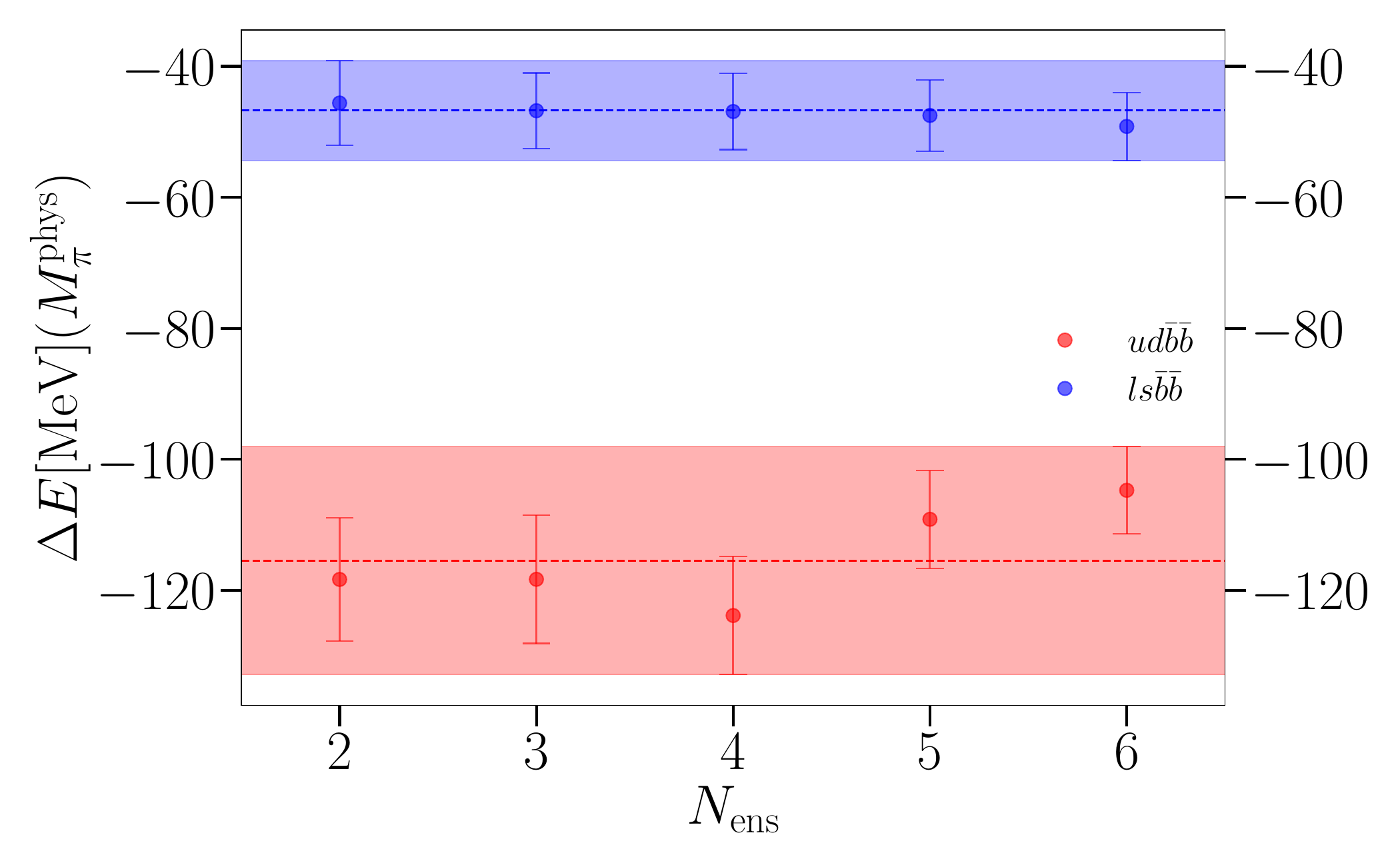}
  \caption{Physical-point extrapolations of the $1^+$ $\udbb$ and $\lsbb$ binding energies
  from fits including $N_{\mathrm{ens}}$ ensembles (see text). Our final results are represented by the coloured bands.}
  \label{fig:N_ens_dependence}
\end{figure}

Figure~\ref{fig:N_ens_dependence} shows the 
resulting extrapolated physical-point values of 
the $1^+$ $\udbb$ and $\lsbb$ binding energies,
$\Delta E(M_{\pi}^{\mathrm{phys}})$, 
as a function of $N_{\mathrm{ens}}$, the
number of ensembles included in the fit. The 
$\lsbb$ binding is seen to be relatively independent 
of $N_{\mathrm{ens}}$, while the $\udbb$ binding 
decreases by about a sigma once the two ensembles
with the heaviest $m_\pi$ are included.

While we expect the fits with smaller
$N_{\mathrm{ens}}$ (and hence smaller maximum
$m_\pi$) to be more reliable, to
be conservative, we take as our final results 
in each channel those represented by the 
central values and half-widths of the colored 
bands shown in Fig.~\ref{fig:N_ens_dependence},
which span the ranges covered by the error bars 
of all six such fits. These are 
\begin{align}\label{eq:phys_bindings1}
  \Delta E_{ud\bar{b}\bar{b}}(m_{\pi}^{\mathrm{phys}}) &= -115(17)~\mathrm{MeV} \\\label{eq:phys_bindings2}
  \Delta E_{ls\bar{b}\bar{b}}(m_{\pi}^{\mathrm{phys}}) &= -46.7(7.6)~\mathrm{MeV}.
\end{align}
The finite volume corrections described above
(and included in Fig.~\ref{fig:N_ens_dependence} and 
Eqs.~\eqref{eq:phys_bindings1} and
~\eqref{eq:phys_bindings2}) decrease the 
$\Delta E_{ud\bar{b}\bar{b}}(m_{\pi}^{\mathrm{phys}})$ 
(i.e., deepen the binding) obtained for each 
$N_{\mathrm{ens}}$ by roughly one standard deviation. 
The net effect is a final $\udbb$ binding, quoted 
in Eq.~\eqref{eq:phys_bindings1}, $7\ \mathrm{MeV}$ 
($\approx 0.4\sigma$) more bound, and with an 
uncertainty $3~\mathrm{MeV}$ larger, than that 
obtained neglecting finite-volume corrections. For 
$\Delta E_{ls\bar{b}\bar{b}}(m_{\pi}^{\mathrm{phys}})$, 
the effects simply increase the uncertainties, without
significantly changing the central values. The result
is a final uncertainty in Eq.~\eqref{eq:phys_bindings2}
$1~\mathrm{MeV}$ larger than that obtained in the
uncorrected case.

\subsection{Heavy mass dependence}
\label{heavymassdependence}
The heavy mass dependence of the binding, 
$\Delta E$, for $\udhh$, $\udbh$, $\lshh$ and 
$\lsbh$ tetraquarks is discussed at length
in~\cite{Francis:2018jyb}, with the 
underlying NRQCD implementation outlined 
in Appendix A of that paper. The variable, 
spin-averaged $\eta_h$-$\Upsilon_h$ mass is 
determined as a function of the input NRQCD parameter
$aM_Q$ by measuring the kinetic mass, with the ratio
of the physical spin-averaged $\eta_b$-$\Upsilon_b$
mass to the resulting kinetic mass taken to define 
the ratio $r_h\equiv m_b/m_h$. We follow 
Ref.~\cite{Francis:2018jyb} in carrying
out this study on the highest-statistics
ensemble available, E5, which has 
$m_\pi =299$ MeV and $m_\pi L=4.4$. We 
also employ the phenomenological fit forms 
used in that paper, here recast using the 
notation $\delta_{l} = M_{B^*} - M_{B}$ and
$\delta_{s} = M_{B^*_s} - M_{B_s}$ for 
the physical heavy bottom-meson splittings.
As usual, with $M^{\mathrm{spin~av.}}_{BB^*}$
the spin average of the $B$ and $B^*$ masses,
$M_B=M^{\mathrm{spin~av.}}_{BB^*} 
- \frac{3}{4} \delta_l$ and 
$M_{B^*}=M^{\mathrm{spin~av.}}_{BB^*} 
+ \frac{1}{4} \delta_l$, with analogous
expressions for $M_{B_s}$ and $M_{B_s^*}$.
These splittings set the overall scales of 
the $m_b/m_h\equiv r_h$-dependent 
contributions to tetraquark binding in 
the various channels from heavy-light
interactions in the associated two-meson
threshold states, represented by the last
terms in the expressions below. The 
mass-dependent factors accompanying 
the splittings in these expressions are
determined by which two mesons form
the two-meson threshold for the channel 
in question. For $\udbh$, for example, the
threshold is $B^*H$ for $m_h<m_b$ ($r_h>1$)
and $H^*B$ for $m_h>m_b$ ($r_h<1$).

With the above notation, the expressions
of Ref.~\cite{Francis:2018jyb} for the
singly-heavy, variable-mass $1^+$ $\udbh$ 
bindings become
\begin{equation}\label{eq:udbh_mh_dependence}
  \begin{split}
    \Delta E^{\udbh} &=\frac{C_0}{r+r_h} +C_1^{ud}+C_2^{ud}(r+r_h)-\Big(\frac{1}{4}r-\frac{3}{4}r_h\Big)\delta_l~~~~~ (r_h>1)\\  
    \Delta E^{\udbh} &=\frac{C_0}{r+r_h} +C_1^{ud}+C_2^{ud}(r+r_h)-\Big(\frac{1}{4}r_h-\frac{3}{4}r\Big)\delta_l~~~~~ (r_h<1),
  \end{split}
\end{equation}
where $r\equiv m_b/m_b =m_h/m_h =1$, and the terms
proportional to $C_0$, $C_1^{ud}$, and $C_2^{ud}$
serve to parameterize the attractive color Coulomb
interaction between the two heavy antiquarks, the 
spin-dependent good-light-diquark attraction, and 
residual $1/m_h$-dependent effects (such as differences
in light-heavy interactions and heavy-quark kinetic 
energy between the tetraquark and two-meson threshold 
states), respectively.{\footnote{The color Coulomb term
in Eq.~(\ref{eq:udbh_mh_dependence}) is proportional to 
the reduced mass of the two heavy quarks, 
$\mu_{bh}=m_b m_h/(m_b+m_h)=m_b/[r_h+1]$, with the 
factor of $m_b$ in the numerator absorbed into the 
definition of $C_0$. Similarly, the term proportional
to $C_2$ reflects an assumed heavy-mass dependence of
the form $(1/m_b)+(1/m_h)=(1/m_b)[1+r_h]$, with
the overall $(1/m_b)$ factor absorbed into $C_2$.}}
The apparently unnecessary notation $r$ for quantities
which are here identically equal to $1$ is introduced
for the purpose of compactness, allowing the
corresponding forms for the doubly-heavy, variable-mass 
$1^+$ $\udhh$ cases to be obtained by replacing all
occurrences of $r$ with $r_h$. For $\lsbh$, the lowest
two-meson thresholds are $B^*H_s$ for $r_h>1$ and 
$H^*B_s$ for $r_h<1$, and the phenomenological forms 
for the singly-heavy, variable-mass $1^+$ $\lsbh$
bindings are
\begin{equation}\label{eq:lsbh_mh_dependence}
  \begin{split}
    \Delta E^{\lsbh} &=\frac{C_0}{r+r_h} +C_1^{ls}+C_2^{ls}(r+r_h)-\Big(\frac{1}{4}\delta_l r-\frac{3}{4}\delta_s r_h\Big)~~~~~ (r_h>1)\\  
    \Delta E^{\lsbh} &=\frac{C_0}{r+r_h} +C_1^{ls}+C_2^{ls}(r+r_h)-\Big(\frac{1}{4}\delta_l r_h-\frac{3}{4}\delta_s r\Big)~~~~~ (r_h<1),
  \end{split}
\end{equation}
with the corresponding doubly-heavy variable-mass forms
again obtained by the replacement $r\rightarrow r_h$. 
The superscripts on the fit parameters $C_1^{ud,ls}$ 
and $C_2^{ud,ls}$ reflect the expected dependence of 
these quantities on the masses of the light quarks
involved. This dependence also means that 
$C_1^{ud,ls}$ and $C_2^{ud,ls}$ will differ from
ensemble to ensemble, with the results obtained
from the fits of the current study specific to 
the E5 ensemble employed. 

The fit forms above are predicated on the 
assumption that the binding in any deeply bound 
doubly heavy tetraquark system will be dominated by 
the combination of the attractive color-Coulomb
interaction between two heavy antiquarks in a color 
$3_c$ configuration and the attractive spin-dependent 
good-light-diquark interaction known to be present in 
the static heavy-quark limit and reflected 
experimentally in the splittings in the singly heavy 
baryon spectrum. Additional contributions, resulting
from deviations of non-color-Coulomb effects from 
their static values, are assumed to be numerically 
subleading, and parametrizable by the terms 
proportional to $C_2^{ud,ls}$. These assumptions 
will become less reliable as the heavy-quark mass 
decreases and one moves farther from the static limit.
The fit forms are also expected to break down for
weakly bound systems, where the wave-function will
become dominated by longer-distance two-meson 
components, reducing the effect of the attractive 
short-distance interactions available in a more
deeply bound, localized system. Note that the 
assumptions underlying these forms, if reliable, 
should satisfy some semi-quantitative self-consistency
tests. For example, the strong increase in the 
strength of the attractive color-Coulomb interaction 
with increasing $am_h$ expected in the $\udhh$ and 
$\lshh$ systems should show up when the heavy-quark
mass is varied to values larger than $am_b$. 
Physically constrained expectations also exist
for $C_1^{ud,ls}$, provided, as assumed, these are 
dominated by the known attractive good-light-diquark
contribution. where we know, from the physical 
$\Sigma_b$-$\Lambda_b$ splitting, that, relative to
the spin average, the spin-dependent 
good-light-diquark attraction is $-3/4$ times the 
$\Sigma_b$-$\Lambda_b$ splitting, or $-145$ MeV.
One would thus expect a physically sensible fit form 
to produce a result $C_1^{ud}\, \simeq\, -145$ MeV for 
an ensemble with physical $m_\pi$ and $m_K$. 
Similarly, from the physical $\Xi_b^\prime$-$\Xi_b$ 
splitting, a physically sensible fit form should
produce a result $C_1^{ls}\, =\, \simeq -104$ MeV 
for an ensemble with physical $m_\pi$ and $m_K$. 
With these physical-point results establishing the
decrease of the good-light-diquark attraction with
increasing light-quark mass, we further expect to find 
smaller negative values of $C_1^{ud}$ and $C_1^{ls}$
for ensembles with heavier-than-physical $m_\pi$, 
such as E5.

Correlated fits of the variable-$r_h$ E5 data are
carried out using 
\textit{lsqfit}~\cite{peter_lepage_2023_7931361}, 
which makes extensive use of 
\textit{gvar}~\cite{peter_lepage_2023_8025535}. 

The results for the parameters $\delta_l$ and 
$\delta_s$, obtained from the combined all-$am_h$ 
fit for $aM_H$, $aM_{H^*}$, $aM_{H_s}$ and 
$aM_{H_s^*}$, are $\delta_l = 41.7(2.7)~\mathrm{MeV}$
and $\delta_s=45.8(1.7)~\mathrm{MeV}$. These 
agree well with the PDG~\cite{Workman:2022ynf} 
values, $\delta_l = 45.42(26)~\mathrm{MeV}$ and 
$\delta_s=46.1(1.5)~\mathrm{MeV}$. Using the
PDG values in place of the fit values has no 
significant impact on our results. As noted
above, the $aM_H$, $aM_{H^*}$, $aM_{H_s}$ and
$aM_{H_s^*}$ results are used to determine
the lowest two-meson thresholds needed for
determining the tetraquark binding energies.

\begin{figure}[!h]
  \includegraphics[width=0.98\textwidth]{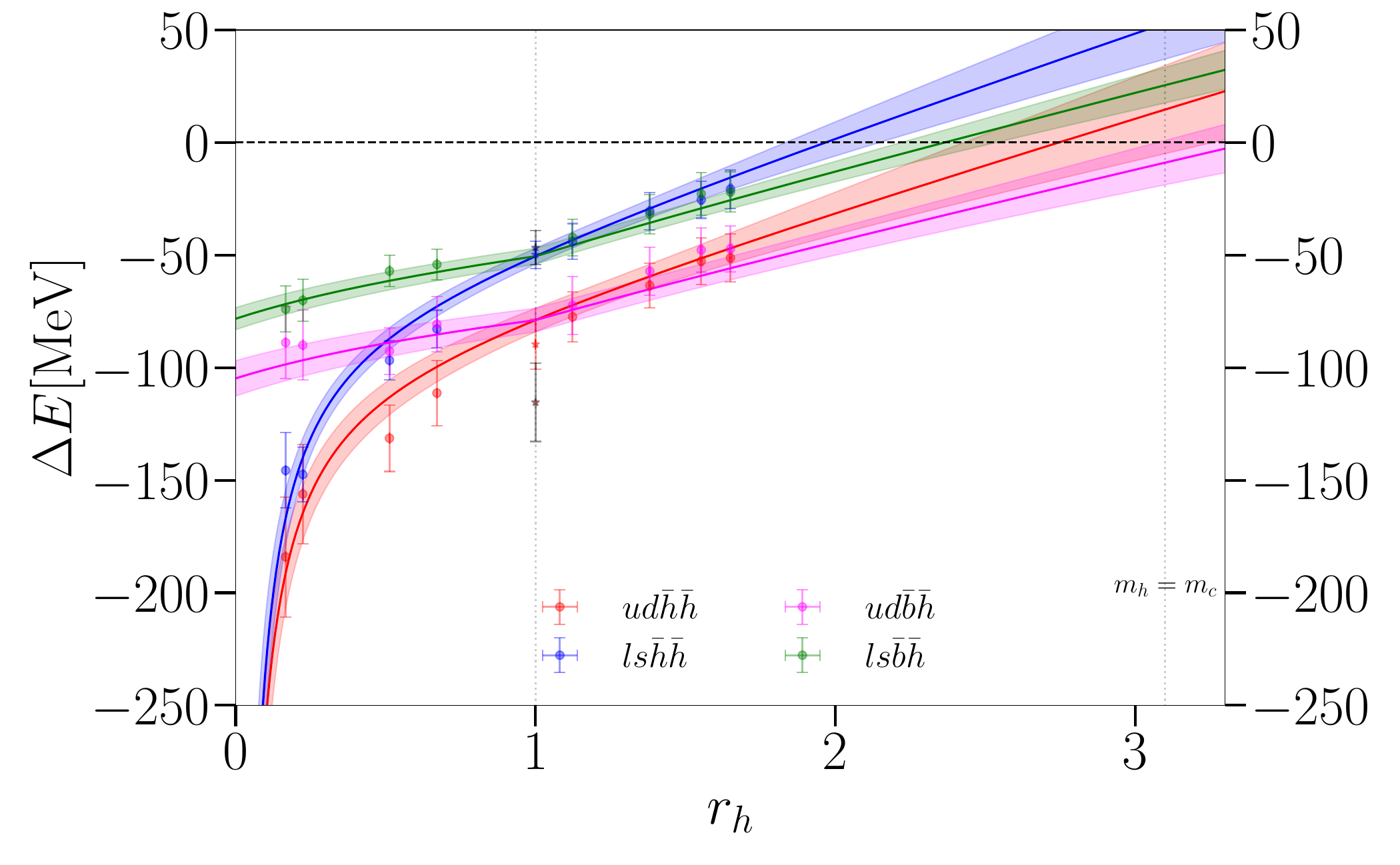}
  \caption{Binding energies of $J^P=1^+$ tetraquark
  states bound below the lowest two-meson threshold on 
  the E5 ensemble, plotted against $r_h$. Coloured 
  bands are the results of correlated fits to 
  Eqs.~\eqref{eq:udbh_mh_dependence} and 
  ~\eqref{eq:lsbh_mh_dependence}. The points at
  $m_b/m_h=1$ with the black error bars are the 
  results of a physical-point continuum extrapolation,
  discussed below in Sec.~\ref{light_mass_dep}.}
  \label{fig:variableb}
\end{figure}

In Fig.~\ref{fig:variableb}, we present the
results for the E5 ensemble $J^P=1^+$ $\udhh$,
$\lshh$, $\udbh$ and $\lsbh$ tetraquark binding 
energies, together with colored bands showing 
the fit to these results produced using the 
$r_h$-dependent phenomenological forms above. 
Table~\ref{tab:var_b_coeffs} gives the resulting
fit parameter values, errors and correlations. 
The fit has a $\chi^2/\mathrm{d.o.f.}$
of 0.55.  
\begin{table}[!h]
  \begin{tabular}{c c c c c }
    \hline
    $C_0$ & $C_1^{ud}$ & $C_2^{ud}$ & $C_1^{ls}$ & $C_2^{ls}$\\
    \hline
    -31.8(6.3) & -102(14) & 14.8(5.1) & -79(12) & 16.7(4.2) \\ 
    1.000 & -0.558 & 0.411 & -0.786 & 0.663  \\ 
     & 1.000 & -0.911 & 0.684 & -0.573  \\ 
     &  & 1.000 & -0.525 & 0.506  \\ 
     &  &  & 1.000 & -0.930  \\ 
     &  &  &  & 1.000  \\
     \hline
  \end{tabular}
\caption{E5 ensemble $J^P=1^+$ tetraquark fit results. 
Line 1: Fit results (in MeV) for the coefficients 
in Eqs.~\eqref{eq:udbh_mh_dependence} 
and~\eqref{eq:lsbh_mh_dependence}. Lines 2-6: the
corresponding correlation matrix. }
\label{tab:var_b_coeffs}
\end{table}

Immediately evident from the figure are the
rapid increases in $\udhh$ and $\lshh$ binding with 
decreasing $r_h$ (increasing $am_h$), associated 
with the unbounded growth of the color-Coulomb
attraction in those cases as $r_h\rightarrow 0_+$ 
($am_h\rightarrow\infty$). The fit results for
$C_1^{ud}$ and $C_1^{ls}$ are also seen to be 
$\sim 20\%$ less attractive than the 
corresponding physical-point values, in keeping 
with the heavier-than-physical $299$ MeV value of
$m_\pi$ on the E5 ensemble and the expectations 
discussed above. The good agreement of the
phenomenological fits with the $r_h$-dependence 
of the lattice results clearly suggests the physical
picture underlying the fit forms has succeeded in 
identifying the main sources responsible for the
observed deep binding in the $\udbb$ and $\lsbb$ 
tetraquark channels. While the region of reliability 
of the NRQCD heavy-quark treatment extends down only 
to $m_h\simeq 0.6m_b$~\cite{Francis:2018jyb} 
($r_h\simeq 1.7$), and hence does not reach 
$m_h=m_c$ ($r_c\simeq 3.1$), this suggests 
extrapolated versions of the fits (bearing in mind 
the additional $\sim 45$ ($\sim 25$) MeV in
good $ud$ ($ls$) diquark binding expected for
physical $m_\pi$ based on the E5 results for
$C_1^{ud}$ and $C_1^{ls}$) may provide useful 
qualitative guides to what is expected as $m_h$ is 
reduced further towards $m_c$. This conclusion is
supported by the observation that a rough by-eye
extrapolation of the NRQCD-based E5 $\udhh$ results 
to $r_h\simeq 3.1$, supplemented by the expected
additional $\sim 45$ MeV in good-light-diquark-source 
binding at physical $m_\pi$ for the points shown in
the figure, leads to the expectation of a binding 
near zero in the physical $1^+$ $\udcc$ tetraquark 
channel, an expectation compatible with the observed 
very weakly bound $T_{cc}$ found by 
LHCb~\cite{LHCb:2021vvq,LHCb:2021auc}. The figure,
further, suggests that (i) with binding in the $1^+$ 
$\lshh$ and $\lsbh$ channels always less than that 
in the $\udhh$ channel and the $T_{cc}$ observed to lie
essentially at threshold, no bound $1^+$ state is
expected in either of the $ls\bar{c}\bar{c}$ or 
$ls\bar{b}\bar{c}$ channels,
and (ii) with the $T_{cc}$ weakly bound and the fit 
showing binding in the $1^+$ $\udbh$ channel greater
than that in the $1^+$ $\udhh$ channel for $r_h>1$, 
a bound $\udbc$ tetraquark is likely to exist in the 
$I=0$, $J^P=1^+$ channel, albeit (given the
preponderance of previous lattice results) with
relatively small binding. The 
latter conclusion is in keeping with expectations 
based on (i) the results of the recent, slightly
heavier-than-physical $m_\pi\simeq 220$ MeV lattice 
scattering study of Ref.~\cite{Alexandrou:2023cqg}, 
which found a genuine bound state (though compatible 
within statistical errors with being a virtual bound 
state) $2.5(2.9)$ MeV below $B^*\bar{D}$ threshold, 
and (ii) the anticipated small increase in binding due 
to increased good-light-diquark attraction at shorter
$B^*\bar{D}$ separations expected at slightly lower, 
physical $m_\pi$. Weak binding in this channel is 
also compatible with the results of
Refs.~\cite{Hudspith:2020tdf,Meinel:2022lzo} which,
though showing no signs of deep binding, could not rule
out a weakly bound state. The weak binding found in
Ref.~\cite{Alexandrou:2023cqg} shows some 
tension with the 
$43\left({}^{+7}_{-6}\right)\left({}^{+24}_{-14}\right)$ 
MeV result of the mixed-action, overlap-valence on 
HISQ-sea study of Ref.~\cite{Padmanath:2023rdu}, 
though the latter employs a small set of
interpolating operators and results from a 
physical-point extrapolation from valence pion 
masses, $0.5$, $0.6$, $0.7$, $1.0$ and $3.0$ GeV,
significantly heavier than considered in 
Ref.~\cite{Alexandrou:2023cqg}. 
The $1^+$ $\udbc$ channel will be discussed  in more
detail in Sec.~\ref{discussion}.

\begin{figure}[!h]
  \includegraphics[width=0.98\textwidth]{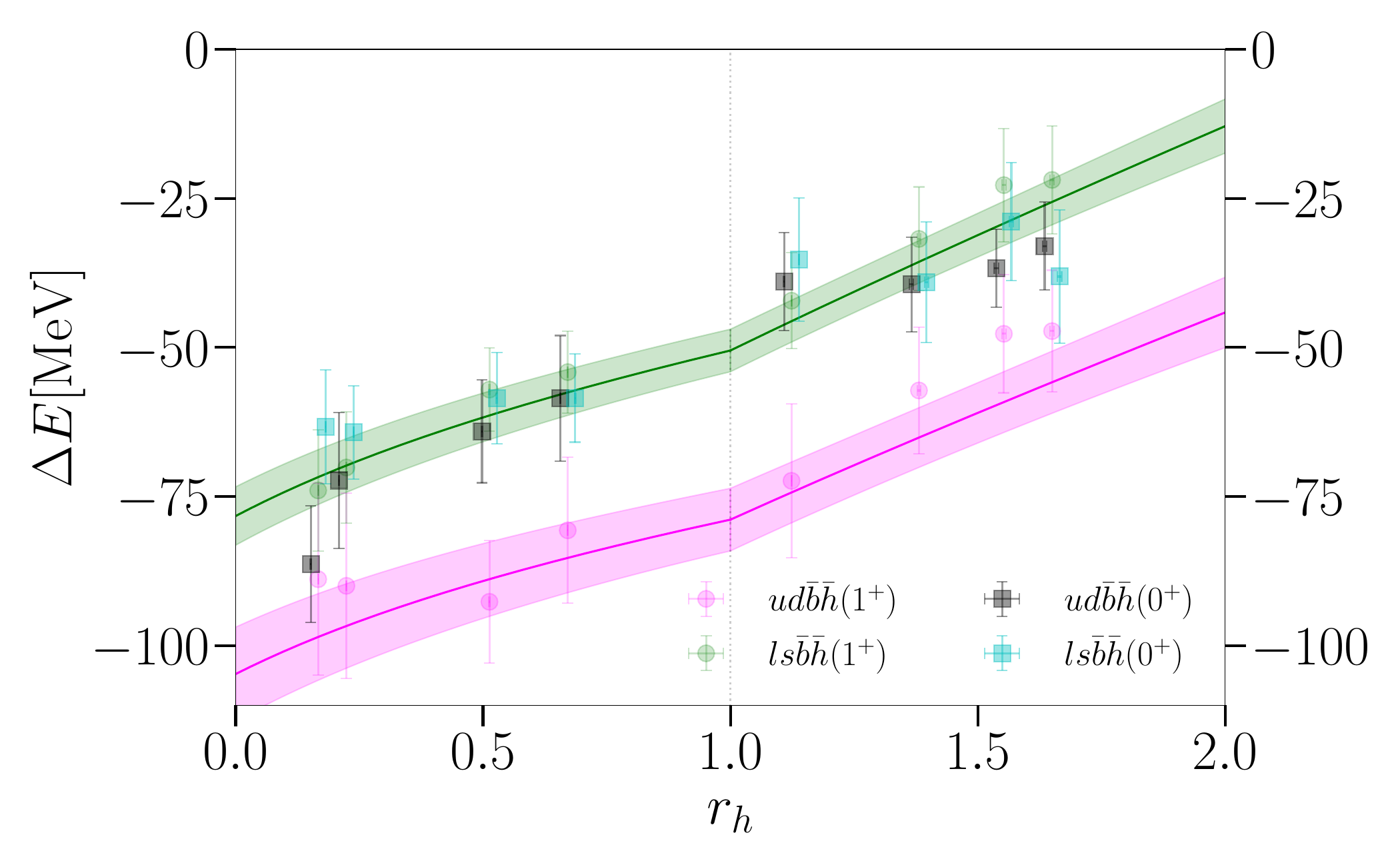}
  \caption{Binding energies of $J^P=1^+$ and $0^+$
  $\udbh$ and $\lsbh$ tetraquark states bound below 
  the lowest two-meson threshold on the E5 ensemble,
  plotted against $r_h$ in the region where we have
  data. Coloured bands are the results of correlated 
  fits to Eqs.~\eqref{eq:udbh_mh_dependence} and 
  ~\eqref{eq:lsbh_mh_dependence}. The $0^+$ 
  results are bolder and slightly offset in $r_h$ 
  for clarity.}
  \label{fig:variablebwith0}
\end{figure}

In Figure~\ref{fig:variablebwith0} we
plot results for tetraquark binding in the
$J^P=0^+$ $\udbh$ and $\lsbh$ channels, channels 
in which tetraquark configurations without 
relative heavy-quark spatial excitation are 
possible when the two heavy quarks are 
different. Also shown for comparison are the
corresponding $1^+$ $\udbh$ and $\lsbh$ results 
and associated phenomenological fit bands from 
Fig.~\ref{fig:variableb}. Tetraquark binding is
found to be similar, within errors, in the $1^+$ and
$0^+$ $\lsbh$ channels. In contrast, $0^+$ binding is found to be somewhat smaller than $1^+$ binding in the $\udbh$ channel.  
The lattice scattering study of
Ref.~\cite{Alexandrou:2023cqg} also found a similar 
comparable, though somewhat reduced, binding in the
$0^+$ than the $1^+$ $\udbc$ channel in a simulation
with $m_h=m_c$ and $m_\pi\simeq 220$ MeV. Given 
that these scattering results, the results of 
Refs.~\cite{Hudspith:2020tdf,Meinel:2022lzo}, and
the qualitative extrapolation of the phenomenological
fit to $m_h\simeq m_c$ all agree on 
at-most-weak-binding in the $1^+$ $\udbc$ channel, 
these observations suggest that, while a deeply bound 
$0^+$ $\udbc$ tetraquark would definitely exist 
were the charm quark to be sufficiently heavier 
than it is in nature, at physical $m_c$ the $0^+$
$\udbc$ state is likely either unbound or only weakly
bound. In the case that the $0^+$ $\udbc$ state 
is weakly bound, the weakness of both the $1^+$ 
and $0^+$ bindings ensures that the mass of the 
$0^+$ state will be less than that of the 
corresponding $1^+$ state. The consequences for
a possible experimental $\udbc$ tetraquark
signal of either this scenario or a scenario where 
the $1^+$ state is bound but the $0^+$ state is not 
are discussed in Sec.~\ref{discussion} below.

\section{Summary and discussion}\label{discussion}
In this paper we have revisited the earlier lattice 
determination of the binding energies of the 
flavour-antitriplet of doubly-bottom $J^P=1^+$
tetraquarks reported in Ref.~\cite{Francis:2016hui},
introducing a number of significant improvements and
producing results which update and supersede those of
Ref.~\cite{Francis:2016hui}. Relative to the previous
studies, our new analysis employs several improvements. First, rather than using local operators 
at the sink, as done in Ref.~\cite{Francis:2016hui},
we implement the improved box-sink construction of
Ref.~\cite{Hudspith:2020tdf}. As shown in
Ref.~\cite{Hudspith:2020tdf}, this construction
significantly reduces excited state contamination
of the ground-state signal, leading to much longer 
ground-state effective-mass plateaus, beginning at 
much earlier Euclidean times, than those obtained in 
the local-sink analysis of Ref.~\cite{Francis:2016hui}.
Second, we add two 
newly generated, larger-volume ensembles with 
$m_\pi<200$ MeV, aimed at reducing possible FV 
effects.
Finally, we employ an expanded sets of interpolating operators 
for the $\udbb$ and $\lsbb$ channels, and approach the correlator fits using a multi-exponential fit, rather than a GEVP.

It is difficult to separate the effects of these different changes out fully, but it is clear from~\cite{Hudspith:2020tdf} that the box sink method is likely the main driver for the shallower bindings we find with respect to previous analyses using local sinks. Whilst the additional pion masses certainly improved our uncertainties and confidence in our findings, it is clear from Fig.~\ref{fig:mpi_dependence} that without these additional data points our results would be broadly similar. Similarly, for the expanded operator basis and multi-exponential fits, it is more the case that these changes give us enhanced confidence in our results, than are directly responsible for any explicit shift in values from earlier analyses.    

Our result for the $I=0$, $J^P=1^+$ $\udbb$ binding,
$115(17)$ MeV, shows some tension with the results of 
two earlier studies, Refs.~\cite{Francis:2016hui}
and ~\cite{Mohanta:2020eed}, which found $189(10)(3)$
MeV and $187(18)$ MeV, respectively, though, as 
noted above, the former result is superseded 
by that of the current, improved analysis. Our new
result agrees within errors with all other 
recent lattice results for this channel, with 
Ref.~\cite{Junnarkar:2018twb} finding $143(34)$ MeV, 
Ref.~\cite{Leskovec:2019ioa} $128(24)(10)$ MeV, 
Ref.~\cite{Hudspith:2023loy} $112(13)$ MeV, 
Ref.~\cite{Aoki:2023nzp} $83(10)(20)$ MeV and 
Ref.~\cite{Alexandrou:2024iwi} 
$100(10)\left( {}^{+51}_{-36}\right)$ MeV. It is 
worth noting that among these the physical-point 
HALQCD result, $83(10)(20)$ MeV~\cite{Aoki:2023nzp}, 
is obtained from a significantly longer physical-point
extrapolation (of results with $m_\pi \sim 410$, $575$
and $700$ MeV) than are the results of 
Refs.~\cite{Junnarkar:2018twb,Leskovec:2019ioa,Hudspith:2023loy,Alexandrou:2024iwi} 
or the current analysis, all of which include ensembles 
with $m_\pi$ at least as low as $\sim 220$ MeV. 
A potentially large systematic error can arise from the use of NRQCD.  It cannot be estimated within our work since we have just one lattice spacing, but a previous study \cite{Hudspith:2021iqu} has found it to be of the order of 10 MeV.
 
We also note that the more recent analysis of
Ref.~\cite{Alexandrou:2024iwi} improves on that of
Ref.~\cite{Leskovec:2019ioa}, including the ``scattering
operators" introduced in Ref.~\cite{Leskovec:2019ioa},
and used there only on the sink side, now also on 
the source side.\footnote{In Ref.~\cite{Leskovec:2019ioa},
these scattering operators (the products of pairs of
independently zero-momentum-projected single-heavy-meson
interpolating operators) were shown to provide improved
access to the threshold two-meson $BB^*$ state signal.} 
Our new result confirms the
existence of the deeply-bound $I=0$, $J^P=1^+$ $\udbb$ 
tetraquark state found in earlier lattice studies 
and joins other recent lattice 
analyses~\cite{Leskovec:2019ioa,Hudspith:2023loy,Alexandrou:2024iwi}
in providing a consistent consensus for its binding energy. 
These results confirm that the state is stable with
respect to both strong and electromagnetic interactions, 
and hence can decay only via the weak interaction. It is 
thus expected to be narrow, long-lived, and have a large 
number of exclusive decay modes, 
each with a small branching fraction, making experimental
detection at current facilities extremely challenging
(see, e.g., the discussion of this point in 
Ref.~\cite{blusklhcbworkshop:2021}). An alternate,
inclusive strategy for detecting the presence
of weakly decaying doubly-bottom {\it hadrons} (via 
searches for $B_c$ mesons originating from displaced
secondary vertices at the LHC) was suggested in
Ref.~\cite{Gershon:2018gda}. Such a signal would, in 
principle, contain contributions from not just the
$J^P=1^+$ $\udbb$ tetraquark of interest, but also
doubly bottom baryons, and hence, on its own, be
insufficient to establish the presence of the $\udbb$
tetraquark signal. It would also, in principle, contain
$J^P=1^+$ $\lsbb$ tetraquark contributions, should the
$\lsbb$ state have a binding energy greater than $46$ MeV
and hence also be able to decay only weakly. In spite of
these complications, a search for such displaced $B_c$
vertex events would be of interest in setting an upper
bound on the production cross-section for weakly decaying
doubly-bottom tetraquark states and determining whether
additional efforts to identify, and separate out, the
doubly bottom baryon component of the signal would be 
worth pursuing, all this, of course, provided a good
understanding of the resolution of the tail of the impact
parameter distribution relative to the primary collision 
vertex can be achieved.{\footnote{Also worth noting in 
this context is the observation of double $\Upsilon$
production by CMS~\cite{CMS:2016liw,CMS:2020qwa}, which 
establishes the existence of a non-negligible double 
$b\bar{b}$ pair production rate, and hence the (at least 
in principle) possibility of the production of a small
number of such weakly decaying doubly bottom hadrons 
at the LHC.}} 

Our result for the $I=1/2$, $J^P=1^+$ $\lsbb$ binding,
$47(8)$ MeV, is (i) significantly lower than the earlier
lattice results, $98(7)(3)$, $87(32)$ and $86(27)(10)$
MeV, obtained in the analyses of 
Refs.~\cite{Francis:2016hui}, ~\cite{Junnarkar:2018twb}, 
and Ref.~\cite{Meinel:2022lzo}, respectively, but 
(ii) in good agreement with the more recent results,
$46(12)$ and $30(3)\left({}^{+31}_{-11}\right)$ MeV, 
obtained in Refs.~\cite{Hudspith:2023loy} and
~\cite{Alexandrou:2024iwi}. As noted above, the latter
analysis represents an improvement on the earlier analysis
of Ref.~\cite{Meinel:2022lzo}, with the ``scattering
operators" of the earlier analysis now included at both
the source and sink, rather than just the sink. The 
analysis of Ref.~\cite{Hudspith:2023loy} also improves 
on earlier analyses, employing extended sink-side
operators which produce much longer effective-mass
plateaus than those found in Refs.~\cite{Francis:2016hui} 
and ~\cite{Junnarkar:2018twb}.{\footnote{See Fig. 1
of Ref.~\cite{Junnarkar:2018twb} and the lower panel of
Fig. 1 of Ref.~\cite{Francis:2016hui} for examples of the
short, late-Euclidean-time effective mass $J^P=1^+$ $\lsbb$
channel plateaus found in those earlier references.}} The
current analysis provides, via the box-sink construction, a
similar improvement for the $J^P=1^+$ $\lsbb$ ground-state
signal. As noted above, our new result, $47(8)$ MeV, 
for the binding in this channel also supersedes that
obtained in Ref.~\cite{Francis:2016hui}. We consider
the improved features of the analyses of
Refs.~\cite{Hudspith:2023loy}, ~\cite{Alexandrou:2024iwi}
and the current analysis, and the good agreement of the
determinations of the $J^P=1^+$ $\lsbb$ binding obtained 
from these improved analyses, to establish the reliability
of the lower range of results for the $\lsbb$ binding they
produce. The binding is, however, still clearly positive,
confirming the strong-interaction-stability of the 
$J^P=1^+$ $\lsbb$ state. With the lowest of the two-meson
thresholds in this channel, $B_s B^{*+}$, lying only
$45.3(3)$ MeV above $B_s B^+$ threshold, these updated
results raise the interesting possibility that the 
$J^P=1^+$ $\lsbb$ tetraquark state might lie above the
lowest two-heavy-pseudoscalar ($B_s B$) threshold and 
hence be able to decay electromagnetically to $B_s B$ 
plus a soft photon. The limited phase space available
to such a decay would restrict the maximum relative
$B_s$-$B$ momentum in the $\lsbb$ tetraquark rest frame
to relatively low values. In contrast to the case of the 
$J^P=1^+$ $\udbb$ state, where only weak decays are
possible and one expects many exclusive decay modes, 
all with small branching fractions, an electromagnetic
$B_s B\gamma$ decay of the $J^P=1^+$ $\lsbb$ state would 
be expected to have an essentially $100\%$ branching
fraction. While the soft photon from the decay would not
be detectable at LHCb, once boosted to the lab frame
the $B_s$ and $B$ pair from the decay would be highly 
collimated and, because of the almost $100\%$ branching
fraction, provide a potentially much higher statistics
target for an experimental doubly bottom tetraquark
search than is the case for any exclusive weak decay 
of the $\udbb$ analogue.

We close with a discussion of the $I=0$, $J^P=1^+$ 
and $0^+$ $\udbc$ channels. Though we have not 
explicitly revisited these channels in this paper,
the results of the updated variable-heavy-mass study 
detailed in Sec.~\ref{heavymassdependence} provide
some additional qualitative information of relevance 
to expectations for possible experimental signatures 
in the bottom-charm sector. 

We first note that the longest-distance, one-pion-exchange
potentials between the lowest two-meson threshold states
are the same for the $I=0$, $J^P=1^+$ $\udcc$, $\udbc$ and
$\udbb$ channels~\cite{Manohar:1992nd}. At shorter 
distances, where the variable-heavy-mass results and
associated phenomenological fit provide additional
qualitative information, the corresponding 
$\bar{D}\bar{D}^*$, $\bar{D}B^*$ and $BB^*$ two-meson 
$s$-wave threshold pairs, in contrast, have access to the 
heavy-mass-dependent attractive $3_c$ color Coulomb 
$\bar{c}\bar{c}$, $\bar{c}\bar{b}$ and $\bar{b}\bar{b}$ 
interactions, as well as to the other heavy-mass-dependent 
interactions collectively parametrized by the $C_2$ term
in the phenomenological fit form. The color Coulomb 
attraction, which is proportional to the heavy-heavy
reduced mass, is more attractive in the $\udbc$ channel
than the $\udcc$ channel, though less attractive than in
the $\udbb$ channel. The shorter-distance, repulsive 
contribution proportional to $C_2$ in the $I=0$, 
$J^P=1^+$, $\udbh$ phenomenological fit form, similarly,
becomes less repulsive (hence increasing the binding) as 
the variable heavy mass, $m_h$, is increased. If we 
imagine starting from the very weakly bound $T_{cc}$ 
state, and changing the $\bar{D}^*$ of the two-meson
threshold in that channel slowly to $B^*$ by dialing up
the mass of the heavy quark in the vector meson of the 
two-meson threshold state, we thus expect (i) a small
increase in binding (reduced kinetic energy) from 
long-distance sources due to the larger $B^*\bar{D}$ 
(c.f. $\bar{D}\bar{D}^*$) reduced mass, and 
(ii) additional small increases in binding from the 
short-distance sources just discussed,
which become more attractive/less repulsive as the 
variable heavy mass is increased. These observations 
provide further support the expectation that a bound 
tetraquark with binding energy at least somewhat greater
than that of the $T_{cc}$ should exist in the $I=0$, 
$J^P=1^+$ $\udbc$ channel. Recent lattice results restrict 
how much larger this increased $1^+$ $\udbc$ binding can
be, with (i) the results of
Refs.~\cite{Hudspith:2020tdf,Pflaumer:2020ogv,Pflaumer:2021ong,Meinel:2022lzo} 
indicating at best weak binding\footnote{In the single-ensemble study of Ref.~\cite{Hudspith:2020tdf}, we were easily able to resolve an $\lsbb$ binding of 36~MeV.} and favoring results on the low side of the 
$43\left({}^{+7}_{-6}\right)\left({}^{+24}_{-14}\right)$ 
MeV range reported in Ref.~\cite{Padmanath:2023rdu}, and 
(ii) the result of the improved, more recent (albeit 
not-yet-physical-point-extrapolated) $m_\pi\sim 220$ MeV
scattering study of Ref.~\cite{Alexandrou:2023cqg}, 
suggesting an even smaller, few to several MeV, value. 
The preponderance of lattice evidence thus points to the
existence of a strong-interaction-stable $I=0$, $J^P=1^+$ 
$\udbc$ state with binding almost certainly less than the
$46$ MeV needed for its mass to lie below $B\bar{D}$ 
threshold and thus be stable also with respect to 
electromagnetic decay. The $I=0$, $J^P=1^+$ $\udbc$ state
is therefore expected to decay electromagnetically.

The lattice scattering study of 
Ref.~\cite{Alexandrou:2023cqg} also points to the
possibility of a bound $I=0$, $J^P=0^+$ $\udbc$ state,
quoting a binding relative to $B\bar{D}$ threshold of 
$0.5(9)$ MeV at $m_\pi\simeq 220$ MeV. If bound at 
physical $m_\pi$, this state would be able to decay only
weakly, with multiple exclusive decay modes, each with
relatively small branching fraction, representing a
challenge for experimental detection. The either very
weakly bound or unbound $0^+$ result of
Ref.~\cite{Alexandrou:2023cqg} is compatible with the
results of Refs.~\cite{Hudspith:2020tdf,Meinel:2022lzo},
which found no sign of a deeply bound state in this
channel, though in some tension with the larger 
$39\left({}^{+6}_{-4}\right)\left({}^{+18}_{-8}\right)$ 
MeV $0^+$ binding found in the alternate
scattering analysis of Ref.~\cite{Radhakrishnan:2024ihu}, 
where fewer interpolating operators, heavier valence 
pion masses, but some finer ensembles were employed. 

A summary of the above discussion of the $\udbc$ sector
is as follows. First, there is now good evidence for 
the existence of a strong-interaction-stable $I=0$, 
$J^P=1^+$ $\udbc$ partner of the $T_{cc}$ with binding
small enough to allow it to decay electromagnetically.
Second, though it is clear that
a bound $I=0$, $J^P=0^+$ $\udbc$ tetraquark would 
exist were the charm-quark mass to be sufficiently 
heavier than it is in nature, without a $J^P=0^+$ analogue of the $T_{cc}$
to anchor the discussion it is not yet clear 
whether binding survives in this channel at physical 
$m_c$ and $m_\pi$. If the $J^P=0^+$ state is {\it not}
bound, then the $1^+$ state should decay essentially
100\% of the time to $B\bar{D}\gamma$, producing
highly collimated $B$-$\bar{D}$ pairs in the lab
frame with a narrow range of invariant masses
lying between $m_B+m_D$ and the $1^+$ tetraquark
mass. If, in contrast, the $J^P=0^+$ state {\it is} 
bound, the $1^+$ state will decay electromagnetically 
to a soft photon plus the $0^+$ state as well as to
$B\bar{D}\gamma$. With at most weak $0^+$ binding, 
the branching fraction of the $B\bar{D}\gamma$ mode 
should, however, still be sizeable. While the 
multiple small branching fraction exclusive modes
of the weak decay of the $0^+$ state in this scenario 
will make detecting not just the production of the $0^+$
state itself, but also the $0^+$-plus-soft photon decay
branch of the $1^+$ state experimentally challenging, the
$B\bar{D}\gamma$ branch will have the same potentially
useful, statistically enhanced signal as in the 
scenario where the $0^+$ state is unbound.

\section*{Acknowledgements}
The authors wish to thank G. P. Lepage for considerable input and advice regarding correlator fitting with his software (described in text).
WP, RL and KM are supported by grants from
the Natural Sciences and Engineering Research Council of Canada. AF acknowledges support by the National Science and Technology Council of Taiwan under grant 111-2112-M-A49-018-MY2.
All computations were carried out using an allocations as part of the Digital Research Alliance of Canada on the
Niagara supercomputer at Scinet.

\appendix

\section{Operator basis}\label{app:op_basis}
The operator basis used in this work is a subset of that used in~\cite{Hudspith:2020tdf}. Copying the generalised notation in that reference for our specific case, our operators are,  
\begin{equation}\label{eq:op_defs}
\begin{gathered}
D(\Gamma_1,\Gamma_2) = (\psi_a^T C\Gamma_1 \phi_b)(\bar{\theta}_a C\Gamma_2 \bar{\omega}^T_b ),\\
E(\Gamma_1,\Gamma_2) = (\psi_a^T C\Gamma_1 \phi_b)(\bar{\theta}_a C\Gamma_2 \bar{\omega}^T_b - \bar{\theta}_b C\Gamma_2 \bar{\omega}^T_a),\\
M(\Gamma_1,\Gamma_2) = (\bar{\theta}\Gamma_1 \psi)(\bar{\omega}\Gamma_2 \phi),\qquad N(\Gamma_1,\Gamma_2) = (\bar{\theta}\Gamma_1 \phi)(\bar{\omega}\Gamma_2 \psi).
\end{gathered}
\end{equation}
The list of operators used in this work for each tetraquark channel is given in Tables~\ref{j0_j1_operator_combos} and~\ref{tab:0_ops}.
\begin{table}[ht]
\begin{tabular}{c|ccc}
\toprule
Type $(\psi \phi \theta \omega)$ & $I(J)^{P}$ & Diquark-Antidiquark & Dimeson \\
\hline
\multirow{2}{*}{$ud\bar{b}\bar{h}$} & \multirow{2}{*}{$0(1)^+$}
& \multirow{2}{*}{ $E(\gamma_5,\gamma_i),E(\gamma_t\gamma_5,\gamma_i\gamma_t)$ } &
$M(\gamma_5,\gamma_i)-N(\gamma_5,\gamma_i)$\\
& &
& $M(I,\gamma_i\gamma_5)-N(I,\gamma_i\gamma_5)$\\
\hline
\multirow{2}{*}{$ud\bar{h}\bar{h}$} & \multirow{2}{*}{$0(1)^+$}
& \multirow{2}{*}{$D(\gamma_5,\gamma_i),\;D(\gamma_t\gamma_5,\gamma_i\gamma_t)$} &
$M(\gamma_5,\gamma_i)-N(\gamma_5,\gamma_i)$\\
& & & $M(I,\gamma_i\gamma_5)-N(I,\gamma_i\gamma_5)$\\
\hline
\multirow{2}{*}{$ls\bar{h}\bar{h}$} & \multirow{2}{*}{$\frac{1}{2}(1)^+$}
& \multirow{2}{*}{$D(\gamma_5,\gamma_i),\;D(\gamma_t\gamma_5,\gamma_i\gamma_t)$} &
$M(\gamma_5,\gamma_i),\;M(I,\gamma_i\gamma_5)$\\
& & & $N(\gamma_5,\gamma_i),\;N(I,\gamma_i\gamma_5)$\\
\hline
\multirow{2}{*}{$ls\bar{b}\bar{h}$} & \multirow{2}{*}{$\frac{1}{2}(1)^+$}
& \multirow{2}{*}{$E(\gamma_5,\gamma_i),\;E(\gamma_t\gamma_5,\gamma_i\gamma_t)$} &
$M(\gamma_5,\gamma_i),\;M(I,\gamma_i\gamma_5)$\\
& & 
& $N(\gamma_5,\gamma_i),\;N(I,\gamma_i\gamma_5)$\\
\botrule
\end{tabular}
\caption{$J^P=1^+$ operators (from Eq.~\eqref{eq:op_defs}) used in this work.}
\label{j0_j1_operator_combos}
\end{table}

\begin{table}[ht]
\begin{tabular}{c|ccc}
\toprule
Type $(\psi \phi \theta \omega)$ & $I(J)^{P}$ & Diquark-Antidiquark & Dimeson \\
\hline
\multirow{2}{*}{$ud\bar{b}\bar{h}$} & \multirow{2}{*}{$0(0)^+$}
& \multirow{2}{*}{$E(\gamma_5,\gamma_5),E(\gamma_t\gamma_5,\gamma_t\gamma_5)$} &
$M(\gamma_5,\gamma_5)-N(\gamma_5,\gamma_5)$\\
& & & $M(I,I)-N(I,I)$\\
\hline
\multirow{2}{*}{$ls\bar{b}\bar{h}$} & \multirow{2}{*}{$\frac{1}{2}(0)^+$}
& \multirow{2}{*}{$E(\gamma_5,\gamma_5),\;E(\gamma_t\gamma_5,\gamma_t\gamma_5)$} &
$M(\gamma_5,\gamma_5),\;M(I,I)$\\
& & & $N(\gamma_5,\gamma_5),\;N(I,I)$\\
\botrule
\end{tabular}
\caption{$J^P=0^+$ operators (from Eq.~\eqref{eq:op_defs}) used in this work.}
\label{tab:0_ops}
\end{table}

\bibliography{sources}{}

\begin{thebibliography}{114}%
\makeatletter
\providecommand \@ifxundefined [1]{%
 \@ifx{#1\undefined}
}%
\providecommand \@ifnum [1]{%
 \ifnum #1\expandafter \@firstoftwo
 \else \expandafter \@secondoftwo
 \fi
}%
\providecommand \@ifx [1]{%
 \ifx #1\expandafter \@firstoftwo
 \else \expandafter \@secondoftwo
 \fi
}%
\providecommand \natexlab [1]{#1}%
\providecommand \enquote  [1]{``#1''}%
\providecommand \bibnamefont  [1]{#1}%
\providecommand \bibfnamefont [1]{#1}%
\providecommand \citenamefont [1]{#1}%
\providecommand \href@noop [0]{\@secondoftwo}%
\providecommand \href [0]{\begingroup \@sanitize@url \@href}%
\providecommand \@href[1]{\@@startlink{#1}\@@href}%
\providecommand \@@href[1]{\endgroup#1\@@endlink}%
\providecommand \@sanitize@url [0]{\catcode `\\12\catcode `\$12\catcode
  `\&12\catcode `\#12\catcode `\^12\catcode `\_12\catcode `\%12\relax}%
\providecommand \@@startlink[1]{}%
\providecommand \@@endlink[0]{}%
\providecommand \url  [0]{\begingroup\@sanitize@url \@url }%
\providecommand \@url [1]{\endgroup\@href {#1}{\urlprefix }}%
\providecommand \urlprefix  [0]{URL }%
\providecommand \Eprint [0]{\href }%
\providecommand \doibase [0]{http://dx.doi.org/}%
\providecommand \selectlanguage [0]{\@gobble}%
\providecommand \bibinfo  [0]{\@secondoftwo}%
\providecommand \bibfield  [0]{\@secondoftwo}%
\providecommand \translation [1]{[#1]}%
\providecommand \BibitemOpen [0]{}%
\providecommand \bibitemStop [0]{}%
\providecommand \bibitemNoStop [0]{.\EOS\space}%
\providecommand \EOS [0]{\spacefactor3000\relax}%
\providecommand \BibitemShut  [1]{\csname bibitem#1\endcsname}%
\let\auto@bib@innerbib\@empty
\bibitem [{\citenamefont {Ader}\ \emph {et~al.}(1982)\citenamefont {Ader},
  \citenamefont {Richard},\ and\ \citenamefont {Taxil}}]{Ader:1981db}%
  \BibitemOpen
  \bibfield  {author} {\bibinfo {author} {\bibfnamefont {J.}~\bibnamefont
  {Ader}}, \bibinfo {author} {\bibfnamefont {J.}~\bibnamefont {Richard}}, \
  and\ \bibinfo {author} {\bibfnamefont {P.}~\bibnamefont {Taxil}},\ }\href
  {\doibase 10.1103/PhysRevD.25.2370} {\bibfield  {journal} {\bibinfo
  {journal} {Phys. Rev. D}\ }\textbf {\bibinfo {volume} {25}},\ \bibinfo
  {pages} {2370} (\bibinfo {year} {1982})}\BibitemShut {NoStop}%
\bibitem [{\citenamefont {Heller}\ and\ \citenamefont
  {Tjon}(1987)}]{Heller:1986bt}%
  \BibitemOpen
  \bibfield  {author} {\bibinfo {author} {\bibfnamefont {L.}~\bibnamefont
  {Heller}}\ and\ \bibinfo {author} {\bibfnamefont {J.}~\bibnamefont {Tjon}},\
  }\href {\doibase 10.1103/PhysRevD.35.969} {\bibfield  {journal} {\bibinfo
  {journal} {Phys. Rev. D}\ }\textbf {\bibinfo {volume} {35}},\ \bibinfo
  {pages} {969} (\bibinfo {year} {1987})}\BibitemShut {NoStop}%
\bibitem [{\citenamefont {Manohar}\ and\ \citenamefont
  {Wise}(1993)}]{Manohar:1992nd}%
  \BibitemOpen
  \bibfield  {author} {\bibinfo {author} {\bibfnamefont {A.~V.}\ \bibnamefont
  {Manohar}}\ and\ \bibinfo {author} {\bibfnamefont {M.~B.}\ \bibnamefont
  {Wise}},\ }\href {\doibase 10.1016/0550-3213(93)90614-U} {\bibfield
  {journal} {\bibinfo  {journal} {Nucl. Phys. B}\ }\textbf {\bibinfo {volume}
  {399}},\ \bibinfo {pages} {17} (\bibinfo {year} {1993})},\ \Eprint
  {http://arxiv.org/abs/hep-ph/9212236} {arXiv:hep-ph/9212236} \BibitemShut
  {NoStop}%
\bibitem [{\citenamefont {Richards}\ \emph {et~al.}(1990)\citenamefont
  {Richards}, \citenamefont {Sinclair},\ and\ \citenamefont
  {Sivers}}]{Richards:1990xf}%
  \BibitemOpen
  \bibfield  {author} {\bibinfo {author} {\bibfnamefont {D.~G.}\ \bibnamefont
  {Richards}}, \bibinfo {author} {\bibfnamefont {D.~K.}\ \bibnamefont
  {Sinclair}}, \ and\ \bibinfo {author} {\bibfnamefont {D.~W.}\ \bibnamefont
  {Sivers}},\ }\href {\doibase 10.1103/PhysRevD.42.3191} {\bibfield  {journal}
  {\bibinfo  {journal} {Phys. Rev.}\ }\textbf {\bibinfo {volume} {D42}},\
  \bibinfo {pages} {3191} (\bibinfo {year} {1990})}\BibitemShut {NoStop}%
\bibitem [{\citenamefont {Mihaly}\ \emph {et~al.}(1997)\citenamefont {Mihaly},
  \citenamefont {Fiebig}, \citenamefont {Markum},\ and\ \citenamefont
  {Rabitsch}}]{Mihaly:1996ue}%
  \BibitemOpen
  \bibfield  {author} {\bibinfo {author} {\bibfnamefont {A.}~\bibnamefont
  {Mihaly}}, \bibinfo {author} {\bibfnamefont {H.~R.}\ \bibnamefont {Fiebig}},
  \bibinfo {author} {\bibfnamefont {H.}~\bibnamefont {Markum}}, \ and\ \bibinfo
  {author} {\bibfnamefont {K.}~\bibnamefont {Rabitsch}},\ }\href {\doibase
  10.1103/PhysRevD.55.3077} {\bibfield  {journal} {\bibinfo  {journal} {Phys.
  Rev.}\ }\textbf {\bibinfo {volume} {D55}},\ \bibinfo {pages} {3077} (\bibinfo
  {year} {1997})}\BibitemShut {NoStop}%
\bibitem [{\citenamefont {Green}\ and\ \citenamefont
  {Pennanen}(1998)}]{Green:1998nt}%
  \BibitemOpen
  \bibfield  {author} {\bibinfo {author} {\bibfnamefont {A.~M.}\ \bibnamefont
  {Green}}\ and\ \bibinfo {author} {\bibfnamefont {P.}~\bibnamefont
  {Pennanen}},\ }\href {\doibase 10.1103/PhysRevC.57.3384} {\bibfield
  {journal} {\bibinfo  {journal} {Phys. Rev.}\ }\textbf {\bibinfo {volume}
  {C57}},\ \bibinfo {pages} {3384} (\bibinfo {year} {1998})},\ \Eprint
  {http://arxiv.org/abs/hep-lat/9804003} {arXiv:hep-lat/9804003 [hep-lat]}
  \BibitemShut {NoStop}%
\bibitem [{\citenamefont {Stewart}\ and\ \citenamefont
  {Koniuk}(1998)}]{Stewart:1998hk}%
  \BibitemOpen
  \bibfield  {author} {\bibinfo {author} {\bibfnamefont {C.}~\bibnamefont
  {Stewart}}\ and\ \bibinfo {author} {\bibfnamefont {R.}~\bibnamefont
  {Koniuk}},\ }\href {\doibase 10.1103/PhysRevD.57.5581} {\bibfield  {journal}
  {\bibinfo  {journal} {Phys. Rev.}\ }\textbf {\bibinfo {volume} {D57}},\
  \bibinfo {pages} {5581} (\bibinfo {year} {1998})},\ \Eprint
  {http://arxiv.org/abs/hep-lat/9803003} {arXiv:hep-lat/9803003 [hep-lat]}
  \BibitemShut {NoStop}%
\bibitem [{\citenamefont {Michael}\ and\ \citenamefont
  {Pennanen}(1999)}]{Michael:1999nq}%
  \BibitemOpen
  \bibfield  {author} {\bibinfo {author} {\bibfnamefont {C.}~\bibnamefont
  {Michael}}\ and\ \bibinfo {author} {\bibfnamefont {P.}~\bibnamefont
  {Pennanen}} (\bibinfo {collaboration} {UKQCD}),\ }\href {\doibase
  10.1103/PhysRevD.60.054012} {\bibfield  {journal} {\bibinfo  {journal} {Phys.
  Rev.}\ }\textbf {\bibinfo {volume} {D60}},\ \bibinfo {pages} {054012}
  (\bibinfo {year} {1999})},\ \Eprint {http://arxiv.org/abs/hep-lat/9901007}
  {arXiv:hep-lat/9901007 [hep-lat]} \BibitemShut {NoStop}%
\bibitem [{\citenamefont {Pennanen}\ \emph {et~al.}(2000)\citenamefont
  {Pennanen}, \citenamefont {Michael},\ and\ \citenamefont
  {Green}}]{Pennanen:1999xi}%
  \BibitemOpen
  \bibfield  {author} {\bibinfo {author} {\bibfnamefont {P.}~\bibnamefont
  {Pennanen}}, \bibinfo {author} {\bibfnamefont {C.}~\bibnamefont {Michael}}, \
  and\ \bibinfo {author} {\bibfnamefont {A.~M.}\ \bibnamefont {Green}}
  (\bibinfo {collaboration} {UKQCD}),\ }\href {\doibase
  10.1016/S0920-5632(00)91622-0} {\bibfield  {journal} {\bibinfo  {journal}
  {Nucl. Phys. Proc. Suppl.}\ }\textbf {\bibinfo {volume} {83}},\ \bibinfo
  {pages} {200} (\bibinfo {year} {2000})},\ \Eprint
  {http://arxiv.org/abs/hep-lat/9908032} {arXiv:hep-lat/9908032 [hep-lat]}
  \BibitemShut {NoStop}%
\bibitem [{\citenamefont {Cook}\ and\ \citenamefont
  {Fiebig}(2002)}]{Cook:2002am}%
  \BibitemOpen
  \bibfield  {author} {\bibinfo {author} {\bibfnamefont {M.~S.}\ \bibnamefont
  {Cook}}\ and\ \bibinfo {author} {\bibfnamefont {H.~R.}\ \bibnamefont
  {Fiebig}},\ }\href@noop {} {\  (\bibinfo {year} {2002})},\ \Eprint
  {http://arxiv.org/abs/hep-lat/0210054} {arXiv:hep-lat/0210054 [hep-lat]}
  \BibitemShut {NoStop}%
\bibitem [{\citenamefont {Doi}\ \emph {et~al.}(2006)\citenamefont {Doi},
  \citenamefont {Takahashi},\ and\ \citenamefont {Suganuma}}]{Doi:2006kx}%
  \BibitemOpen
  \bibfield  {author} {\bibinfo {author} {\bibfnamefont {T.}~\bibnamefont
  {Doi}}, \bibinfo {author} {\bibfnamefont {T.~T.}\ \bibnamefont {Takahashi}},
  \ and\ \bibinfo {author} {\bibfnamefont {H.}~\bibnamefont {Suganuma}},\
  }\href {\doibase 10.1063/1.2220239} {\bibfield  {journal} {\bibinfo
  {journal} {AIP Conf. Proc.}\ }\textbf {\bibinfo {volume} {842}},\ \bibinfo
  {pages} {246} (\bibinfo {year} {2006})},\ \Eprint
  {http://arxiv.org/abs/hep-lat/0601008} {arXiv:hep-lat/0601008} \BibitemShut
  {NoStop}%
\bibitem [{\citenamefont {Detmold}\ \emph {et~al.}(2007)\citenamefont
  {Detmold}, \citenamefont {Orginos},\ and\ \citenamefont
  {Savage}}]{Detmold:2007wk}%
  \BibitemOpen
  \bibfield  {author} {\bibinfo {author} {\bibfnamefont {W.}~\bibnamefont
  {Detmold}}, \bibinfo {author} {\bibfnamefont {K.}~\bibnamefont {Orginos}}, \
  and\ \bibinfo {author} {\bibfnamefont {M.~J.}\ \bibnamefont {Savage}},\
  }\href {\doibase 10.1103/PhysRevD.76.114503} {\bibfield  {journal} {\bibinfo
  {journal} {Phys. Rev.}\ }\textbf {\bibinfo {volume} {D76}},\ \bibinfo {pages}
  {114503} (\bibinfo {year} {2007})},\ \Eprint
  {http://arxiv.org/abs/hep-lat/0703009} {arXiv:hep-lat/0703009 [HEP-LAT]}
  \BibitemShut {NoStop}%
\bibitem [{\citenamefont {Carlson}\ \emph {et~al.}(1988)\citenamefont
  {Carlson}, \citenamefont {Heller},\ and\ \citenamefont
  {Tjon}}]{Carlson:1987hh}%
  \BibitemOpen
  \bibfield  {author} {\bibinfo {author} {\bibfnamefont {J.}~\bibnamefont
  {Carlson}}, \bibinfo {author} {\bibfnamefont {L.}~\bibnamefont {Heller}}, \
  and\ \bibinfo {author} {\bibfnamefont {J.}~\bibnamefont {Tjon}},\ }\href
  {\doibase 10.1103/PhysRevD.37.744} {\bibfield  {journal} {\bibinfo  {journal}
  {Phys. Rev. D}\ }\textbf {\bibinfo {volume} {37}},\ \bibinfo {pages} {744}
  (\bibinfo {year} {1988})}\BibitemShut {NoStop}%
\bibitem [{\citenamefont {Zouzou}\ \emph {et~al.}(1986)\citenamefont {Zouzou},
  \citenamefont {Silvestre-Brac}, \citenamefont {Gignoux},\ and\ \citenamefont
  {Richard}}]{Zouzou:1986qh}%
  \BibitemOpen
  \bibfield  {author} {\bibinfo {author} {\bibfnamefont {S.}~\bibnamefont
  {Zouzou}}, \bibinfo {author} {\bibfnamefont {B.}~\bibnamefont
  {Silvestre-Brac}}, \bibinfo {author} {\bibfnamefont {C.}~\bibnamefont
  {Gignoux}}, \ and\ \bibinfo {author} {\bibfnamefont {J.}~\bibnamefont
  {Richard}},\ }\href {\doibase 10.1007/BF01557611} {\bibfield  {journal}
  {\bibinfo  {journal} {Z. Phys. C}\ }\textbf {\bibinfo {volume} {30}},\
  \bibinfo {pages} {457} (\bibinfo {year} {1986})}\BibitemShut {NoStop}%
\bibitem [{\citenamefont {Lipkin}(1986)}]{Lipkin:1986dw}%
  \BibitemOpen
  \bibfield  {author} {\bibinfo {author} {\bibfnamefont {H.~J.}\ \bibnamefont
  {Lipkin}},\ }\href {\doibase 10.1016/0370-2693(86)90843-9} {\bibfield
  {journal} {\bibinfo  {journal} {Phys. Lett. B}\ }\textbf {\bibinfo {volume}
  {172}},\ \bibinfo {pages} {242} (\bibinfo {year} {1986})}\BibitemShut
  {NoStop}%
\bibitem [{\citenamefont {Silvestre-Brac}\ and\ \citenamefont
  {Semay}(1993)}]{SilvestreBrac:1993ss}%
  \BibitemOpen
  \bibfield  {author} {\bibinfo {author} {\bibfnamefont {B.}~\bibnamefont
  {Silvestre-Brac}}\ and\ \bibinfo {author} {\bibfnamefont {C.}~\bibnamefont
  {Semay}},\ }\href {\doibase 10.1007/BF01565058} {\bibfield  {journal}
  {\bibinfo  {journal} {Z. Phys. C}\ }\textbf {\bibinfo {volume} {57}},\
  \bibinfo {pages} {273} (\bibinfo {year} {1993})}\BibitemShut {NoStop}%
\bibitem [{\citenamefont {Semay}\ and\ \citenamefont
  {Silvestre-Brac}(1994)}]{Semay:1994ht}%
  \BibitemOpen
  \bibfield  {author} {\bibinfo {author} {\bibfnamefont {C.}~\bibnamefont
  {Semay}}\ and\ \bibinfo {author} {\bibfnamefont {B.}~\bibnamefont
  {Silvestre-Brac}},\ }\href {\doibase 10.1007/BF01413104} {\bibfield
  {journal} {\bibinfo  {journal} {Z. Phys. C}\ }\textbf {\bibinfo {volume}
  {61}},\ \bibinfo {pages} {271} (\bibinfo {year} {1994})}\BibitemShut
  {NoStop}%
\bibitem [{\citenamefont {Pepin}\ \emph {et~al.}(1997)\citenamefont {Pepin},
  \citenamefont {Stancu}, \citenamefont {Genovese},\ and\ \citenamefont
  {Richard}}]{Pepin:1996id}%
  \BibitemOpen
  \bibfield  {author} {\bibinfo {author} {\bibfnamefont {S.}~\bibnamefont
  {Pepin}}, \bibinfo {author} {\bibfnamefont {F.}~\bibnamefont {Stancu}},
  \bibinfo {author} {\bibfnamefont {M.}~\bibnamefont {Genovese}}, \ and\
  \bibinfo {author} {\bibfnamefont {J.}~\bibnamefont {Richard}},\ }\href
  {\doibase 10.1016/S0370-2693(96)01597-3} {\bibfield  {journal} {\bibinfo
  {journal} {Phys. Lett. B}\ }\textbf {\bibinfo {volume} {393}},\ \bibinfo
  {pages} {119} (\bibinfo {year} {1997})},\ \Eprint
  {http://arxiv.org/abs/hep-ph/9609348} {arXiv:hep-ph/9609348} \BibitemShut
  {NoStop}%
\bibitem [{\citenamefont {Brink}\ and\ \citenamefont
  {Stancu}(1998)}]{Brink:1998as}%
  \BibitemOpen
  \bibfield  {author} {\bibinfo {author} {\bibfnamefont {D.}~\bibnamefont
  {Brink}}\ and\ \bibinfo {author} {\bibfnamefont {F.}~\bibnamefont {Stancu}},\
  }\href {\doibase 10.1103/PhysRevD.57.6778} {\bibfield  {journal} {\bibinfo
  {journal} {Phys. Rev. D}\ }\textbf {\bibinfo {volume} {57}},\ \bibinfo
  {pages} {6778} (\bibinfo {year} {1998})}\BibitemShut {NoStop}%
\bibitem [{\citenamefont {Vijande}\ \emph {et~al.}(2004)\citenamefont
  {Vijande}, \citenamefont {Fernandez}, \citenamefont {Valcarce},\ and\
  \citenamefont {Silvestre-Brac}}]{Vijande:2003ki}%
  \BibitemOpen
  \bibfield  {author} {\bibinfo {author} {\bibfnamefont {J.}~\bibnamefont
  {Vijande}}, \bibinfo {author} {\bibfnamefont {F.}~\bibnamefont {Fernandez}},
  \bibinfo {author} {\bibfnamefont {A.}~\bibnamefont {Valcarce}}, \ and\
  \bibinfo {author} {\bibfnamefont {B.}~\bibnamefont {Silvestre-Brac}},\ }\href
  {\doibase 10.1140/epja/i2003-10128-9} {\bibfield  {journal} {\bibinfo
  {journal} {Eur. Phys. J. A}\ }\textbf {\bibinfo {volume} {19}},\ \bibinfo
  {pages} {383} (\bibinfo {year} {2004})},\ \Eprint
  {http://arxiv.org/abs/hep-ph/0310007} {arXiv:hep-ph/0310007} \BibitemShut
  {NoStop}%
\bibitem [{\citenamefont {Janc}\ and\ \citenamefont
  {Rosina}(2004)}]{Janc:2004qn}%
  \BibitemOpen
  \bibfield  {author} {\bibinfo {author} {\bibfnamefont {D.}~\bibnamefont
  {Janc}}\ and\ \bibinfo {author} {\bibfnamefont {M.}~\bibnamefont {Rosina}},\
  }\href {\doibase 10.1007/s00601-004-0068-9} {\bibfield  {journal} {\bibinfo
  {journal} {Few Body Syst.}\ }\textbf {\bibinfo {volume} {35}},\ \bibinfo
  {pages} {175} (\bibinfo {year} {2004})},\ \Eprint
  {http://arxiv.org/abs/hep-ph/0405208} {arXiv:hep-ph/0405208} \BibitemShut
  {NoStop}%
\bibitem [{\citenamefont {Vijande}\ \emph {et~al.}(2007)\citenamefont
  {Vijande}, \citenamefont {Weissman}, \citenamefont {Valcarce},\ and\
  \citenamefont {Barnea}}]{Vijande:2007rf}%
  \BibitemOpen
  \bibfield  {author} {\bibinfo {author} {\bibfnamefont {J.}~\bibnamefont
  {Vijande}}, \bibinfo {author} {\bibfnamefont {E.}~\bibnamefont {Weissman}},
  \bibinfo {author} {\bibfnamefont {A.}~\bibnamefont {Valcarce}}, \ and\
  \bibinfo {author} {\bibfnamefont {N.}~\bibnamefont {Barnea}},\ }\href
  {\doibase 10.1103/PhysRevD.76.094027} {\bibfield  {journal} {\bibinfo
  {journal} {Phys. Rev. D}\ }\textbf {\bibinfo {volume} {76}},\ \bibinfo
  {pages} {094027} (\bibinfo {year} {2007})},\ \Eprint
  {http://arxiv.org/abs/0710.2516} {arXiv:0710.2516 [hep-ph]} \BibitemShut
  {NoStop}%
\bibitem [{\citenamefont {Ebert}\ \emph {et~al.}(2007)\citenamefont {Ebert},
  \citenamefont {Faustov}, \citenamefont {Galkin},\ and\ \citenamefont
  {Lucha}}]{Ebert:2007rn}%
  \BibitemOpen
  \bibfield  {author} {\bibinfo {author} {\bibfnamefont {D.}~\bibnamefont
  {Ebert}}, \bibinfo {author} {\bibfnamefont {R.}~\bibnamefont {Faustov}},
  \bibinfo {author} {\bibfnamefont {V.}~\bibnamefont {Galkin}}, \ and\ \bibinfo
  {author} {\bibfnamefont {W.}~\bibnamefont {Lucha}},\ }\href {\doibase
  10.1103/PhysRevD.76.114015} {\bibfield  {journal} {\bibinfo  {journal} {Phys.
  Rev. D}\ }\textbf {\bibinfo {volume} {76}},\ \bibinfo {pages} {114015}
  (\bibinfo {year} {2007})},\ \Eprint {http://arxiv.org/abs/0706.3853}
  {arXiv:0706.3853 [hep-ph]} \BibitemShut {NoStop}%
\bibitem [{\citenamefont {Zhang}\ \emph {et~al.}(2008)\citenamefont {Zhang},
  \citenamefont {Zhang},\ and\ \citenamefont {Zhang}}]{Zhang:2007mu}%
  \BibitemOpen
  \bibfield  {author} {\bibinfo {author} {\bibfnamefont {M.}~\bibnamefont
  {Zhang}}, \bibinfo {author} {\bibfnamefont {H.}~\bibnamefont {Zhang}}, \ and\
  \bibinfo {author} {\bibfnamefont {Z.}~\bibnamefont {Zhang}},\ }\href
  {\doibase 10.1088/0253-6102/50/2/31} {\bibfield  {journal} {\bibinfo
  {journal} {Commun. Theor. Phys.}\ }\textbf {\bibinfo {volume} {50}},\
  \bibinfo {pages} {437} (\bibinfo {year} {2008})},\ \Eprint
  {http://arxiv.org/abs/0711.1029} {arXiv:0711.1029 [nucl-th]} \BibitemShut
  {NoStop}%
\bibitem [{\citenamefont {Vijande}\ \emph {et~al.}(2009)\citenamefont
  {Vijande}, \citenamefont {Valcarce},\ and\ \citenamefont
  {Barnea}}]{Vijande:2009kj}%
  \BibitemOpen
  \bibfield  {author} {\bibinfo {author} {\bibfnamefont {J.}~\bibnamefont
  {Vijande}}, \bibinfo {author} {\bibfnamefont {A.}~\bibnamefont {Valcarce}}, \
  and\ \bibinfo {author} {\bibfnamefont {N.}~\bibnamefont {Barnea}},\ }\href
  {\doibase 10.1103/PhysRevD.79.074010} {\bibfield  {journal} {\bibinfo
  {journal} {Phys. Rev. D}\ }\textbf {\bibinfo {volume} {79}},\ \bibinfo
  {pages} {074010} (\bibinfo {year} {2009})},\ \Eprint
  {http://arxiv.org/abs/0903.2949} {arXiv:0903.2949 [hep-ph]} \BibitemShut
  {NoStop}%
\bibitem [{\citenamefont {Yang}\ \emph {et~al.}(2009)\citenamefont {Yang},
  \citenamefont {Deng}, \citenamefont {Ping},\ and\ \citenamefont
  {Goldman}}]{Yang:2009zzp}%
  \BibitemOpen
  \bibfield  {author} {\bibinfo {author} {\bibfnamefont {Y.}~\bibnamefont
  {Yang}}, \bibinfo {author} {\bibfnamefont {C.}~\bibnamefont {Deng}}, \bibinfo
  {author} {\bibfnamefont {J.}~\bibnamefont {Ping}}, \ and\ \bibinfo {author}
  {\bibfnamefont {T.}~\bibnamefont {Goldman}},\ }\href {\doibase
  10.1103/PhysRevD.80.114023} {\bibfield  {journal} {\bibinfo  {journal} {Phys.
  Rev. D}\ }\textbf {\bibinfo {volume} {80}},\ \bibinfo {pages} {114023}
  (\bibinfo {year} {2009})}\BibitemShut {NoStop}%
\bibitem [{\citenamefont {Carames}\ \emph {et~al.}(2011)\citenamefont
  {Carames}, \citenamefont {Valcarce},\ and\ \citenamefont
  {Vijande}}]{Carames:2011zz}%
  \BibitemOpen
  \bibfield  {author} {\bibinfo {author} {\bibfnamefont {T.}~\bibnamefont
  {Carames}}, \bibinfo {author} {\bibfnamefont {A.}~\bibnamefont {Valcarce}}, \
  and\ \bibinfo {author} {\bibfnamefont {J.}~\bibnamefont {Vijande}},\ }\href
  {\doibase 10.1016/j.physletb.2011.04.023} {\bibfield  {journal} {\bibinfo
  {journal} {Phys. Lett. B}\ }\textbf {\bibinfo {volume} {699}},\ \bibinfo
  {pages} {291} (\bibinfo {year} {2011})}\BibitemShut {NoStop}%
\bibitem [{\citenamefont {Silbar}\ and\ \citenamefont
  {Goldman}(2014)}]{Silbar:2013dda}%
  \BibitemOpen
  \bibfield  {author} {\bibinfo {author} {\bibfnamefont {R.~R.}\ \bibnamefont
  {Silbar}}\ and\ \bibinfo {author} {\bibfnamefont {T.}~\bibnamefont
  {Goldman}},\ }\href {\doibase 10.1142/S0218301314500918} {\bibfield
  {journal} {\bibinfo  {journal} {Int. J. Mod. Phys. E}\ }\textbf {\bibinfo
  {volume} {23}},\ \bibinfo {pages} {1450091} (\bibinfo {year} {2014})},\
  \Eprint {http://arxiv.org/abs/1304.5480} {arXiv:1304.5480 [nucl-th]}
  \BibitemShut {NoStop}%
\bibitem [{\citenamefont {Karliner}\ and\ \citenamefont
  {Rosner}(2017)}]{Karliner:2017qjm}%
  \BibitemOpen
  \bibfield  {author} {\bibinfo {author} {\bibfnamefont {M.}~\bibnamefont
  {Karliner}}\ and\ \bibinfo {author} {\bibfnamefont {J.~L.}\ \bibnamefont
  {Rosner}},\ }\href {\doibase 10.1103/PhysRevLett.119.202001} {\bibfield
  {journal} {\bibinfo  {journal} {Phys. Rev. Lett.}\ }\textbf {\bibinfo
  {volume} {119}},\ \bibinfo {pages} {202001} (\bibinfo {year} {2017})},\
  \Eprint {http://arxiv.org/abs/1707.07666} {arXiv:1707.07666 [hep-ph]}
  \BibitemShut {NoStop}%
\bibitem [{\citenamefont {Caram\'es}\ \emph {et~al.}(2019)\citenamefont
  {Caram\'es}, \citenamefont {Vijande},\ and\ \citenamefont
  {Valcarce}}]{Caramees:2018oue}%
  \BibitemOpen
  \bibfield  {author} {\bibinfo {author} {\bibfnamefont {T.~F.}\ \bibnamefont
  {Caram\'es}}, \bibinfo {author} {\bibfnamefont {J.}~\bibnamefont {Vijande}},
  \ and\ \bibinfo {author} {\bibfnamefont {A.}~\bibnamefont {Valcarce}},\
  }\href {\doibase 10.1103/PhysRevD.99.014006} {\bibfield  {journal} {\bibinfo
  {journal} {Phys. Rev. D}\ }\textbf {\bibinfo {volume} {99}},\ \bibinfo
  {pages} {014006} (\bibinfo {year} {2019})},\ \Eprint
  {http://arxiv.org/abs/1812.08991} {arXiv:1812.08991 [hep-ph]} \BibitemShut
  {NoStop}%
\bibitem [{\citenamefont {Deng}\ \emph {et~al.}(2020)\citenamefont {Deng},
  \citenamefont {Chen},\ and\ \citenamefont {Ping}}]{Deng:2018kly}%
  \BibitemOpen
  \bibfield  {author} {\bibinfo {author} {\bibfnamefont {C.}~\bibnamefont
  {Deng}}, \bibinfo {author} {\bibfnamefont {H.}~\bibnamefont {Chen}}, \ and\
  \bibinfo {author} {\bibfnamefont {J.}~\bibnamefont {Ping}},\ }\href {\doibase
  10.1140/epja/s10050-019-00012-y} {\bibfield  {journal} {\bibinfo  {journal}
  {Eur. Phys. J. A}\ }\textbf {\bibinfo {volume} {56}},\ \bibinfo {pages} {9}
  (\bibinfo {year} {2020})},\ \Eprint {http://arxiv.org/abs/1811.06462}
  {arXiv:1811.06462 [hep-ph]} \BibitemShut {NoStop}%
\bibitem [{\citenamefont {Park}\ \emph {et~al.}(2019)\citenamefont {Park},
  \citenamefont {Noh},\ and\ \citenamefont {Lee}}]{Park:2018wjk}%
  \BibitemOpen
  \bibfield  {author} {\bibinfo {author} {\bibfnamefont {W.}~\bibnamefont
  {Park}}, \bibinfo {author} {\bibfnamefont {S.}~\bibnamefont {Noh}}, \ and\
  \bibinfo {author} {\bibfnamefont {S.~H.}\ \bibnamefont {Lee}},\ }\href
  {\doibase 10.1016/j.nuclphysa.2018.12.019} {\bibfield  {journal} {\bibinfo
  {journal} {Nucl. Phys. A}\ }\textbf {\bibinfo {volume} {983}},\ \bibinfo
  {pages} {1} (\bibinfo {year} {2019})},\ \Eprint
  {http://arxiv.org/abs/1809.05257} {arXiv:1809.05257 [nucl-th]} \BibitemShut
  {NoStop}%
\bibitem [{\citenamefont {Yang}\ \emph {et~al.}(2020)\citenamefont {Yang},
  \citenamefont {Ping},\ and\ \citenamefont {Segovia}}]{Yang:2019itm}%
  \BibitemOpen
  \bibfield  {author} {\bibinfo {author} {\bibfnamefont {G.}~\bibnamefont
  {Yang}}, \bibinfo {author} {\bibfnamefont {J.}~\bibnamefont {Ping}}, \ and\
  \bibinfo {author} {\bibfnamefont {J.}~\bibnamefont {Segovia}},\ }\href
  {\doibase 10.1103/PhysRevD.101.014001} {\bibfield  {journal} {\bibinfo
  {journal} {Phys. Rev. D}\ }\textbf {\bibinfo {volume} {101}},\ \bibinfo
  {pages} {014001} (\bibinfo {year} {2020})},\ \Eprint
  {http://arxiv.org/abs/1911.00215} {arXiv:1911.00215 [hep-ph]} \BibitemShut
  {NoStop}%
\bibitem [{\citenamefont {Hernández}\ \emph {et~al.}(2020)\citenamefont
  {Hernández}, \citenamefont {Vijande}, \citenamefont {Valcarce},\ and\
  \citenamefont {Richard}}]{Hernandez:2019eox}%
  \BibitemOpen
  \bibfield  {author} {\bibinfo {author} {\bibfnamefont {E.}~\bibnamefont
  {Hernández}}, \bibinfo {author} {\bibfnamefont {J.}~\bibnamefont {Vijande}},
  \bibinfo {author} {\bibfnamefont {A.}~\bibnamefont {Valcarce}}, \ and\
  \bibinfo {author} {\bibfnamefont {J.-M.}\ \bibnamefont {Richard}},\ }\href
  {\doibase 10.1016/j.physletb.2019.135073} {\bibfield  {journal} {\bibinfo
  {journal} {Phys. Lett. B}\ }\textbf {\bibinfo {volume} {800}},\ \bibinfo
  {pages} {135073} (\bibinfo {year} {2020})},\ \Eprint
  {http://arxiv.org/abs/1910.13394} {arXiv:1910.13394 [hep-ph]} \BibitemShut
  {NoStop}%
\bibitem [{\citenamefont {Tan}\ \emph {et~al.}(2020)\citenamefont {Tan},
  \citenamefont {Lu},\ and\ \citenamefont {Ping}}]{Tan:2020ldi}%
  \BibitemOpen
  \bibfield  {author} {\bibinfo {author} {\bibfnamefont {Y.}~\bibnamefont
  {Tan}}, \bibinfo {author} {\bibfnamefont {W.}~\bibnamefont {Lu}}, \ and\
  \bibinfo {author} {\bibfnamefont {J.}~\bibnamefont {Ping}},\ }\href {\doibase
  10.1140/epjp/s13360-020-00741-w} {\bibfield  {journal} {\bibinfo  {journal}
  {Eur. Phys. J. Plus}\ }\textbf {\bibinfo {volume} {135}},\ \bibinfo {pages}
  {716} (\bibinfo {year} {2020})},\ \Eprint {http://arxiv.org/abs/2004.02106}
  {arXiv:2004.02106 [hep-ph]} \BibitemShut {NoStop}%
\bibitem [{\citenamefont {L\"u}\ \emph {et~al.}(2020)\citenamefont {L\"u},
  \citenamefont {Chen},\ and\ \citenamefont {Dong}}]{Lu:2020rog}%
  \BibitemOpen
  \bibfield  {author} {\bibinfo {author} {\bibfnamefont {Q.-F.}\ \bibnamefont
  {L\"u}}, \bibinfo {author} {\bibfnamefont {D.-Y.}\ \bibnamefont {Chen}}, \
  and\ \bibinfo {author} {\bibfnamefont {Y.-B.}\ \bibnamefont {Dong}},\ }\href
  {\doibase 10.1103/PhysRevD.102.034012} {\bibfield  {journal} {\bibinfo
  {journal} {Phys. Rev. D}\ }\textbf {\bibinfo {volume} {102}},\ \bibinfo
  {pages} {034012} (\bibinfo {year} {2020})},\ \Eprint
  {http://arxiv.org/abs/2006.08087} {arXiv:2006.08087 [hep-ph]} \BibitemShut
  {NoStop}%
\bibitem [{\citenamefont {Cheng}\ \emph {et~al.}(2021)\citenamefont {Cheng},
  \citenamefont {Li}, \citenamefont {Liu}, \citenamefont {Si},\ and\
  \citenamefont {Yao}}]{Cheng:2020wxa}%
  \BibitemOpen
  \bibfield  {author} {\bibinfo {author} {\bibfnamefont {J.-B.}\ \bibnamefont
  {Cheng}}, \bibinfo {author} {\bibfnamefont {S.-Y.}\ \bibnamefont {Li}},
  \bibinfo {author} {\bibfnamefont {Y.-R.}\ \bibnamefont {Liu}}, \bibinfo
  {author} {\bibfnamefont {Z.-G.}\ \bibnamefont {Si}}, \ and\ \bibinfo {author}
  {\bibfnamefont {T.}~\bibnamefont {Yao}},\ }\href {\doibase
  10.1088/1674-1137/abde2f} {\bibfield  {journal} {\bibinfo  {journal} {Chin.
  Phys. C}\ }\textbf {\bibinfo {volume} {45}},\ \bibinfo {pages} {043102}
  (\bibinfo {year} {2021})},\ \Eprint {http://arxiv.org/abs/2008.00737}
  {arXiv:2008.00737 [hep-ph]} \BibitemShut {NoStop}%
\bibitem [{\citenamefont {Meng}\ \emph {et~al.}(2021)\citenamefont {Meng},
  \citenamefont {Hiyama}, \citenamefont {Hosaka}, \citenamefont {Oka},
  \citenamefont {Gubler}, \citenamefont {Can}, \citenamefont {Takahashi},\ and\
  \citenamefont {Zong}}]{Meng:2020knc}%
  \BibitemOpen
  \bibfield  {author} {\bibinfo {author} {\bibfnamefont {Q.}~\bibnamefont
  {Meng}}, \bibinfo {author} {\bibfnamefont {E.}~\bibnamefont {Hiyama}},
  \bibinfo {author} {\bibfnamefont {A.}~\bibnamefont {Hosaka}}, \bibinfo
  {author} {\bibfnamefont {M.}~\bibnamefont {Oka}}, \bibinfo {author}
  {\bibfnamefont {P.}~\bibnamefont {Gubler}}, \bibinfo {author} {\bibfnamefont
  {K.~U.}\ \bibnamefont {Can}}, \bibinfo {author} {\bibfnamefont {T.~T.}\
  \bibnamefont {Takahashi}}, \ and\ \bibinfo {author} {\bibfnamefont {H.~S.}\
  \bibnamefont {Zong}},\ }\href {\doibase 10.1016/j.physletb.2021.136095}
  {\bibfield  {journal} {\bibinfo  {journal} {Phys. Lett. B}\ }\textbf
  {\bibinfo {volume} {814}},\ \bibinfo {pages} {136095} (\bibinfo {year}
  {2021})},\ \Eprint {http://arxiv.org/abs/2009.14493} {arXiv:2009.14493
  [nucl-th]} \BibitemShut {NoStop}%
\bibitem [{\citenamefont {Noh}\ \emph {et~al.}(2021)\citenamefont {Noh},
  \citenamefont {Park},\ and\ \citenamefont {Lee}}]{Noh:2021lqs}%
  \BibitemOpen
  \bibfield  {author} {\bibinfo {author} {\bibfnamefont {S.}~\bibnamefont
  {Noh}}, \bibinfo {author} {\bibfnamefont {W.}~\bibnamefont {Park}}, \ and\
  \bibinfo {author} {\bibfnamefont {S.~H.}\ \bibnamefont {Lee}},\ }\href
  {\doibase 10.1103/PhysRevD.103.114009} {\bibfield  {journal} {\bibinfo
  {journal} {Phys. Rev. D}\ }\textbf {\bibinfo {volume} {103}},\ \bibinfo
  {pages} {114009} (\bibinfo {year} {2021})},\ \Eprint
  {http://arxiv.org/abs/2102.09614} {arXiv:2102.09614 [hep-ph]} \BibitemShut
  {NoStop}%
\bibitem [{\citenamefont {Ohkoda}\ \emph {et~al.}(2012)\citenamefont {Ohkoda},
  \citenamefont {Yamaguchi}, \citenamefont {Yasui}, \citenamefont {Sudoh},\
  and\ \citenamefont {Hosaka}}]{Ohkoda:2012hv}%
  \BibitemOpen
  \bibfield  {author} {\bibinfo {author} {\bibfnamefont {S.}~\bibnamefont
  {Ohkoda}}, \bibinfo {author} {\bibfnamefont {Y.}~\bibnamefont {Yamaguchi}},
  \bibinfo {author} {\bibfnamefont {S.}~\bibnamefont {Yasui}}, \bibinfo
  {author} {\bibfnamefont {K.}~\bibnamefont {Sudoh}}, \ and\ \bibinfo {author}
  {\bibfnamefont {A.}~\bibnamefont {Hosaka}},\ }\href {\doibase
  10.1103/PhysRevD.86.034019} {\bibfield  {journal} {\bibinfo  {journal} {Phys.
  Rev. D}\ }\textbf {\bibinfo {volume} {86}},\ \bibinfo {pages} {034019}
  (\bibinfo {year} {2012})},\ \Eprint {http://arxiv.org/abs/1202.0760}
  {arXiv:1202.0760 [hep-ph]} \BibitemShut {NoStop}%
\bibitem [{\citenamefont {Czarnecki}\ \emph {et~al.}(2018)\citenamefont
  {Czarnecki}, \citenamefont {Leng},\ and\ \citenamefont
  {Voloshin}}]{Czarnecki:2017vco}%
  \BibitemOpen
  \bibfield  {author} {\bibinfo {author} {\bibfnamefont {A.}~\bibnamefont
  {Czarnecki}}, \bibinfo {author} {\bibfnamefont {B.}~\bibnamefont {Leng}}, \
  and\ \bibinfo {author} {\bibfnamefont {M.}~\bibnamefont {Voloshin}},\ }\href
  {\doibase 10.1016/j.physletb.2018.01.034} {\bibfield  {journal} {\bibinfo
  {journal} {Phys. Lett. B}\ }\textbf {\bibinfo {volume} {778}},\ \bibinfo
  {pages} {233} (\bibinfo {year} {2018})},\ \Eprint
  {http://arxiv.org/abs/1708.04594} {arXiv:1708.04594 [hep-ph]} \BibitemShut
  {NoStop}%
\bibitem [{\citenamefont {Eichten}\ and\ \citenamefont
  {Quigg}(2017)}]{Eichten:2017ffp}%
  \BibitemOpen
  \bibfield  {author} {\bibinfo {author} {\bibfnamefont {E.~J.}\ \bibnamefont
  {Eichten}}\ and\ \bibinfo {author} {\bibfnamefont {C.}~\bibnamefont
  {Quigg}},\ }\href {\doibase 10.1103/PhysRevLett.119.202002} {\bibfield
  {journal} {\bibinfo  {journal} {Phys. Rev. Lett.}\ }\textbf {\bibinfo
  {volume} {119}},\ \bibinfo {pages} {202002} (\bibinfo {year} {2017})},\
  \Eprint {http://arxiv.org/abs/1707.09575} {arXiv:1707.09575 [hep-ph]}
  \BibitemShut {NoStop}%
\bibitem [{\citenamefont {Braaten}\ \emph {et~al.}(2021)\citenamefont
  {Braaten}, \citenamefont {He},\ and\ \citenamefont
  {Mohapatra}}]{Braaten:2020nwp}%
  \BibitemOpen
  \bibfield  {author} {\bibinfo {author} {\bibfnamefont {E.}~\bibnamefont
  {Braaten}}, \bibinfo {author} {\bibfnamefont {L.-P.}\ \bibnamefont {He}}, \
  and\ \bibinfo {author} {\bibfnamefont {A.}~\bibnamefont {Mohapatra}},\ }\href
  {\doibase 10.1103/PhysRevD.103.016001} {\bibfield  {journal} {\bibinfo
  {journal} {Phys. Rev. D}\ }\textbf {\bibinfo {volume} {103}},\ \bibinfo
  {pages} {016001} (\bibinfo {year} {2021})},\ \Eprint
  {http://arxiv.org/abs/2006.08650} {arXiv:2006.08650 [hep-ph]} \BibitemShut
  {NoStop}%
\bibitem [{\citenamefont {Shifman}\ \emph
  {et~al.}(1979{\natexlab{a}})\citenamefont {Shifman}, \citenamefont
  {Vainshtein},\ and\ \citenamefont {Zakharov}}]{Shifman:1978bx}%
  \BibitemOpen
  \bibfield  {author} {\bibinfo {author} {\bibfnamefont {M.~A.}\ \bibnamefont
  {Shifman}}, \bibinfo {author} {\bibfnamefont {A.}~\bibnamefont {Vainshtein}},
  \ and\ \bibinfo {author} {\bibfnamefont {V.~I.}\ \bibnamefont {Zakharov}},\
  }\href {\doibase 10.1016/0550-3213(79)90022-1} {\bibfield  {journal}
  {\bibinfo  {journal} {Nucl. Phys. B}\ }\textbf {\bibinfo {volume} {147}},\
  \bibinfo {pages} {385} (\bibinfo {year} {1979}{\natexlab{a}})}\BibitemShut
  {NoStop}%
\bibitem [{\citenamefont {Shifman}\ \emph
  {et~al.}(1979{\natexlab{b}})\citenamefont {Shifman}, \citenamefont
  {Vainshtein},\ and\ \citenamefont {Zakharov}}]{Shifman:1978by}%
  \BibitemOpen
  \bibfield  {author} {\bibinfo {author} {\bibfnamefont {M.~A.}\ \bibnamefont
  {Shifman}}, \bibinfo {author} {\bibfnamefont {A.}~\bibnamefont {Vainshtein}},
  \ and\ \bibinfo {author} {\bibfnamefont {V.~I.}\ \bibnamefont {Zakharov}},\
  }\href {\doibase 10.1016/0550-3213(79)90023-3} {\bibfield  {journal}
  {\bibinfo  {journal} {Nucl. Phys. B}\ }\textbf {\bibinfo {volume} {147}},\
  \bibinfo {pages} {448} (\bibinfo {year} {1979}{\natexlab{b}})}\BibitemShut
  {NoStop}%
\bibitem [{\citenamefont {Navarra}\ \emph {et~al.}(2007)\citenamefont
  {Navarra}, \citenamefont {Nielsen},\ and\ \citenamefont
  {Lee}}]{Navarra:2007yw}%
  \BibitemOpen
  \bibfield  {author} {\bibinfo {author} {\bibfnamefont {F.~S.}\ \bibnamefont
  {Navarra}}, \bibinfo {author} {\bibfnamefont {M.}~\bibnamefont {Nielsen}}, \
  and\ \bibinfo {author} {\bibfnamefont {S.~H.}\ \bibnamefont {Lee}},\ }\href
  {\doibase 10.1016/j.physletb.2007.04.010} {\bibfield  {journal} {\bibinfo
  {journal} {Phys. Lett. B}\ }\textbf {\bibinfo {volume} {649}},\ \bibinfo
  {pages} {166} (\bibinfo {year} {2007})},\ \Eprint
  {http://arxiv.org/abs/hep-ph/0703071} {arXiv:hep-ph/0703071} \BibitemShut
  {NoStop}%
\bibitem [{\citenamefont {Du}\ \emph {et~al.}(2013)\citenamefont {Du},
  \citenamefont {Chen}, \citenamefont {Chen},\ and\ \citenamefont
  {Zhu}}]{Du:2012wp}%
  \BibitemOpen
  \bibfield  {author} {\bibinfo {author} {\bibfnamefont {M.-L.}\ \bibnamefont
  {Du}}, \bibinfo {author} {\bibfnamefont {W.}~\bibnamefont {Chen}}, \bibinfo
  {author} {\bibfnamefont {X.-L.}\ \bibnamefont {Chen}}, \ and\ \bibinfo
  {author} {\bibfnamefont {S.-L.}\ \bibnamefont {Zhu}},\ }\href {\doibase
  10.1103/PhysRevD.87.014003} {\bibfield  {journal} {\bibinfo  {journal} {Phys.
  Rev. D}\ }\textbf {\bibinfo {volume} {87}},\ \bibinfo {pages} {014003}
  (\bibinfo {year} {2013})},\ \Eprint {http://arxiv.org/abs/1209.5134}
  {arXiv:1209.5134 [hep-ph]} \BibitemShut {NoStop}%
\bibitem [{\citenamefont {Chen}\ \emph {et~al.}(2014)\citenamefont {Chen},
  \citenamefont {Steele},\ and\ \citenamefont {Zhu}}]{Chen:2013aba}%
  \BibitemOpen
  \bibfield  {author} {\bibinfo {author} {\bibfnamefont {W.}~\bibnamefont
  {Chen}}, \bibinfo {author} {\bibfnamefont {T.}~\bibnamefont {Steele}}, \ and\
  \bibinfo {author} {\bibfnamefont {S.-L.}\ \bibnamefont {Zhu}},\ }\href
  {\doibase 10.1103/PhysRevD.89.054037} {\bibfield  {journal} {\bibinfo
  {journal} {Phys. Rev. D}\ }\textbf {\bibinfo {volume} {89}},\ \bibinfo
  {pages} {054037} (\bibinfo {year} {2014})},\ \Eprint
  {http://arxiv.org/abs/1310.8337} {arXiv:1310.8337 [hep-ph]} \BibitemShut
  {NoStop}%
\bibitem [{\citenamefont {Wang}(2018)}]{Wang:2017uld}%
  \BibitemOpen
  \bibfield  {author} {\bibinfo {author} {\bibfnamefont {Z.-G.}\ \bibnamefont
  {Wang}},\ }\href {\doibase 10.5506/APhysPolB.49.1781} {\bibfield  {journal}
  {\bibinfo  {journal} {Acta Phys. Polon. B}\ }\textbf {\bibinfo {volume}
  {49}},\ \bibinfo {pages} {1781} (\bibinfo {year} {2018})},\ \Eprint
  {http://arxiv.org/abs/1708.04545} {arXiv:1708.04545 [hep-ph]} \BibitemShut
  {NoStop}%
\bibitem [{\citenamefont {Agaev}\ \emph
  {et~al.}(2019{\natexlab{a}})\citenamefont {Agaev}, \citenamefont {Azizi},
  \citenamefont {Barsbay},\ and\ \citenamefont {Sundu}}]{Agaev:2018khe}%
  \BibitemOpen
  \bibfield  {author} {\bibinfo {author} {\bibfnamefont {S.}~\bibnamefont
  {Agaev}}, \bibinfo {author} {\bibfnamefont {K.}~\bibnamefont {Azizi}},
  \bibinfo {author} {\bibfnamefont {B.}~\bibnamefont {Barsbay}}, \ and\
  \bibinfo {author} {\bibfnamefont {H.}~\bibnamefont {Sundu}},\ }\href
  {\doibase 10.1103/PhysRevD.99.033002} {\bibfield  {journal} {\bibinfo
  {journal} {Phys. Rev. D}\ }\textbf {\bibinfo {volume} {99}},\ \bibinfo
  {pages} {033002} (\bibinfo {year} {2019}{\natexlab{a}})},\ \Eprint
  {http://arxiv.org/abs/1809.07791} {arXiv:1809.07791 [hep-ph]} \BibitemShut
  {NoStop}%
\bibitem [{\citenamefont {Agaev}\ \emph {et~al.}(2020)\citenamefont {Agaev},
  \citenamefont {Azizi},\ and\ \citenamefont {Sundu}}]{Agaev:2019kkz}%
  \BibitemOpen
  \bibfield  {author} {\bibinfo {author} {\bibfnamefont {S.~S.}\ \bibnamefont
  {Agaev}}, \bibinfo {author} {\bibfnamefont {K.}~\bibnamefont {Azizi}}, \ and\
  \bibinfo {author} {\bibfnamefont {H.}~\bibnamefont {Sundu}},\ }\href
  {\doibase 10.1016/j.nuclphysb.2019.114890} {\bibfield  {journal} {\bibinfo
  {journal} {Nucl. Phys. B}\ }\textbf {\bibinfo {volume} {951}},\ \bibinfo
  {pages} {114890} (\bibinfo {year} {2020})},\ \Eprint
  {http://arxiv.org/abs/1905.07591} {arXiv:1905.07591 [hep-ph]} \BibitemShut
  {NoStop}%
\bibitem [{\citenamefont {Agaev}\ \emph
  {et~al.}(2019{\natexlab{b}})\citenamefont {Agaev}, \citenamefont {Azizi},\
  and\ \citenamefont {Sundu}}]{Agaev:2019wkk}%
  \BibitemOpen
  \bibfield  {author} {\bibinfo {author} {\bibfnamefont {S.}~\bibnamefont
  {Agaev}}, \bibinfo {author} {\bibfnamefont {K.}~\bibnamefont {Azizi}}, \ and\
  \bibinfo {author} {\bibfnamefont {H.}~\bibnamefont {Sundu}},\ }\href
  {\doibase 10.1103/PhysRevD.100.094020} {\bibfield  {journal} {\bibinfo
  {journal} {Phys. Rev. D}\ }\textbf {\bibinfo {volume} {100}},\ \bibinfo
  {pages} {094020} (\bibinfo {year} {2019}{\natexlab{b}})},\ \Eprint
  {http://arxiv.org/abs/1907.04017} {arXiv:1907.04017 [hep-ph]} \BibitemShut
  {NoStop}%
\bibitem [{\citenamefont {Tang}\ \emph {et~al.}(2020)\citenamefont {Tang},
  \citenamefont {Wan}, \citenamefont {Maltman},\ and\ \citenamefont
  {Qiao}}]{Tang:2019nwv}%
  \BibitemOpen
  \bibfield  {author} {\bibinfo {author} {\bibfnamefont {L.}~\bibnamefont
  {Tang}}, \bibinfo {author} {\bibfnamefont {B.-D.}\ \bibnamefont {Wan}},
  \bibinfo {author} {\bibfnamefont {K.}~\bibnamefont {Maltman}}, \ and\
  \bibinfo {author} {\bibfnamefont {C.-F.}\ \bibnamefont {Qiao}},\ }\href
  {\doibase 10.1103/PhysRevD.101.094032} {\bibfield  {journal} {\bibinfo
  {journal} {Phys. Rev. D}\ }\textbf {\bibinfo {volume} {101}},\ \bibinfo
  {pages} {094032} (\bibinfo {year} {2020})},\ \Eprint
  {http://arxiv.org/abs/1911.10951} {arXiv:1911.10951 [hep-ph]} \BibitemShut
  {NoStop}%
\bibitem [{\citenamefont {Wang}\ and\ \citenamefont
  {Chen}(2020)}]{Wang:2020jgb}%
  \BibitemOpen
  \bibfield  {author} {\bibinfo {author} {\bibfnamefont {Q.-N.}\ \bibnamefont
  {Wang}}\ and\ \bibinfo {author} {\bibfnamefont {W.}~\bibnamefont {Chen}},\
  }\href {\doibase 10.1140/epjc/s10052-020-7938-2} {\bibfield  {journal}
  {\bibinfo  {journal} {Eur. Phys. J. C}\ }\textbf {\bibinfo {volume} {80}},\
  \bibinfo {pages} {389} (\bibinfo {year} {2020})},\ \Eprint
  {http://arxiv.org/abs/2002.04243} {arXiv:2002.04243 [hep-ph]} \BibitemShut
  {NoStop}%
\bibitem [{\citenamefont {Agaev}\ \emph {et~al.}(2021)\citenamefont {Agaev},
  \citenamefont {Azizi}, \citenamefont {Barsbay},\ and\ \citenamefont
  {Sundu}}]{Agaev:2020zag}%
  \BibitemOpen
  \bibfield  {author} {\bibinfo {author} {\bibfnamefont {S.~S.}\ \bibnamefont
  {Agaev}}, \bibinfo {author} {\bibfnamefont {K.}~\bibnamefont {Azizi}},
  \bibinfo {author} {\bibfnamefont {B.}~\bibnamefont {Barsbay}}, \ and\
  \bibinfo {author} {\bibfnamefont {H.}~\bibnamefont {Sundu}},\ }\href
  {\doibase 10.1088/1674-1137/abc16d} {\bibfield  {journal} {\bibinfo
  {journal} {Chin. Phys. C}\ }\textbf {\bibinfo {volume} {45}},\ \bibinfo
  {pages} {013105} (\bibinfo {year} {2021})},\ \Eprint
  {http://arxiv.org/abs/2002.04553} {arXiv:2002.04553 [hep-ph]} \BibitemShut
  {NoStop}%
\bibitem [{\citenamefont {Hudspith}\ \emph {et~al.}(2020)\citenamefont
  {Hudspith}, \citenamefont {Colquhoun}, \citenamefont {Francis}, \citenamefont
  {Lewis},\ and\ \citenamefont {Maltman}}]{Hudspith:2020tdf}%
  \BibitemOpen
  \bibfield  {author} {\bibinfo {author} {\bibfnamefont {R.~J.}\ \bibnamefont
  {Hudspith}}, \bibinfo {author} {\bibfnamefont {B.}~\bibnamefont {Colquhoun}},
  \bibinfo {author} {\bibfnamefont {A.}~\bibnamefont {Francis}}, \bibinfo
  {author} {\bibfnamefont {R.}~\bibnamefont {Lewis}}, \ and\ \bibinfo {author}
  {\bibfnamefont {K.}~\bibnamefont {Maltman}},\ }\href {\doibase
  10.1103/PhysRevD.102.114506} {\bibfield  {journal} {\bibinfo  {journal}
  {Phys. Rev. D}\ }\textbf {\bibinfo {volume} {102}},\ \bibinfo {pages}
  {114506} (\bibinfo {year} {2020})},\ \Eprint
  {http://arxiv.org/abs/2006.14294} {arXiv:2006.14294 [hep-lat]} \BibitemShut
  {NoStop}%
\bibitem [{\citenamefont {Bicudo}\ \emph {et~al.}(2016)\citenamefont {Bicudo},
  \citenamefont {Cichy}, \citenamefont {Peters},\ and\ \citenamefont
  {Wagner}}]{Bicudo:2015kna}%
  \BibitemOpen
  \bibfield  {author} {\bibinfo {author} {\bibfnamefont {P.}~\bibnamefont
  {Bicudo}}, \bibinfo {author} {\bibfnamefont {K.}~\bibnamefont {Cichy}},
  \bibinfo {author} {\bibfnamefont {A.}~\bibnamefont {Peters}}, \ and\ \bibinfo
  {author} {\bibfnamefont {M.}~\bibnamefont {Wagner}},\ }\href {\doibase
  10.1103/PhysRevD.93.034501} {\bibfield  {journal} {\bibinfo  {journal} {Phys.
  Rev. D}\ }\textbf {\bibinfo {volume} {93}},\ \bibinfo {pages} {034501}
  (\bibinfo {year} {2016})},\ \Eprint {http://arxiv.org/abs/1510.03441}
  {arXiv:1510.03441 [hep-lat]} \BibitemShut {NoStop}%
\bibitem [{\citenamefont {Bicudo}\ \emph
  {et~al.}(2017{\natexlab{a}})\citenamefont {Bicudo}, \citenamefont
  {Scheunert},\ and\ \citenamefont {Wagner}}]{Bicudo:2016ooe}%
  \BibitemOpen
  \bibfield  {author} {\bibinfo {author} {\bibfnamefont {P.}~\bibnamefont
  {Bicudo}}, \bibinfo {author} {\bibfnamefont {J.}~\bibnamefont {Scheunert}}, \
  and\ \bibinfo {author} {\bibfnamefont {M.}~\bibnamefont {Wagner}},\ }\href
  {\doibase 10.1103/PhysRevD.95.034502} {\bibfield  {journal} {\bibinfo
  {journal} {Phys. Rev. D}\ }\textbf {\bibinfo {volume} {95}},\ \bibinfo
  {pages} {034502} (\bibinfo {year} {2017}{\natexlab{a}})},\ \Eprint
  {http://arxiv.org/abs/1612.02758} {arXiv:1612.02758 [hep-lat]} \BibitemShut
  {NoStop}%
\bibitem [{\citenamefont {Bicudo}\ \emph
  {et~al.}(2017{\natexlab{b}})\citenamefont {Bicudo}, \citenamefont {Cardoso},
  \citenamefont {Peters}, \citenamefont {Pflaumer},\ and\ \citenamefont
  {Wagner}}]{Bicudo:2017szl}%
  \BibitemOpen
  \bibfield  {author} {\bibinfo {author} {\bibfnamefont {P.}~\bibnamefont
  {Bicudo}}, \bibinfo {author} {\bibfnamefont {M.}~\bibnamefont {Cardoso}},
  \bibinfo {author} {\bibfnamefont {A.}~\bibnamefont {Peters}}, \bibinfo
  {author} {\bibfnamefont {M.}~\bibnamefont {Pflaumer}}, \ and\ \bibinfo
  {author} {\bibfnamefont {M.}~\bibnamefont {Wagner}},\ }\href {\doibase
  10.1103/PhysRevD.96.054510} {\bibfield  {journal} {\bibinfo  {journal} {Phys.
  Rev. D}\ }\textbf {\bibinfo {volume} {96}},\ \bibinfo {pages} {054510}
  (\bibinfo {year} {2017}{\natexlab{b}})},\ \Eprint
  {http://arxiv.org/abs/1704.02383} {arXiv:1704.02383 [hep-lat]} \BibitemShut
  {NoStop}%
\bibitem [{\citenamefont {Francis}\ \emph {et~al.}(2017)\citenamefont
  {Francis}, \citenamefont {Hudspith}, \citenamefont {Lewis},\ and\
  \citenamefont {Maltman}}]{Francis:2016hui}%
  \BibitemOpen
  \bibfield  {author} {\bibinfo {author} {\bibfnamefont {A.}~\bibnamefont
  {Francis}}, \bibinfo {author} {\bibfnamefont {R.~J.}\ \bibnamefont
  {Hudspith}}, \bibinfo {author} {\bibfnamefont {R.}~\bibnamefont {Lewis}}, \
  and\ \bibinfo {author} {\bibfnamefont {K.}~\bibnamefont {Maltman}},\ }\href
  {\doibase 10.1103/PhysRevLett.118.142001} {\bibfield  {journal} {\bibinfo
  {journal} {Phys. Rev. Lett.}\ }\textbf {\bibinfo {volume} {118}},\ \bibinfo
  {pages} {142001} (\bibinfo {year} {2017})},\ \Eprint
  {http://arxiv.org/abs/1607.05214} {arXiv:1607.05214 [hep-lat]} \BibitemShut
  {NoStop}%
\bibitem [{\citenamefont {Junnarkar}\ \emph {et~al.}(2019)\citenamefont
  {Junnarkar}, \citenamefont {Mathur},\ and\ \citenamefont
  {Padmanath}}]{Junnarkar:2018twb}%
  \BibitemOpen
  \bibfield  {author} {\bibinfo {author} {\bibfnamefont {P.}~\bibnamefont
  {Junnarkar}}, \bibinfo {author} {\bibfnamefont {N.}~\bibnamefont {Mathur}}, \
  and\ \bibinfo {author} {\bibfnamefont {M.}~\bibnamefont {Padmanath}},\ }\href
  {\doibase 10.1103/PhysRevD.99.034507} {\bibfield  {journal} {\bibinfo
  {journal} {Phys. Rev. D}\ }\textbf {\bibinfo {volume} {99}},\ \bibinfo
  {pages} {034507} (\bibinfo {year} {2019})},\ \Eprint
  {http://arxiv.org/abs/1810.12285} {arXiv:1810.12285 [hep-lat]} \BibitemShut
  {NoStop}%
\bibitem [{\citenamefont {Leskovec}\ \emph {et~al.}(2019)\citenamefont
  {Leskovec}, \citenamefont {Meinel}, \citenamefont {Pflaumer},\ and\
  \citenamefont {Wagner}}]{Leskovec:2019ioa}%
  \BibitemOpen
  \bibfield  {author} {\bibinfo {author} {\bibfnamefont {L.}~\bibnamefont
  {Leskovec}}, \bibinfo {author} {\bibfnamefont {S.}~\bibnamefont {Meinel}},
  \bibinfo {author} {\bibfnamefont {M.}~\bibnamefont {Pflaumer}}, \ and\
  \bibinfo {author} {\bibfnamefont {M.}~\bibnamefont {Wagner}},\ }\href
  {\doibase 10.1103/PhysRevD.100.014503} {\bibfield  {journal} {\bibinfo
  {journal} {Phys. Rev.}\ }\textbf {\bibinfo {volume} {D100}},\ \bibinfo
  {pages} {014503} (\bibinfo {year} {2019})},\ \Eprint
  {http://arxiv.org/abs/1904.04197} {arXiv:1904.04197 [hep-lat]} \BibitemShut
  {NoStop}%
\bibitem [{\citenamefont {Mohanta}\ and\ \citenamefont
  {Basak}(2020)}]{Mohanta:2020eed}%
  \BibitemOpen
  \bibfield  {author} {\bibinfo {author} {\bibfnamefont {P.}~\bibnamefont
  {Mohanta}}\ and\ \bibinfo {author} {\bibfnamefont {S.}~\bibnamefont
  {Basak}},\ }\href {\doibase 10.1103/PhysRevD.102.094516} {\bibfield
  {journal} {\bibinfo  {journal} {Phys. Rev. D}\ }\textbf {\bibinfo {volume}
  {102}},\ \bibinfo {pages} {094516} (\bibinfo {year} {2020})},\ \Eprint
  {http://arxiv.org/abs/2008.11146} {arXiv:2008.11146 [hep-lat]} \BibitemShut
  {NoStop}%
\bibitem [{\citenamefont {Pflaumer}\ \emph {et~al.}(2020)\citenamefont
  {Pflaumer}, \citenamefont {Leskovec}, \citenamefont {Meinel},\ and\
  \citenamefont {Wagner}}]{Pflaumer:2020ogv}%
  \BibitemOpen
  \bibfield  {author} {\bibinfo {author} {\bibfnamefont {M.}~\bibnamefont
  {Pflaumer}}, \bibinfo {author} {\bibfnamefont {L.}~\bibnamefont {Leskovec}},
  \bibinfo {author} {\bibfnamefont {S.}~\bibnamefont {Meinel}}, \ and\ \bibinfo
  {author} {\bibfnamefont {M.}~\bibnamefont {Wagner}},\ }in\ \href@noop {}
  {\emph {\bibinfo {booktitle} {{Asia-Pacific Symposium for Lattice Field
  Theory}}}}\ (\bibinfo {year} {2020})\ \Eprint
  {http://arxiv.org/abs/2009.10538} {arXiv:2009.10538 [hep-lat]} \BibitemShut
  {NoStop}%
\bibitem [{\citenamefont {Pflaumer}\ \emph {et~al.}(2022)\citenamefont
  {Pflaumer}, \citenamefont {Leskovec}, \citenamefont {Meinel},\ and\
  \citenamefont {Wagner}}]{Pflaumer:2021ong}%
  \BibitemOpen
  \bibfield  {author} {\bibinfo {author} {\bibfnamefont {M.}~\bibnamefont
  {Pflaumer}}, \bibinfo {author} {\bibfnamefont {L.}~\bibnamefont {Leskovec}},
  \bibinfo {author} {\bibfnamefont {S.}~\bibnamefont {Meinel}}, \ and\ \bibinfo
  {author} {\bibfnamefont {M.}~\bibnamefont {Wagner}},\ }\href {\doibase
  10.22323/1.396.0392} {\bibfield  {journal} {\bibinfo  {journal} {PoS}\
  }\textbf {\bibinfo {volume} {LATTICE2021}},\ \bibinfo {pages} {392} (\bibinfo
  {year} {2022})},\ \Eprint {http://arxiv.org/abs/2108.10704} {arXiv:2108.10704
  [hep-lat]} \BibitemShut {NoStop}%
\bibitem [{\citenamefont {Wagner}\ \emph {et~al.}(2023)\citenamefont {Wagner},
  \citenamefont {Alexandrou}, \citenamefont {Finkenrath}, \citenamefont
  {Leontiou}, \citenamefont {Meinel},\ and\ \citenamefont
  {Pflaumer}}]{Wagner:2022bff}%
  \BibitemOpen
  \bibfield  {author} {\bibinfo {author} {\bibfnamefont {M.}~\bibnamefont
  {Wagner}}, \bibinfo {author} {\bibfnamefont {C.}~\bibnamefont {Alexandrou}},
  \bibinfo {author} {\bibfnamefont {J.}~\bibnamefont {Finkenrath}}, \bibinfo
  {author} {\bibfnamefont {T.}~\bibnamefont {Leontiou}}, \bibinfo {author}
  {\bibfnamefont {S.}~\bibnamefont {Meinel}}, \ and\ \bibinfo {author}
  {\bibfnamefont {M.}~\bibnamefont {Pflaumer}},\ }\href {\doibase
  10.22323/1.430.0270} {\bibfield  {journal} {\bibinfo  {journal} {PoS}\
  }\textbf {\bibinfo {volume} {LATTICE2022}},\ \bibinfo {pages} {270} (\bibinfo
  {year} {2023})},\ \Eprint {http://arxiv.org/abs/2210.09281} {arXiv:2210.09281
  [hep-lat]} \BibitemShut {NoStop}%
\bibitem [{\citenamefont {Pflaumer}\ \emph {et~al.}(2023)\citenamefont
  {Pflaumer}, \citenamefont {Alexandrou}, \citenamefont {Finkenrath},
  \citenamefont {Leontiou}, \citenamefont {Meinel},\ and\ \citenamefont
  {Wagner}}]{Pflaumer:2022lgp}%
  \BibitemOpen
  \bibfield  {author} {\bibinfo {author} {\bibfnamefont {M.}~\bibnamefont
  {Pflaumer}}, \bibinfo {author} {\bibfnamefont {C.}~\bibnamefont
  {Alexandrou}}, \bibinfo {author} {\bibfnamefont {J.}~\bibnamefont
  {Finkenrath}}, \bibinfo {author} {\bibfnamefont {T.}~\bibnamefont
  {Leontiou}}, \bibinfo {author} {\bibfnamefont {S.}~\bibnamefont {Meinel}}, \
  and\ \bibinfo {author} {\bibfnamefont {M.}~\bibnamefont {Wagner}},\ }\href
  {\doibase 10.22323/1.430.0075} {\bibfield  {journal} {\bibinfo  {journal}
  {PoS}\ }\textbf {\bibinfo {volume} {LATTICE2022}},\ \bibinfo {pages} {075}
  (\bibinfo {year} {2023})},\ \Eprint {http://arxiv.org/abs/2211.00951}
  {arXiv:2211.00951 [hep-lat]} \BibitemShut {NoStop}%
\bibitem [{\citenamefont {Colquhoun}\ \emph
  {et~al.}(2022{\natexlab{a}})\citenamefont {Colquhoun}, \citenamefont
  {Francis}, \citenamefont {Hudspith}, \citenamefont {Lewis},\ and\
  \citenamefont {Maltman}}]{Colquhoun:2022dte}%
  \BibitemOpen
  \bibfield  {author} {\bibinfo {author} {\bibfnamefont {B.}~\bibnamefont
  {Colquhoun}}, \bibinfo {author} {\bibfnamefont {A.}~\bibnamefont {Francis}},
  \bibinfo {author} {\bibfnamefont {R.~J.}\ \bibnamefont {Hudspith}}, \bibinfo
  {author} {\bibfnamefont {R.}~\bibnamefont {Lewis}}, \ and\ \bibinfo {author}
  {\bibfnamefont {K.}~\bibnamefont {Maltman}},\ }\href {\doibase
  10.22323/1.396.0144} {\bibfield  {journal} {\bibinfo  {journal} {PoS}\
  }\textbf {\bibinfo {volume} {LATTICE2021}},\ \bibinfo {pages} {144} (\bibinfo
  {year} {2022}{\natexlab{a}})}\BibitemShut {NoStop}%
\bibitem [{\citenamefont {Colquhoun}\ \emph
  {et~al.}(2022{\natexlab{b}})\citenamefont {Colquhoun}, \citenamefont
  {Francis}, \citenamefont {Hudspith}, \citenamefont {Lewis},\ and\
  \citenamefont {Maltman}}]{Colquhoun:2022sip}%
  \BibitemOpen
  \bibfield  {author} {\bibinfo {author} {\bibfnamefont {B.}~\bibnamefont
  {Colquhoun}}, \bibinfo {author} {\bibfnamefont {A.}~\bibnamefont {Francis}},
  \bibinfo {author} {\bibfnamefont {R.}~\bibnamefont {Hudspith}}, \bibinfo
  {author} {\bibfnamefont {R.}~\bibnamefont {Lewis}}, \ and\ \bibinfo {author}
  {\bibfnamefont {K.}~\bibnamefont {Maltman}},\ }\href {\doibase
  10.31349/SuplRevMexFis.3.0308044} {\bibfield  {journal} {\bibinfo  {journal}
  {Rev. Mex. Fis. Suppl.}\ }\textbf {\bibinfo {volume} {3}},\ \bibinfo {pages}
  {0308044} (\bibinfo {year} {2022}{\natexlab{b}})}\BibitemShut {NoStop}%
\bibitem [{\citenamefont {Aoki}\ \emph {et~al.}(2023)\citenamefont {Aoki},
  \citenamefont {Aoki},\ and\ \citenamefont {Inoue}}]{Aoki:2023nzp}%
  \BibitemOpen
  \bibfield  {author} {\bibinfo {author} {\bibfnamefont {T.}~\bibnamefont
  {Aoki}}, \bibinfo {author} {\bibfnamefont {S.}~\bibnamefont {Aoki}}, \ and\
  \bibinfo {author} {\bibfnamefont {T.}~\bibnamefont {Inoue}},\ }\href
  {\doibase 10.1103/PhysRevD.108.054502} {\bibfield  {journal} {\bibinfo
  {journal} {Phys. Rev. D}\ }\textbf {\bibinfo {volume} {108}},\ \bibinfo
  {pages} {054502} (\bibinfo {year} {2023})},\ \Eprint
  {http://arxiv.org/abs/2306.03565} {arXiv:2306.03565 [hep-lat]} \BibitemShut
  {NoStop}%
\bibitem [{\citenamefont {Hudspith}\ and\ \citenamefont
  {Mohler}(2023)}]{Hudspith:2023loy}%
  \BibitemOpen
  \bibfield  {author} {\bibinfo {author} {\bibfnamefont {R.~J.}\ \bibnamefont
  {Hudspith}}\ and\ \bibinfo {author} {\bibfnamefont {D.}~\bibnamefont
  {Mohler}},\ }\href {\doibase 10.1103/PhysRevD.107.114510} {\bibfield
  {journal} {\bibinfo  {journal} {Phys. Rev. D}\ }\textbf {\bibinfo {volume}
  {107}},\ \bibinfo {pages} {114510} (\bibinfo {year} {2023})},\ \Eprint
  {http://arxiv.org/abs/2303.17295} {arXiv:2303.17295 [hep-lat]} \BibitemShut
  {NoStop}%
\bibitem [{\citenamefont {Mueller}\ \emph {et~al.}(2023)\citenamefont
  {Mueller}, \citenamefont {Bicudo}, \citenamefont {Krstic~Marinkovic},\ and\
  \citenamefont {Wagner}}]{Mueller:2023wzd}%
  \BibitemOpen
  \bibfield  {author} {\bibinfo {author} {\bibfnamefont {L.}~\bibnamefont
  {Mueller}}, \bibinfo {author} {\bibfnamefont {P.}~\bibnamefont {Bicudo}},
  \bibinfo {author} {\bibfnamefont {M.}~\bibnamefont {Krstic~Marinkovic}}, \
  and\ \bibinfo {author} {\bibfnamefont {M.}~\bibnamefont {Wagner}}\ }(\bibinfo
  {year} {2023})\ \Eprint {http://arxiv.org/abs/2312.17060} {arXiv:2312.17060
  [hep-lat]} \BibitemShut {NoStop}%
\bibitem [{\citenamefont {Alexandrou}\ \emph {et~al.}(2024)\citenamefont
  {Alexandrou}, \citenamefont {Finkenrath}, \citenamefont {Leontiou},
  \citenamefont {Meinel}, \citenamefont {Pflaumer},\ and\ \citenamefont
  {Wagner}}]{Alexandrou:2024iwi}%
  \BibitemOpen
  \bibfield  {author} {\bibinfo {author} {\bibfnamefont {C.}~\bibnamefont
  {Alexandrou}}, \bibinfo {author} {\bibfnamefont {J.}~\bibnamefont
  {Finkenrath}}, \bibinfo {author} {\bibfnamefont {T.}~\bibnamefont
  {Leontiou}}, \bibinfo {author} {\bibfnamefont {S.}~\bibnamefont {Meinel}},
  \bibinfo {author} {\bibfnamefont {M.}~\bibnamefont {Pflaumer}}, \ and\
  \bibinfo {author} {\bibfnamefont {M.}~\bibnamefont {Wagner}},\ }\href@noop {}
  {\  (\bibinfo {year} {2024})},\ \Eprint {http://arxiv.org/abs/2404.03588}
  {arXiv:2404.03588 [hep-lat]} \BibitemShut {NoStop}%
\bibitem [{\citenamefont {Meinel}\ \emph {et~al.}(2022)\citenamefont {Meinel},
  \citenamefont {Pflaumer},\ and\ \citenamefont {Wagner}}]{Meinel:2022lzo}%
  \BibitemOpen
  \bibfield  {author} {\bibinfo {author} {\bibfnamefont {S.}~\bibnamefont
  {Meinel}}, \bibinfo {author} {\bibfnamefont {M.}~\bibnamefont {Pflaumer}}, \
  and\ \bibinfo {author} {\bibfnamefont {M.}~\bibnamefont {Wagner}},\ }\href
  {\doibase 10.1103/PhysRevD.106.034507} {\bibfield  {journal} {\bibinfo
  {journal} {Phys. Rev. D}\ }\textbf {\bibinfo {volume} {106}},\ \bibinfo
  {pages} {034507} (\bibinfo {year} {2022})},\ \Eprint
  {http://arxiv.org/abs/2205.13982} {arXiv:2205.13982 [hep-lat]} \BibitemShut
  {NoStop}%
\bibitem [{\citenamefont {Mathur}\ and\ \citenamefont
  {Padmanath}(2022)}]{Mathur:2021gqn}%
  \BibitemOpen
  \bibfield  {author} {\bibinfo {author} {\bibfnamefont {N.}~\bibnamefont
  {Mathur}}\ and\ \bibinfo {author} {\bibfnamefont {M.}~\bibnamefont
  {Padmanath}},\ }\href {\doibase 10.22323/1.396.0443} {\bibfield  {journal}
  {\bibinfo  {journal} {PoS}\ }\textbf {\bibinfo {volume} {LATTICE2021}},\
  \bibinfo {pages} {443} (\bibinfo {year} {2022})},\ \Eprint
  {http://arxiv.org/abs/2111.01147} {arXiv:2111.01147 [hep-lat]} \BibitemShut
  {NoStop}%
\bibitem [{\citenamefont {Padmanath}\ \emph {et~al.}(2023)\citenamefont
  {Padmanath}, \citenamefont {Radhakrishnan},\ and\ \citenamefont
  {Mathur}}]{Padmanath:2023rdu}%
  \BibitemOpen
  \bibfield  {author} {\bibinfo {author} {\bibfnamefont {M.}~\bibnamefont
  {Padmanath}}, \bibinfo {author} {\bibfnamefont {A.}~\bibnamefont
  {Radhakrishnan}}, \ and\ \bibinfo {author} {\bibfnamefont {N.}~\bibnamefont
  {Mathur}},\ }\href@noop {} {\  (\bibinfo {year} {2023})},\ \Eprint
  {http://arxiv.org/abs/2307.14128} {arXiv:2307.14128 [hep-lat]} \BibitemShut
  {NoStop}%
\bibitem [{\citenamefont {Radhakrishnan}\ \emph {et~al.}(2024)\citenamefont
  {Radhakrishnan}, \citenamefont {Padmanath},\ and\ \citenamefont
  {Mathur}}]{Radhakrishnan:2024ihu}%
  \BibitemOpen
  \bibfield  {author} {\bibinfo {author} {\bibfnamefont {A.}~\bibnamefont
  {Radhakrishnan}}, \bibinfo {author} {\bibfnamefont {M.}~\bibnamefont
  {Padmanath}}, \ and\ \bibinfo {author} {\bibfnamefont {N.}~\bibnamefont
  {Mathur}},\ }\href@noop {} {\  (\bibinfo {year} {2024})},\ \Eprint
  {http://arxiv.org/abs/2404.08109} {arXiv:2404.08109 [hep-lat]} \BibitemShut
  {NoStop}%
\bibitem [{\citenamefont {Alexandrou}\ \emph {et~al.}(2023)\citenamefont
  {Alexandrou}, \citenamefont {Finkenrath}, \citenamefont {Leontiou},
  \citenamefont {Meinel}, \citenamefont {Pflaumer},\ and\ \citenamefont
  {Wagner}}]{Alexandrou:2023cqg}%
  \BibitemOpen
  \bibfield  {author} {\bibinfo {author} {\bibfnamefont {C.}~\bibnamefont
  {Alexandrou}}, \bibinfo {author} {\bibfnamefont {J.}~\bibnamefont
  {Finkenrath}}, \bibinfo {author} {\bibfnamefont {T.}~\bibnamefont
  {Leontiou}}, \bibinfo {author} {\bibfnamefont {S.}~\bibnamefont {Meinel}},
  \bibinfo {author} {\bibfnamefont {M.}~\bibnamefont {Pflaumer}}, \ and\
  \bibinfo {author} {\bibfnamefont {M.}~\bibnamefont {Wagner}},\ }\href@noop {}
  {\  (\bibinfo {year} {2023})},\ \Eprint {http://arxiv.org/abs/2312.02925}
  {arXiv:2312.02925 [hep-lat]} \BibitemShut {NoStop}%
\bibitem [{\citenamefont {Aaij}\ \emph
  {et~al.}(2022{\natexlab{a}})\citenamefont {Aaij} \emph
  {et~al.}}]{LHCb:2021vvq}%
  \BibitemOpen
  \bibfield  {author} {\bibinfo {author} {\bibfnamefont {R.}~\bibnamefont
  {Aaij}} \emph {et~al.} (\bibinfo {collaboration} {LHCb}),\ }\href {\doibase
  10.1038/s41567-022-01614-y} {\bibfield  {journal} {\bibinfo  {journal}
  {Nature Phys.}\ }\textbf {\bibinfo {volume} {18}},\ \bibinfo {pages} {751}
  (\bibinfo {year} {2022}{\natexlab{a}})},\ \Eprint
  {http://arxiv.org/abs/2109.01038} {arXiv:2109.01038 [hep-ex]} \BibitemShut
  {NoStop}%
\bibitem [{\citenamefont {Aaij}\ \emph
  {et~al.}(2022{\natexlab{b}})\citenamefont {Aaij} \emph
  {et~al.}}]{LHCb:2021auc}%
  \BibitemOpen
  \bibfield  {author} {\bibinfo {author} {\bibfnamefont {R.}~\bibnamefont
  {Aaij}} \emph {et~al.} (\bibinfo {collaboration} {LHCb}),\ }\href {\doibase
  10.1038/s41467-022-30206-w} {\bibfield  {journal} {\bibinfo  {journal}
  {Nature Commun.}\ }\textbf {\bibinfo {volume} {13}},\ \bibinfo {pages} {3351}
  (\bibinfo {year} {2022}{\natexlab{b}})},\ \Eprint
  {http://arxiv.org/abs/2109.01056} {arXiv:2109.01056 [hep-ex]} \BibitemShut
  {NoStop}%
\bibitem [{\citenamefont {Cheung}\ \emph {et~al.}(2017)\citenamefont {Cheung},
  \citenamefont {Thomas}, \citenamefont {Dudek},\ and\ \citenamefont
  {Edwards}}]{Cheung:2017tnt}%
  \BibitemOpen
  \bibfield  {author} {\bibinfo {author} {\bibfnamefont {G.~K.~C.}\
  \bibnamefont {Cheung}}, \bibinfo {author} {\bibfnamefont {C.~E.}\
  \bibnamefont {Thomas}}, \bibinfo {author} {\bibfnamefont {J.~J.}\
  \bibnamefont {Dudek}}, \ and\ \bibinfo {author} {\bibfnamefont {R.~G.}\
  \bibnamefont {Edwards}} (\bibinfo {collaboration} {Hadron Spectrum}),\ }\href
  {\doibase 10.1007/JHEP11(2017)033} {\bibfield  {journal} {\bibinfo  {journal}
  {JHEP}\ }\textbf {\bibinfo {volume} {11}},\ \bibinfo {pages} {033} (\bibinfo
  {year} {2017})},\ \Eprint {http://arxiv.org/abs/1709.01417} {arXiv:1709.01417
  [hep-lat]} \BibitemShut {NoStop}%
\bibitem [{\citenamefont {Francis}\ \emph {et~al.}(2019)\citenamefont
  {Francis}, \citenamefont {Hudspith}, \citenamefont {Lewis},\ and\
  \citenamefont {Maltman}}]{Francis:2018jyb}%
  \BibitemOpen
  \bibfield  {author} {\bibinfo {author} {\bibfnamefont {A.}~\bibnamefont
  {Francis}}, \bibinfo {author} {\bibfnamefont {R.~J.}\ \bibnamefont
  {Hudspith}}, \bibinfo {author} {\bibfnamefont {R.}~\bibnamefont {Lewis}}, \
  and\ \bibinfo {author} {\bibfnamefont {K.}~\bibnamefont {Maltman}},\ }\href
  {\doibase 10.1103/PhysRevD.99.054505} {\bibfield  {journal} {\bibinfo
  {journal} {Phys. Rev. D}\ }\textbf {\bibinfo {volume} {99}},\ \bibinfo
  {pages} {054505} (\bibinfo {year} {2019})},\ \Eprint
  {http://arxiv.org/abs/1810.10550} {arXiv:1810.10550 [hep-lat]} \BibitemShut
  {NoStop}%
\bibitem [{\citenamefont {Padmanath}\ and\ \citenamefont
  {Prelovsek}(2022)}]{Padmanath:2022cvl}%
  \BibitemOpen
  \bibfield  {author} {\bibinfo {author} {\bibfnamefont {M.}~\bibnamefont
  {Padmanath}}\ and\ \bibinfo {author} {\bibfnamefont {S.}~\bibnamefont
  {Prelovsek}},\ }\href {\doibase 10.1103/PhysRevLett.129.032002} {\bibfield
  {journal} {\bibinfo  {journal} {Phys. Rev. Lett.}\ }\textbf {\bibinfo
  {volume} {129}},\ \bibinfo {pages} {032002} (\bibinfo {year} {2022})},\
  \Eprint {http://arxiv.org/abs/2202.10110} {arXiv:2202.10110 [hep-lat]}
  \BibitemShut {NoStop}%
\bibitem [{\citenamefont {Chen}\ \emph {et~al.}(2022)\citenamefont {Chen},
  \citenamefont {Shi}, \citenamefont {Chen}, \citenamefont {Gong},
  \citenamefont {Liu}, \citenamefont {Sun},\ and\ \citenamefont
  {Zhang}}]{Chen:2022vpo}%
  \BibitemOpen
  \bibfield  {author} {\bibinfo {author} {\bibfnamefont {S.}~\bibnamefont
  {Chen}}, \bibinfo {author} {\bibfnamefont {C.}~\bibnamefont {Shi}}, \bibinfo
  {author} {\bibfnamefont {Y.}~\bibnamefont {Chen}}, \bibinfo {author}
  {\bibfnamefont {M.}~\bibnamefont {Gong}}, \bibinfo {author} {\bibfnamefont
  {Z.}~\bibnamefont {Liu}}, \bibinfo {author} {\bibfnamefont {W.}~\bibnamefont
  {Sun}}, \ and\ \bibinfo {author} {\bibfnamefont {R.}~\bibnamefont {Zhang}},\
  }\href {\doibase 10.1016/j.physletb.2022.137391} {\bibfield  {journal}
  {\bibinfo  {journal} {Phys. Lett. B}\ }\textbf {\bibinfo {volume} {833}},\
  \bibinfo {pages} {137391} (\bibinfo {year} {2022})},\ \Eprint
  {http://arxiv.org/abs/2206.06185} {arXiv:2206.06185 [hep-lat]} \BibitemShut
  {NoStop}%
\bibitem [{\citenamefont {Lyu}\ \emph {et~al.}(2023)\citenamefont {Lyu},
  \citenamefont {Aoki}, \citenamefont {Doi}, \citenamefont {Hatsuda},
  \citenamefont {Ikeda},\ and\ \citenamefont {Meng}}]{Lyu:2023xro}%
  \BibitemOpen
  \bibfield  {author} {\bibinfo {author} {\bibfnamefont {Y.}~\bibnamefont
  {Lyu}}, \bibinfo {author} {\bibfnamefont {S.}~\bibnamefont {Aoki}}, \bibinfo
  {author} {\bibfnamefont {T.}~\bibnamefont {Doi}}, \bibinfo {author}
  {\bibfnamefont {T.}~\bibnamefont {Hatsuda}}, \bibinfo {author} {\bibfnamefont
  {Y.}~\bibnamefont {Ikeda}}, \ and\ \bibinfo {author} {\bibfnamefont
  {J.}~\bibnamefont {Meng}},\ }\href {\doibase 10.1103/PhysRevLett.131.161901}
  {\bibfield  {journal} {\bibinfo  {journal} {Phys. Rev. Lett.}\ }\textbf
  {\bibinfo {volume} {131}},\ \bibinfo {pages} {161901} (\bibinfo {year}
  {2023})},\ \Eprint {http://arxiv.org/abs/2302.04505} {arXiv:2302.04505
  [hep-lat]} \BibitemShut {NoStop}%
\bibitem [{\citenamefont {Ortiz-Pacheco}\ \emph {et~al.}(2023)\citenamefont
  {Ortiz-Pacheco}, \citenamefont {Collins}, \citenamefont {Leskovec},
  \citenamefont {Padmanath},\ and\ \citenamefont
  {Prelovsek}}]{Ortiz-Pacheco:2023ble}%
  \BibitemOpen
  \bibfield  {author} {\bibinfo {author} {\bibfnamefont {E.}~\bibnamefont
  {Ortiz-Pacheco}}, \bibinfo {author} {\bibfnamefont {S.}~\bibnamefont
  {Collins}}, \bibinfo {author} {\bibfnamefont {L.}~\bibnamefont {Leskovec}},
  \bibinfo {author} {\bibfnamefont {M.}~\bibnamefont {Padmanath}}, \ and\
  \bibinfo {author} {\bibfnamefont {S.}~\bibnamefont {Prelovsek}},\ }\href
  {\doibase 10.22323/1.453.0052} {\  (\bibinfo {year} {2023}),\
  10.22323/1.453.0052},\ \Eprint {http://arxiv.org/abs/2312.13441}
  {arXiv:2312.13441 [hep-lat]} \BibitemShut {NoStop}%
\bibitem [{\citenamefont {Collins}\ \emph {et~al.}(2024)\citenamefont
  {Collins}, \citenamefont {Nefediev}, \citenamefont {Padmanath},\ and\
  \citenamefont {Prelovsek}}]{Collins:2024sfi}%
  \BibitemOpen
  \bibfield  {author} {\bibinfo {author} {\bibfnamefont {S.}~\bibnamefont
  {Collins}}, \bibinfo {author} {\bibfnamefont {A.}~\bibnamefont {Nefediev}},
  \bibinfo {author} {\bibfnamefont {M.}~\bibnamefont {Padmanath}}, \ and\
  \bibinfo {author} {\bibfnamefont {S.}~\bibnamefont {Prelovsek}},\ }\href@noop
  {} {\  (\bibinfo {year} {2024})},\ \Eprint {http://arxiv.org/abs/2402.14715}
  {arXiv:2402.14715 [hep-lat]} \BibitemShut {NoStop}%
\bibitem [{\citenamefont {Whyte}\ \emph {et~al.}(2024)\citenamefont {Whyte},
  \citenamefont {Wilson},\ and\ \citenamefont {Thomas}}]{Whyte:2024ihh}%
  \BibitemOpen
  \bibfield  {author} {\bibinfo {author} {\bibfnamefont {T.}~\bibnamefont
  {Whyte}}, \bibinfo {author} {\bibfnamefont {D.~J.}\ \bibnamefont {Wilson}}, \
  and\ \bibinfo {author} {\bibfnamefont {C.~E.}\ \bibnamefont {Thomas}},\
  }\href@noop {} {\  (\bibinfo {year} {2024})},\ \Eprint
  {http://arxiv.org/abs/2405.15741} {arXiv:2405.15741 [hep-lat]} \BibitemShut
  {NoStop}%
\bibitem [{\citenamefont {Aoki}\ and\ \citenamefont
  {Aoki}(2023)}]{Aoki:2022xxq}%
  \BibitemOpen
  \bibfield  {author} {\bibinfo {author} {\bibfnamefont {S.}~\bibnamefont
  {Aoki}}\ and\ \bibinfo {author} {\bibfnamefont {T.}~\bibnamefont {Aoki}},\
  }\href {\doibase 10.22323/1.430.0049} {\bibfield  {journal} {\bibinfo
  {journal} {PoS}\ }\textbf {\bibinfo {volume} {LATTICE2022}},\ \bibinfo
  {pages} {049} (\bibinfo {year} {2023})},\ \Eprint
  {http://arxiv.org/abs/2212.00202} {arXiv:2212.00202 [hep-lat]} \BibitemShut
  {NoStop}%
\bibitem [{\citenamefont {Hudspith}(2015)}]{Hudspith:2014oja}%
  \BibitemOpen
  \bibfield  {author} {\bibinfo {author} {\bibfnamefont {R.~J.}\ \bibnamefont
  {Hudspith}} (\bibinfo {collaboration} {RBC, UKQCD}),\ }\href {\doibase
  10.1016/j.cpc.2014.10.017} {\bibfield  {journal} {\bibinfo  {journal}
  {Comput. Phys. Commun.}\ }\textbf {\bibinfo {volume} {187}},\ \bibinfo
  {pages} {115} (\bibinfo {year} {2015})},\ \Eprint
  {http://arxiv.org/abs/1405.5812} {arXiv:1405.5812 [hep-lat]} \BibitemShut
  {NoStop}%
\bibitem [{\citenamefont {Hudspith}\ and\ \citenamefont
  {Mohler}(2022)}]{Hudspith:2021iqu}%
  \BibitemOpen
  \bibfield  {author} {\bibinfo {author} {\bibfnamefont {R.~J.}\ \bibnamefont
  {Hudspith}}\ and\ \bibinfo {author} {\bibfnamefont {D.}~\bibnamefont
  {Mohler}},\ }\href {\doibase 10.1103/PhysRevD.106.034508} {\bibfield
  {journal} {\bibinfo  {journal} {Phys. Rev. D}\ }\textbf {\bibinfo {volume}
  {106}},\ \bibinfo {pages} {034508} (\bibinfo {year} {2022})},\ \Eprint
  {http://arxiv.org/abs/2112.01997} {arXiv:2112.01997 [hep-lat]} \BibitemShut
  {NoStop}%
\bibitem [{\citenamefont {Aoki}\ \emph
  {et~al.}(2009{\natexlab{a}})\citenamefont {Aoki} \emph
  {et~al.}}]{Aoki:2008sm}%
  \BibitemOpen
  \bibfield  {author} {\bibinfo {author} {\bibfnamefont {S.}~\bibnamefont
  {Aoki}} \emph {et~al.} (\bibinfo {collaboration} {PACS-CS}),\ }\href
  {\doibase 10.1103/PhysRevD.79.034503} {\bibfield  {journal} {\bibinfo
  {journal} {Phys. Rev. D}\ }\textbf {\bibinfo {volume} {79}},\ \bibinfo
  {pages} {034503} (\bibinfo {year} {2009}{\natexlab{a}})},\ \Eprint
  {http://arxiv.org/abs/0807.1661} {arXiv:0807.1661 [hep-lat]} \BibitemShut
  {NoStop}%
\bibitem [{\citenamefont {Aoki}\ \emph {et~al.}(2010)\citenamefont {Aoki} \emph
  {et~al.}}]{Aoki:2009ix}%
  \BibitemOpen
  \bibfield  {author} {\bibinfo {author} {\bibfnamefont {S.}~\bibnamefont
  {Aoki}} \emph {et~al.} (\bibinfo {collaboration} {PACS-CS}),\ }\href
  {\doibase 10.1103/PhysRevD.81.074503} {\bibfield  {journal} {\bibinfo
  {journal} {Phys. Rev. D}\ }\textbf {\bibinfo {volume} {81}},\ \bibinfo
  {pages} {074503} (\bibinfo {year} {2010})},\ \Eprint
  {http://arxiv.org/abs/0911.2561} {arXiv:0911.2561 [hep-lat]} \BibitemShut
  {NoStop}%
\bibitem [{\citenamefont {Aoki}\ \emph
  {et~al.}(2009{\natexlab{b}})\citenamefont {Aoki} \emph
  {et~al.}}]{PACS-CS:2008bkb}%
  \BibitemOpen
  \bibfield  {author} {\bibinfo {author} {\bibfnamefont {S.}~\bibnamefont
  {Aoki}} \emph {et~al.} (\bibinfo {collaboration} {PACS-CS}),\ }\href
  {\doibase 10.1103/PhysRevD.79.034503} {\bibfield  {journal} {\bibinfo
  {journal} {Phys. Rev. D}\ }\textbf {\bibinfo {volume} {79}},\ \bibinfo
  {pages} {034503} (\bibinfo {year} {2009}{\natexlab{b}})},\ \Eprint
  {http://arxiv.org/abs/0807.1661} {arXiv:0807.1661 [hep-lat]} \BibitemShut
  {NoStop}%
\bibitem [{\citenamefont {Lang}\ \emph {et~al.}(2014)\citenamefont {Lang},
  \citenamefont {Leskovec}, \citenamefont {Mohler}, \citenamefont {Prelovsek},\
  and\ \citenamefont {Woloshyn}}]{Lang:2014yfa}%
  \BibitemOpen
  \bibfield  {author} {\bibinfo {author} {\bibfnamefont {C.~B.}\ \bibnamefont
  {Lang}}, \bibinfo {author} {\bibfnamefont {L.}~\bibnamefont {Leskovec}},
  \bibinfo {author} {\bibfnamefont {D.}~\bibnamefont {Mohler}}, \bibinfo
  {author} {\bibfnamefont {S.}~\bibnamefont {Prelovsek}}, \ and\ \bibinfo
  {author} {\bibfnamefont {R.~M.}\ \bibnamefont {Woloshyn}},\ }\href {\doibase
  10.1103/PhysRevD.90.034510} {\bibfield  {journal} {\bibinfo  {journal} {Phys.
  Rev. D}\ }\textbf {\bibinfo {volume} {90}},\ \bibinfo {pages} {034510}
  (\bibinfo {year} {2014})},\ \Eprint {http://arxiv.org/abs/1403.8103}
  {arXiv:1403.8103 [hep-lat]} \BibitemShut {NoStop}%
\bibitem [{\citenamefont {Namekawa}\ \emph {et~al.}(2011)\citenamefont
  {Namekawa} \emph {et~al.}}]{PACS-CS:2011ngu}%
  \BibitemOpen
  \bibfield  {author} {\bibinfo {author} {\bibfnamefont {Y.}~\bibnamefont
  {Namekawa}} \emph {et~al.} (\bibinfo {collaboration} {PACS-CS}),\ }\href
  {\doibase 10.1103/PhysRevD.84.074505} {\bibfield  {journal} {\bibinfo
  {journal} {Phys. Rev. D}\ }\textbf {\bibinfo {volume} {84}},\ \bibinfo
  {pages} {074505} (\bibinfo {year} {2011})},\ \Eprint
  {http://arxiv.org/abs/1104.4600} {arXiv:1104.4600 [hep-lat]} \BibitemShut
  {NoStop}%
\bibitem [{\citenamefont {Luscher}\ and\ \citenamefont
  {Schaefer}(2013)}]{Luscher:2012av}%
  \BibitemOpen
  \bibfield  {author} {\bibinfo {author} {\bibfnamefont {M.}~\bibnamefont
  {Luscher}}\ and\ \bibinfo {author} {\bibfnamefont {S.}~\bibnamefont
  {Schaefer}},\ }\href {\doibase 10.1016/j.cpc.2012.10.003} {\bibfield
  {journal} {\bibinfo  {journal} {Comput. Phys. Commun.}\ }\textbf {\bibinfo
  {volume} {184}},\ \bibinfo {pages} {519} (\bibinfo {year} {2013})},\ \Eprint
  {http://arxiv.org/abs/1206.2809} {arXiv:1206.2809 [hep-lat]} \BibitemShut
  {NoStop}%
\bibitem [{ope()}]{openQCD}%
  \BibitemOpen
  \href@noop {} {}\bibinfo {howpublished}
  {\url{http://luscher.web.cern.ch/luscher/openQCD/.}}\BibitemShut {Stop}%
\bibitem [{\citenamefont {Luscher}(2007)}]{Luscher:2007se}%
  \BibitemOpen
  \bibfield  {author} {\bibinfo {author} {\bibfnamefont {M.}~\bibnamefont
  {Luscher}},\ }\href {\doibase 10.1088/1126-6708/2007/07/081} {\bibfield
  {journal} {\bibinfo  {journal} {JHEP}\ }\textbf {\bibinfo {volume} {07}},\
  \bibinfo {pages} {081} (\bibinfo {year} {2007})},\ \Eprint
  {http://arxiv.org/abs/0706.2298} {arXiv:0706.2298 [hep-lat]} \BibitemShut
  {NoStop}%
\bibitem [{\citenamefont {Hasenbusch}\ and\ \citenamefont
  {Jansen}(2003)}]{Hasenbusch:2002ai}%
  \BibitemOpen
  \bibfield  {author} {\bibinfo {author} {\bibfnamefont {M.}~\bibnamefont
  {Hasenbusch}}\ and\ \bibinfo {author} {\bibfnamefont {K.}~\bibnamefont
  {Jansen}},\ }\href {\doibase 10.1016/S0550-3213(03)00227-X} {\bibfield
  {journal} {\bibinfo  {journal} {Nucl. Phys. B}\ }\textbf {\bibinfo {volume}
  {659}},\ \bibinfo {pages} {299} (\bibinfo {year} {2003})},\ \Eprint
  {http://arxiv.org/abs/hep-lat/0211042} {arXiv:hep-lat/0211042} \BibitemShut
  {NoStop}%
\bibitem [{\citenamefont {Brower}\ \emph {et~al.}(1997)\citenamefont {Brower},
  \citenamefont {Ivanenko}, \citenamefont {Levi},\ and\ \citenamefont
  {Orginos}}]{Brower:1995vx}%
  \BibitemOpen
  \bibfield  {author} {\bibinfo {author} {\bibfnamefont {R.~C.}\ \bibnamefont
  {Brower}}, \bibinfo {author} {\bibfnamefont {T.}~\bibnamefont {Ivanenko}},
  \bibinfo {author} {\bibfnamefont {A.~R.}\ \bibnamefont {Levi}}, \ and\
  \bibinfo {author} {\bibfnamefont {K.~N.}\ \bibnamefont {Orginos}},\ }\href
  {\doibase 10.1016/S0550-3213(96)00579-2} {\bibfield  {journal} {\bibinfo
  {journal} {Nucl. Phys. B}\ }\textbf {\bibinfo {volume} {484}},\ \bibinfo
  {pages} {353} (\bibinfo {year} {1997})},\ \Eprint
  {http://arxiv.org/abs/hep-lat/9509012} {arXiv:hep-lat/9509012} \BibitemShut
  {NoStop}%
\bibitem [{\citenamefont {Clark}\ and\ \citenamefont
  {Kennedy}(2004)}]{Clark:2003na}%
  \BibitemOpen
  \bibfield  {author} {\bibinfo {author} {\bibfnamefont {M.~A.}\ \bibnamefont
  {Clark}}\ and\ \bibinfo {author} {\bibfnamefont {A.~D.}\ \bibnamefont
  {Kennedy}},\ }\href {\doibase 10.1016/S0920-5632(03)02732-4} {\bibfield
  {journal} {\bibinfo  {journal} {Nucl. Phys. B Proc. Suppl.}\ }\textbf
  {\bibinfo {volume} {129}},\ \bibinfo {pages} {850} (\bibinfo {year}
  {2004})},\ \Eprint {http://arxiv.org/abs/hep-lat/0309084}
  {arXiv:hep-lat/0309084} \BibitemShut {NoStop}%
\bibitem [{\citenamefont {Frezzotti}\ and\ \citenamefont
  {Jansen}(1999)}]{Frezzotti:1998eu}%
  \BibitemOpen
  \bibfield  {author} {\bibinfo {author} {\bibfnamefont {R.}~\bibnamefont
  {Frezzotti}}\ and\ \bibinfo {author} {\bibfnamefont {K.}~\bibnamefont
  {Jansen}},\ }\href {\doibase 10.1016/S0550-3213(99)00321-1} {\bibfield
  {journal} {\bibinfo  {journal} {Nucl. Phys. B}\ }\textbf {\bibinfo {volume}
  {555}},\ \bibinfo {pages} {395} (\bibinfo {year} {1999})},\ \Eprint
  {http://arxiv.org/abs/hep-lat/9808011} {arXiv:hep-lat/9808011} \BibitemShut
  {NoStop}%
\bibitem [{\citenamefont {Lepage}\ \emph {et~al.}(2002)\citenamefont {Lepage},
  \citenamefont {Clark}, \citenamefont {Davies}, \citenamefont {Hornbostel},
  \citenamefont {Mackenzie}, \citenamefont {Morningstar},\ and\ \citenamefont
  {Trottier}}]{Lepage:2001ym}%
  \BibitemOpen
  \bibfield  {author} {\bibinfo {author} {\bibfnamefont {G.~P.}\ \bibnamefont
  {Lepage}}, \bibinfo {author} {\bibfnamefont {B.}~\bibnamefont {Clark}},
  \bibinfo {author} {\bibfnamefont {C.~T.~H.}\ \bibnamefont {Davies}}, \bibinfo
  {author} {\bibfnamefont {K.}~\bibnamefont {Hornbostel}}, \bibinfo {author}
  {\bibfnamefont {P.~B.}\ \bibnamefont {Mackenzie}}, \bibinfo {author}
  {\bibfnamefont {C.}~\bibnamefont {Morningstar}}, \ and\ \bibinfo {author}
  {\bibfnamefont {H.}~\bibnamefont {Trottier}} (\bibinfo {collaboration}
  {HPQCD}),\ }\href {\doibase 10.1016/S0920-5632(01)01638-3} {\bibfield
  {journal} {\bibinfo  {journal} {Nucl. Phys. B Proc. Suppl.}\ }\textbf
  {\bibinfo {volume} {106}},\ \bibinfo {pages} {12} (\bibinfo {year} {2002})},\
  \Eprint {http://arxiv.org/abs/hep-lat/0110175} {arXiv:hep-lat/0110175}
  \BibitemShut {NoStop}%
\bibitem [{\citenamefont {Lepage}(2021)}]{peter_lepage_2021_5733391}%
  \BibitemOpen
  \bibfield  {author} {\bibinfo {author} {\bibfnamefont {G.~P.}\ \bibnamefont
  {Lepage}},\ }\href {\doibase 10.5281/zenodo.5733391} {\enquote {\bibinfo
  {title} {Corrfitter version 8.2 (github.com/gplepage/corrfitter),
  10.5281/zenodo.5733391},}\ } (\bibinfo {year} {2021})\BibitemShut {NoStop}%
\bibitem [{\citenamefont {Lepage}\ and\ \citenamefont
  {Gohlke}(2023)}]{peter_lepage_2023_7931361}%
  \BibitemOpen
  \bibfield  {author} {\bibinfo {author} {\bibfnamefont {G.~P.}\ \bibnamefont
  {Lepage}}\ and\ \bibinfo {author} {\bibfnamefont {C.}~\bibnamefont
  {Gohlke}},\ }\href {\doibase 10.5281/zenodo.7931361} {\enquote {\bibinfo
  {title} {Lsqfit version 13.0.1 (github.com/gplepage/lsqfit),
  10.5281/zenodo.7931361},}\ } (\bibinfo {year} {2023})\BibitemShut {NoStop}%
\bibitem [{\citenamefont {Lepage}\ \emph {et~al.}(2023)\citenamefont {Lepage},
  \citenamefont {Gohlke},\ and\ \citenamefont
  {Hackett}}]{peter_lepage_2023_8025535}%
  \BibitemOpen
  \bibfield  {author} {\bibinfo {author} {\bibfnamefont {G.~P.}\ \bibnamefont
  {Lepage}}, \bibinfo {author} {\bibfnamefont {C.}~\bibnamefont {Gohlke}}, \
  and\ \bibinfo {author} {\bibfnamefont {D.}~\bibnamefont {Hackett}},\ }\href
  {\doibase 10.5281/zenodo.8025535} {\enquote {\bibinfo {title} {Gvar version
  11.11.11 (github.com/gplepage/gvar), 10.5281/zenodo.8025535},}\ } (\bibinfo
  {year} {2023})\BibitemShut {NoStop}%
\bibitem [{\citenamefont {Dowdall}\ \emph {et~al.}(2019)\citenamefont
  {Dowdall}, \citenamefont {Davies}, \citenamefont {Horgan}, \citenamefont
  {Lepage}, \citenamefont {Monahan}, \citenamefont {Shigemitsu},\ and\
  \citenamefont {Wingate}}]{Dowdall:2019bea}%
  \BibitemOpen
  \bibfield  {author} {\bibinfo {author} {\bibfnamefont {R.~J.}\ \bibnamefont
  {Dowdall}}, \bibinfo {author} {\bibfnamefont {C.~T.~H.}\ \bibnamefont
  {Davies}}, \bibinfo {author} {\bibfnamefont {R.~R.}\ \bibnamefont {Horgan}},
  \bibinfo {author} {\bibfnamefont {G.~P.}\ \bibnamefont {Lepage}}, \bibinfo
  {author} {\bibfnamefont {C.~J.}\ \bibnamefont {Monahan}}, \bibinfo {author}
  {\bibfnamefont {J.}~\bibnamefont {Shigemitsu}}, \ and\ \bibinfo {author}
  {\bibfnamefont {M.}~\bibnamefont {Wingate}},\ }\href {\doibase
  10.1103/PhysRevD.100.094508} {\bibfield  {journal} {\bibinfo  {journal}
  {Phys. Rev. D}\ }\textbf {\bibinfo {volume} {100}},\ \bibinfo {pages}
  {094508} (\bibinfo {year} {2019})},\ \Eprint
  {http://arxiv.org/abs/1907.01025} {arXiv:1907.01025 [hep-lat]} \BibitemShut
  {NoStop}%
\bibitem [{\citenamefont {D'Agostini}(1994)}]{DAGOSTINI1994306}%
  \BibitemOpen
  \bibfield  {author} {\bibinfo {author} {\bibfnamefont {G.}~\bibnamefont
  {D'Agostini}},\ }\href {\doibase
  https://doi.org/10.1016/0168-9002(94)90719-6} {\bibfield  {journal} {\bibinfo
   {journal} {Nuclear Instruments and Methods in Physics Research Section A:
  Accelerators, Spectrometers, Detectors and Associated Equipment}\ }\textbf
  {\bibinfo {volume} {346}},\ \bibinfo {pages} {306} (\bibinfo {year}
  {1994})}\BibitemShut {NoStop}%
\bibitem [{\citenamefont {Workman}\ and\ \citenamefont
  {Others}(2022)}]{Workman:2022ynf}%
  \BibitemOpen
  \bibfield  {author} {\bibinfo {author} {\bibfnamefont {R.~L.}\ \bibnamefont
  {Workman}}\ and\ \bibinfo {author} {\bibnamefont {Others}} (\bibinfo
  {collaboration} {Particle Data Group}),\ }\href {\doibase
  10.1093/ptep/ptac097} {\bibfield  {journal} {\bibinfo  {journal} {PTEP}\
  }\textbf {\bibinfo {volume} {2022}},\ \bibinfo {pages} {083C01} (\bibinfo
  {year} {2022})}\BibitemShut {NoStop}%
\bibitem [{\citenamefont {Blusk}()}]{blusklhcbworkshop:2021}%
  \BibitemOpen
  \bibfield  {author} {\bibinfo {author} {\bibfnamefont {S.}~\bibnamefont
  {Blusk}},\ }\href@noop {} {\bibinfo  {journal} {{\it Beyond $T_{cc}$:
  $T_{bc}$ and $T_{bb}$}, LHCb Online Mini-Workshop: $T_{cc}$ and Beyond,
  September 14, 2021}\ }\BibitemShut {NoStop}%
\bibitem [{\citenamefont {Gershon}\ and\ \citenamefont
  {Poluektov}(2019)}]{Gershon:2018gda}%
  \BibitemOpen
\bibfield  {journal} {  }\bibfield  {author} {\bibinfo {author} {\bibfnamefont
  {T.}~\bibnamefont {Gershon}}\ and\ \bibinfo {author} {\bibfnamefont
  {A.}~\bibnamefont {Poluektov}},\ }\href {\doibase 10.1007/JHEP01(2019)019}
  {\bibfield  {journal} {\bibinfo  {journal} {JHEP}\ }\textbf {\bibinfo
  {volume} {01}},\ \bibinfo {pages} {019} (\bibinfo {year} {2019})},\ \Eprint
  {http://arxiv.org/abs/1810.06657} {arXiv:1810.06657 [hep-ph]} \BibitemShut
  {NoStop}%
\bibitem [{\citenamefont {Khachatryan}\ \emph {et~al.}(2017)\citenamefont
  {Khachatryan} \emph {et~al.}}]{CMS:2016liw}%
  \BibitemOpen
  \bibfield  {author} {\bibinfo {author} {\bibfnamefont {V.}~\bibnamefont
  {Khachatryan}} \emph {et~al.} (\bibinfo {collaboration} {CMS}),\ }\href
  {\doibase 10.1007/JHEP05(2017)013} {\bibfield  {journal} {\bibinfo  {journal}
  {JHEP}\ }\textbf {\bibinfo {volume} {05}},\ \bibinfo {pages} {013} (\bibinfo
  {year} {2017})},\ \Eprint {http://arxiv.org/abs/1610.07095} {arXiv:1610.07095
  [hep-ex]} \BibitemShut {NoStop}%
\bibitem [{\citenamefont {Sirunyan}\ \emph {et~al.}(2020)\citenamefont
  {Sirunyan} \emph {et~al.}}]{CMS:2020qwa}%
  \BibitemOpen
  \bibfield  {author} {\bibinfo {author} {\bibfnamefont {A.~M.}\ \bibnamefont
  {Sirunyan}} \emph {et~al.} (\bibinfo {collaboration} {CMS}),\ }\href
  {\doibase 10.1016/j.physletb.2020.135578} {\bibfield  {journal} {\bibinfo
  {journal} {Phys. Lett. B}\ }\textbf {\bibinfo {volume} {808}},\ \bibinfo
  {pages} {135578} (\bibinfo {year} {2020})},\ \Eprint
  {http://arxiv.org/abs/2002.06393} {arXiv:2002.06393 [hep-ex]} \BibitemShut
  {NoStop}%
\end{thebibliography}%

\end{document}